\documentclass[showpacs,twocolumn,aps,prb]{revtex4-1}
\usepackage{graphicx,amsmath,amsfonts,amssymb}
\usepackage{subfigure}
\usepackage{bm}

\usepackage[
colorlinks=true,
linkcolor=blue,
urlcolor=blue,
citecolor=blue,
]{hyperref}

\begin{document}
	\title{Extrinsic nonlinear acoustic valley Hall effect in the massive Dirac materials}
	\author{Jia-Liang Wan$^{1}$}
	\author{Ying-Li Wu$^{1}$}
	\author{Ke-Qiu Chen$^{1}$}
	\author{Xiao-Qin Yu$^{1}$}
	\email{yuxiaoqin@hnu.edu.cn}
	\affiliation{$^{1}$ School of Physics and Electronics, Hunan University, Changsha 410082, China}
	

	\begin{abstract}

The nonlinear acoustic valley Hall effect (AVHE), a recently discovered novel acoustically driven phenomena, has sparked extensive interests in valleytronics.
So far, only the intrinsic contributions from band structure (Berry curvature or asymmetric
energy dispersions) to nonlinear AVHE have been investigated. Here, we theoretically investigate the nonlinear AVHE from both intrinsic and extrinsic contributions in two-dimensional (2D) hexagonal massive Dirac materials with disorders based on the Boltzmann formalism and also concretely analyse the behaviours of nonlinear AVHE in disordered monolayer MoS$_2$. It's found that the extrinsic contributions (side jump and skew scattering) can also give rise to a \textit{pure} nonlinear AVHE in the 2D hexagonal massive Dirac materials. Remarkably, the extrinsic mechanisms dominate the nonlinear AVHE in the disordered monolayer MoS$_{2}$.
	
	\end{abstract}
\maketitle
\section{Introduction}\label{introduction}
The valley, an extra degree of freedom of electron, in two-dimensional (2D) crystal with honeycomb
lattice structure shows a potential to store and carry information and gives birth to a tremendously active field: valleytronics~\cite{review2016valleytronics,review2018valleytronics}, whose primary focus is to control and manipulate the valley degree of freedom efficiently. Various kinds of means, including the electrical, optical, thermal and magnetic ones, have been explored to manipulate the valley. Recently, the resuscitation of the acoustoelectric  effect (AEE) \cite{Sukhachov2020,Bhalla2022,Zhao2022PRL,Mou2024NL,Su2024,Parm1953,Weinreich1957} in low-dimensional systems (LDS) brings new opportunities to valleytronics.

 AEE describes the phenomenon in which the surface acoustic wave (SAW, a mechanical wave) drives the carrier, rather than electric field or temperature gradient~\cite{Parm1953,Weinreich1957}.
 Apart from the traditional deformation potential mechanism~\cite{Parm1953,Weinreich1957,Weinreich1959}, current works about AEE in LDS~\cite{Willett1990,Fal1993,Hern2018,PerDelsing2019Rev,Zhang2011} focus on the piezoelectric mechanism of interaction between SAWs and electrons. When propagating the SAWs through LDS placing on the piezoelectric substrate, the generated effective overall electric field, which includes an in-plane piezoelectric field from the piezoelectric substrate and an induced electric field stemming from the fluctuations of the electron density, will drag carriers and give rise to acoustoelectric (AE) current. Recent studies show that, in addition to electron, the valley degree of freedom can also  be \textit{acoustically} manipulated and a nonlinear valley current would be generated through  propagating SAWs in 2D hexagonal Dirac materials\cite{Kalameitsev2019PRL,Sonowal2020PRB,Ominato2022PRB,Wan2024PRB}, attracting great attentions in valleytronics.

\begin{figure}[ht]
	\centering
	\subfigure{
	\includegraphics[width=1\linewidth]{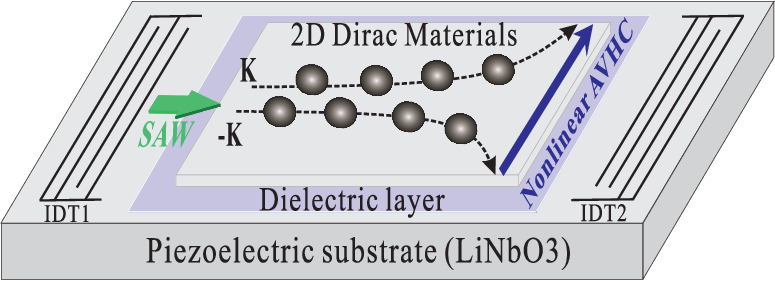}}
\caption{Illustration of the generation of the nonlinear VAHE through SAW in the time-reversal symmetric 2D Dirac materials.}\label{Fig1}
		\end{figure}

 A new valley AEE, namely the nonlinear acoustic valley Hall effect (AVHE) [Fig.~\ref{Fig1}], referring to the generation of a nonlinear transverse valley current as a second-order response to the longitudinal SAW-induced field, was firstly predicted in TMDs placed on a piezoelectric substrate and identified to stem from the warping effect of the Fermi surface and nontrival Berry phase \cite{Kalameitsev2019PRL}. Subsequently, the nonlinear AVHE has also been found to exit in the strained graphene monolayer owing to the band tilting effect of Dirac cones \cite{Wan2024PRB}. Those works only focus on the intrinsic contributions (Berry curvature or asymmetric energy dispersion) from the band structure and have not consider the extrinsic mechanisms (side jump and skew scattering) from the disorder, which have been confirmed to also play an significant role in nonlinear Hall transports, such as the nonlinear Hall effect\cite{ xc2019PRB,du2019disorder,du2021quantum}, nonlinear thermal Hall effect\cite{Zhou2022NTHE} and the nonlinear Nernst effect\cite{Papaj2021PRB}. We notice a recently reported generic valley Hall effect driving by photon (electromagnetic wave) \cite{GlazovPRB2020} has investigated both intrinsic and extrinsic contributions for a parabolic energy dispersion and might describe the nonlinear AVHE approximately but only when the dielectric screening can be neglected. 
 In AEE experiment, the screening effect often needs to be considered owing to the presence of dielectric materials.

In this paper, we will derive the general formulas of the nonlinear response coefficients for the AEE based on the Boltzmann formalism in presence of the disorder with considering the screening effect and the both drift and diffusive current. Both intrinsic and extrinsic contributions to nonlinear AEE will be taken into account. The formulas can be applied to different models (including the asymmetric energy dispersion) to calculate various nonlinear AEEs. We apply the formulas to a tiny model (i.e. massive Dirac model) to investigate the nonlinear AVHE in 2D hexagonal massive Dirac materials and also concretely analyse the behaviours of nonlinear AVHE in disordered monolayer MoS$_{2}$, a typical 2D hexagonal Dirac material with massive Dirac cones. Our study shows that extrinsic mechanisms can also give rise to a pure nonlinear AVHE in monolayer MoS$_{2}$ in presence of disorder and the signal of the nonlinear AVHE stemming from the side-jump contribution is even twice larger than that from the intrinsic contribution. 
Additionally, the sign of AVHE from side-jump contribution is opposite to those from both intrinsic and skew-scattering contributions.

The paper is organized as follows. Within the semiclassical framework of the electron dynamics, the general formulas of the nonlinear AEE response coefficients with considering the disorder scattering are derived in Sec.~\ref{Theoretical review}. Through the derived formulas, the nonlinear acoustic valley Hall coefficients (NAVHC), which quantize the nonlinear AVHE, from intrinsic, side-jump and skew-scattering contributions are determined for the 2D massive Dirac model in Sec.~\ref{NAVHE}, respectively.  The behaviors of NAVHC in the monolayer  $\mathrm{MoS_2}$ originating from both intrinsic and extrinsic mechanisms are analyzed in Sec.\ref{R and D}. Finally, we give a conclusion in Sec.\ref{conlusions}.

\section{The Theoretical derivation}\label{Theoretical review}
 When taking into account the extrinsic contributions (side jump and skew scattering) originating from the disorder scattering on the electron transport properties, it is convenient to adopt the semiclassical framework of the electron dynamics based on Boltzmann kinetics\cite{du2019disorder,isobe2020SA,Qiang2023PRB}. We start by deriving the nonequilibrium Fermi distribution function (NFDF) $ f_{\mathrm{\mathbf{k}}}\equiv f_{\mathrm{\mathbf{k}}}(\mathbf{r},t)$ through the Boltzmann transport equation (BTE). In presence of a SAW-induced electric field and the disorder scattering, BTE can be expressed as
\begin{align}
	\frac{\partial f_{\mathrm{\mathbf{k}}}}{\partial t}+\dot{\mathbf{r}}\cdot\frac{\partial f_{\mathrm{\mathbf{k}}}}{\partial\mathbf{r}}+\dot{\mathbf{k}}\cdot\frac{\partial f_{\mathrm{\mathbf{k}}}}{\partial\mathbf{k}}=I_{c}[f_{\mathrm{\mathbf{k}}}]\label{BTE}
\end{align}
with
\begin{equation}
	\ensuremath{\dot{\mathbf{r}}}=  \mathbf{v}_{\mathbf{k}}-\dot{\mathbf{k}}\times\bm{\Omega}_{\mathbf{k}}+\mathbf{v}_{\mathbf{k}}^
{\mathrm{sj}},\,\,\,\,\,
	\dot{\mathbf{k}}=-\frac{e}{\hbar}\tilde{\mathbf{E}}(\mathbf{r},t),
\label{rk}
\end{equation}	
where $e$ and $\hbar$ are electron charge and the Plank constant, respectively, $\mathbf{v}_{\mathbf{k}}=\partial \varepsilon_\mathbf{k}/(\hbar\partial\mathbf{k})$ refers to the band velocity, $\bm{\Omega}_\mathbf{k}$ denotes the Berry curvature, $\mathbf{v}_{\mathbf{k}}^{\mathrm{sj}}$ (where superscript ``{sj}" refers to side jump) represents the side-jump velocity originating from
the disorder-induced coordinate shift $\delta \mathbf{r_{kk^\prime}}$ [Eq.~\eqref{App-B-shif}], $\tilde{\mathbf{E}}(\mathbf{r},t)=\frac{1}{2}\vec{\tilde{\mathcal{E}}}e^{i(\mathbf{q}\cdot\mathbf{r}-\omega t)}+c.c$ represents the SAW-induced overall electric field, which includes the in-plane piezoelectric field ${\mathbf{E}}(\mathbf{r},t)$ and the induced electric
field $\mathbf{E}_{i}(\mathbf{r},t)$, with $\vec{\tilde{\mathcal{E}}}$ indicating amplitude vector, and the collision term $I_{c}[f_{\mathrm{\mathbf{k}}}]$ indicates the elastic disorder scattering from static defects or impurities and can be expressed as
\begin{align}
	I_{c}[f_{\mathrm{\mathbf{k}}}]=-\sum_{\mathbf{k}^{\prime}}
\left[w_{\mathbf{k}\mathbf{k}^{\prime}}f_{\mathrm{\mathbf{k}}}-
w_{\mathbf{k}^{\prime}\mathbf{k}}f_{\mathrm{\mathbf{k}}^{\prime}}\right],\label{Ictotal}
\end{align}
where $w_{\mathbf{k}\mathbf{k}^{\prime}}$ presents the scattering rate from the initial state $\mathbf{k}$ into final state $\mathbf{k^\prime}$, which can be determined by the Fermi golden rule: $w_{\mathbf{k}\mathbf{k}^{\prime}}= \frac{2\pi}{\hbar}|T_{\mathbf{k}\mathbf{k}^{\prime}}|^{2}\delta
 (\varepsilon_{\mathbf{k}}-\varepsilon_{\mathbf{k}^{\prime}})$ with $T_{\mathbf{k}\mathbf{k}^{\prime}}$ indicating scattering matrix. Generally, the scattering rate  $w_{\mathbf{k}\mathbf{k}^{\prime}}$ is not symmetric when interchanging the initial and final states  i.e., $w_{\mathbf{k}^{\prime}\mathbf{k}}\neq w_{\mathbf{k}\mathbf{k}^{\prime}}$. Therefore, it is convenient to formally decompose $w_{\mathbf{k}\mathbf{k}^{\prime}}$ into a symmetric part $w_{\mathbf{kk^\prime}}^{\mathrm{S}}$ and a antisymmetrical part $w_{\mathbf{kk^\prime}}^{\mathrm{A}}$ as
\begin{align}
w_{\mathbf{k}\mathbf{k}^{\prime}}={w}_{\mathbf{kk^\prime}}^{\mathrm{S}}
+w_{\mathbf{kk^\prime}}^{\mathrm{A}},\label{Wkk}
\end{align}
with ${w}_{\mathbf{kk^\prime}}^{\mathrm{S}}=
(w_{\mathbf{k}\mathbf{k}^{\prime}}+
w_{\mathbf{k}^{\prime}\mathbf{k}})/2={w}_{\mathbf{k^\prime}\mathbf{k}}^{\mathrm{S}}$ and
${w}_{\mathbf{kk^\prime}}^{\mathrm{A}}=(w_{\mathbf{k}\mathbf{k}^{\prime}}-
w_{\mathbf{k}^{\prime}\mathbf{k}})/2=-{w}_{\mathbf{k^\prime}\mathbf{k}}^{\mathrm{A}}$.
The antisymmetric scattering rate $w_{\mathbf{kk^\prime}}^{\mathrm{A}}$ is, actually, responsible for the skew-scattering process.
In addition, owing to the side jump effect, an extra modification [$e\tilde{\mathbf{E}}
(\mathbf{r},t)\cdot\mathbf{O}_{\mathbf{k}\mathbf{k}^{\prime}}$] should be taken into
the symmetric scattering rate $w_{\mathbf{kk^\prime}}^{\mathrm{S}}$ as~\cite{du2019disorder,Qiang2023PRB}
\begin{align}
w_{\mathbf{k}\mathbf{k}^{\prime}}^{\mathrm{S}}\Rightarrow	\tilde{w}_{\mathbf{k}\mathbf{k}^{\prime}}^{\mathrm{S}}\approx w_{\mathbf{k}\mathbf{k}^{\prime}}^{\mathrm{S}}+e\tilde{\mathbf{E}}
(\mathbf{r},t)\cdot\mathbf{O}_{\mathbf{k}\mathbf{k}^{\prime}}\label{WSkk}
\end{align}
with $\mathbf{O}_{\mathbf{k}\mathbf{k}^{\prime}}= \frac{2\pi}{\hbar}|T_{\mathbf{k}\mathbf{k}^{\prime}}|^{2}\delta\mathbf{r}_{\mathbf{k}
 \mathbf{k}^{\prime}}\frac{\partial}{\partial\varepsilon_{\mathbf{k}}}\delta
 (\varepsilon_{\mathbf{k}}-\varepsilon_{\mathbf{k}^{\prime}})$.
Therefore, based on Eqs.~\eqref{Ictotal}\eqref{Wkk}\eqref{WSkk}, the collision term can be approximately decomposed into the intrinsic (in), side-jump (sj), and skew-scattering (sk) parts as
\begin{align}
	I_{c}[f_{\mathrm{\mathbf{k}}}]=  I_{c}^{\mathrm{in}}[f_{\mathrm{\mathbf{k}}}]+I_{c}^{\mathrm{sj}}[f_{\mathrm{\mathbf{k}}}]+I_{c}^{\mathrm{sk}}[f_{\mathrm{\mathbf{k}}}]
\end{align}
with
\begin{align}
	I_{c}^{\mathrm{in}}[f_{\mathrm{\mathbf{k}}}]= & -\sum_{\mathbf{k}^{\prime}}w_{\mathbf{k}\mathbf{k}^{\prime}}^{\mathrm{S}}
(f_{\mathrm{\mathbf{k}}}-f_{\mathrm{\mathbf{k}}^{\prime}}),\notag\\
	I_{c}^{\mathrm{sj}}[f_{\mathrm{\mathbf{k}}}]= & -\sum_{\mathbf{k}^{\prime}}e\mathbf{\tilde{E}}(\mathbf{r},t)
\cdot\mathbf{O}_{\mathbf{k}\mathbf{k}^{\prime}}
(f_{\mathrm{\mathbf{k}}}-f_{\mathrm{\mathbf{k}}^{\prime}}),\notag\\
	I_{c}^{\mathrm{sk}}[f_{\mathrm{\mathbf{k}}}]= & -\sum_{\mathbf{k}^{\prime}}w_{\mathbf{k}^{\prime}\mathbf{k}}^{\mathrm{A}}
(f_{\mathrm{\mathbf{k}}}+f_{\mathrm{\mathbf{k}}^{\prime}}).\label{IcInSjSk}
\end{align}
Furthermore, when formally decomposing the Fermi distribution function as $f_\mathbf{k}=f_\mathbf{k}^{\mathrm{in}}+\delta f_\mathbf{k}^{\mathrm{sj}}+\delta f_\mathbf{k}^{\mathrm{sk}}$ in the presence of the SAW-induced electric field and disorder, where $\delta f_\mathbf{k}^{\mathrm{sj}}$ /$\delta f_\mathbf{k}^{\mathrm{sk}}$  is the side-jump/ skew-scattering induced modificaiton, the BTE in Eq.~\eqref{BTE} can be, consequently, divided into three parts as

\begin{equation}
\begin{aligned}
\left[\partial_{t}+\mathbf{v}_{\mathbf{k}}\cdot\partial_{\mathbf{r}}+\dot{\mathbf{k}}
\cdot\partial_{\mathbf{k}}\right]f_{\mathrm{\mathbf{k}}}^{\mathrm{in}}
&=I_{c}^{\mathrm{in}}[ f_{\mathrm{\mathbf{k}}}^{\mathrm{in}}],\\
\left[\partial_{t}+\left(\mathbf{v}_{\mathbf{k}}+\mathbf{v}_{\mathbf{k}}^{\mathrm{sj}}\right)
\cdot\partial_{\mathbf{r}}+\dot{\mathbf{k}}\cdot\partial_{\mathbf{k}}
\right]\delta  f_{\mathrm{\mathbf{k}}}^{\mathrm{sj}}&=I_{c}^{\mathrm{in}}[\delta f_{\mathrm{\mathbf{k}}}^{\mathrm{sj}}]
+I_{c}^{\mathrm{sj}}
[f_{\mathrm{\mathbf{k}}}^{\mathrm{in}}],\\
\left[\partial_{t}+\mathbf{v}_{\mathbf{k}}\cdot\partial_{\mathbf{r}}+
\dot{\mathbf{k}}\cdot\partial_{\mathbf{k}}\right]\delta f_{\mathrm{\mathbf{k}}}^{\mathrm{sk}}
&=I_{c}^{\mathrm{in}}[\delta f_{\mathrm{\mathbf{k}}}^{\mathrm{sk}}]+I_{c}^{\mathrm{sk}}
[f_{\mathrm{\mathbf{k}}}^{\mathrm{in}}],
\end{aligned}
\label{BTE-three}
\end{equation}
where the terms mixing side-jump and skew-scattering contributions have been neglected. We are interested in the response up to the second order in the SAW-induced overall electric field $\tilde{\mathbf{E}}(\mathbf{r},t)$ and, hence, have approximated the nonequilibrium distribution functions as $f_{\mathrm{\mathbf{k}}}^{\mathrm{in}}=  f_{\mathrm{\mathbf{k}}}^{(0)}+\delta f_{\mathrm{\mathbf{k}}}^{\mathrm{in},1}+
\delta f_{\mathrm{\mathbf{k}}}^{\mathrm{in},2}$,
$\delta f_{\mathrm{\mathbf{k}}}^{\mathrm{sj}}=\delta f_{\mathrm{\mathbf{k}}}^{\mathrm{sj,1}}+\delta f_{\mathrm{\mathbf{k}}}^{\mathrm{sj,2}}$ and $\delta f_{\mathrm{\mathbf{k}}}^{\mathrm{sk}}=\delta f_{\mathrm{\mathbf{k}}}^{\mathrm{sk,1}}+\delta f_{\mathrm{\mathbf{k}}}^{\mathrm{sk,2}}$ with the high-order terms in $\tilde{\mathbf{E}}(\mathbf{r},t)$ understood to vanish. In addition, we focus on the direct part of the AE current. Hence, only the stationary part of the second-order distribution functions, namely independent on the space and time, needs to be considered. After a series of careful derivation (see details in Appendix \ref{NFDF}), the formulas for the amplitudes of the first-order distribution functions
[$\delta f_{\mathrm{\mathbf{k},{am}}}^{\mathrm{in,1}}$,
$\delta f_{\mathrm{\mathbf{k},{am}}}^{\mathrm{sj,1}}$,
$\delta f_{\mathrm{\mathbf{k},{am}}}^{\mathrm{sk,1}}$] (where the subscript ``{am}" presents amplitude)  and the stationary parts of the second-order distribution functions[
$\delta f_{\mathrm{\mathbf{k},\text{dc}}}^{\mathrm{in,2}}$,
$\delta f_{\mathrm{\mathbf{k},\text{dc}}}^{\mathrm{sj,2}}$,
$\delta f_{\mathrm{\mathbf{k},\text{dc}}}^{\mathrm{sk,2}}$] (where the subscript ``{dc}" refer to  direct) can been determined and given in Eqs.~\eqref{APP-A-f1f2}\eqref{Appendix-A-NE}\eqref{AP-FSj2dc}\eqref{AP-FSj01-2}\eqref{AP-FSjdc} and \eqref{AP-FSk1}, respectively.

The total acoustic current as a response to the surface acoustic waves in presence of a nontrivial Berry curvature $\Omega_{\mathbf{k}}$ and disorder is given by~\cite{Xiao2010Rev}
\begin{equation}
\mathbf{J}=-e\int [d\mathbf{k}]\mathbf{v}_{\mathbf{k}}f_\mathbf{k}-\frac{e^{2}}{\hbar}\int[d\mathbf{k}]
\left[\tilde{\mathbf{E}}(\mathbf{r},t)
\times\Omega_{\mathbf{k}}\right]f_\mathbf{k},
\label{cur}
\end{equation}
where $\int[d\mathbf{k}]$ is shorthand for $\int d\mathbf{k}/(2\pi)^d$ with $d$ dimension. Since the nonequilibrium Fermi distribution $f_\mathbf{k}$ can be decomposed as $f_\mathbf{k}=f_\mathbf{k}^{\mathrm{in}}+\delta f_\mathbf{k}^{\mathrm{sj}}+\delta f_\mathbf{k}^{\mathrm{sk}}$ in presence of SAW-induced electric field and disorder, the total acoustic current in Eq.~(\ref{cur}), thus, can be decomposed into three parts as
\begin{equation}
\mathbf{J}=\mathbf{J}^\text{in}+\mathbf{J}^\text{sj}+\mathbf{J}^\text{sk},
\label{Jtot}
\end{equation}
corresponding to the intrinsic (in), side-jump (sj), and skew-scattering (sk) contributions to the acoustic current, respectively. Accompanying the formulas of first-order nonequilibrium distribution and the stationary parts of second-order nonequilibrium distribution in the SAW-induced field determined in Appendix \ref{NFDF}, the nonlinear dc AE current $\mathbf{j}_\text{dc}^{\text{nl}}$ (where the superscript ``{nl}"  refers to nonlinear) in the  $\alpha$ direction, as a response to the second-order in the SAW-induced field, can be expressed as $\mathbf{j}_{\text{dc},\alpha}^{\text{nl}}\equiv\tilde{\chi}_{\alpha\beta\gamma}\tilde{\mathcal{E}}_{\beta}^{*}
\tilde{\mathcal{E}}_{\gamma}$. Moreover, since the amplitude of the SAW-induced field $\vec{\tilde{\mathcal{E}}}$ is found to be dependent on the amplitude of piezoelectric field $\vec{\mathcal{E}}$ as $\vec{\tilde{\mathcal{E}}}=\vec{\mathcal{E}}/g\left(\mathbf{q},\omega\right)$ with $g\left(\mathbf{q},\omega\right)$ representing the dielectric function given in Eq.~\eqref{App-A-G}, it is convenient to rewrite the nonlinear dc AE current as
\begin{equation}
\mathbf{j}_{\text{dc},\alpha}^{\text{nl}}\equiv\chi_{\alpha\beta\gamma}
\mathcal{E}_{\beta}^{*}
\mathcal{E}_{\gamma}
\label{mainjdc}
\end{equation}
where $\chi_{\alpha\beta\gamma}$ is a coefficient characterizing the nonlinear AEE and the amplitude of piezoelectric field $\mathcal{E}_{\gamma}=\mathcal{E}^{*}_{\gamma}=\omega\varphi_\text{SAW}/v_{s}$ linearly depends on both the frequency $\omega$ and the acoustic wave piezoelectric potential amplitude $\varphi_\text{SAW}$ but inversely depends on the velocity $v_{s}$ of SAW. After a series of derivation in Appendix \ref{AE-coeff},  the coefficients due to the intrinsic, side-jump, and skew-scattering mechanisms are found to be, respectively

\begin{align}
	\chi_{\alpha\beta\gamma}^{\mathrm{in}}&= \frac{e^{2}}{2\hbar}|F|^{2}\mathrm{Re}\sum_\mathbf{k}\left[\partial_{\beta}(v_{\alpha}
\tau_{\mathbf{k}})-\epsilon_{\alpha\beta\eta}\Omega_{\text{\ensuremath{\eta}}}\right]G_{\gamma}
,\nonumber\\
\chi_{\alpha\beta\gamma}^{\mathrm{sj}}&=\chi_{\alpha\beta\gamma}^{\text{sj-v}}+
\chi_{\alpha\beta\gamma}^{\text{sj-mff}}+\chi_{\alpha\beta\gamma}^{\text{
sj-je}}\nonumber,\\
\chi_{\alpha\beta\gamma}^{\mathrm{sk}}&=\chi_{\alpha\beta\gamma}^{\text{sk-mff}}+
\chi_{\alpha\beta\gamma}^{\text{sk-je}},\label{chinsksk}
\end{align}
where $F$ represents the auxiliary function given in Eq.~\eqref{App-A-F}, $\tau_{\mathbf{k}}$ is the momentum-dependent relaxation time, $\partial_{\beta}$ is the shorthand for $\partial/\partial k_{\beta}$, $\epsilon_{\alpha\beta\eta}$ is Levi-Civita symbol, and  the superscripts ``{v}" represents velocity, ``{mff}" denotes modified Fermi function, ``{je}" indicates joint effect, respectively. The decomposed components $\chi_{\alpha\beta\gamma}^{\text{sj-v}}$, $\chi_{\alpha\beta\gamma}^{\text{sj-mff}}$, and $\chi_{\alpha\beta\gamma}^{\text{sj-je}}$  represent the nonlinear side-jump AE coefficient $\chi_{\alpha\beta\gamma}^{\mathrm{sj}}$ stemming from the side-jump velocity, the side-jump-modified NFDF, and the joint effect of anomalous velocity item and the side-jump effect, respectively. The components $\chi_{\alpha\beta\gamma}^{\text{sk-mff}}$ and $\chi_{\alpha\beta\gamma}^{\text{sk-je}}$ for nonlinear skew-scattering AE coefficient $\chi_{\alpha\beta\gamma}^{\mathrm{sk}}$ attribute to the skew-scattering-modified-NFDF contribution and the contribution from joint effect of anomalous velocity and skew scattering, respectively. The explicit expressions for the relevant side-jump nonlinear AE coefficients ($\chi_{\alpha\beta\gamma}^{\text{sj-v}}$, $\chi_{\alpha\beta\gamma}^{\text{sj-mff}}$, $\chi_{\alpha\beta\gamma}^{\text{sj-je}}$) and skew-scattering nonlinear AE coefficients ($\chi_{\alpha\beta\gamma}^{\text{sk-mff}}$, $\chi_{\alpha\beta\gamma}^{\text{sk-je}}$) are given in Eq.~\eqref{AP-coeff-sj} and Eq.~\eqref{AP-coef-Sk-ad}, respectively.

\section{Valley Acoustoelectric Effect for the massive Dirac materials}\label{NAVHE}

 A tiny model for the 2D hexagonal massive Dirac materials can be described by the massive Dirac model, namely
\begin{align}
	\hat{H}_{0}=  v_F \hbar(\tau_{v}k_{x}\sigma_{x}+k_{y}\sigma_{y})+m\sigma_{z},\label{Hami}
\end{align}
where $v_F$ is the Fermi velocity, $\tau_{v}=\pm1$ refers to the valley index, $\hat{\bm{\sigma}}$ represents the Pauli matrices, and the massive term $2m$ indicates the energy gap.
For simplicity, we only focus on $n$-doped systems. The energy dispersion of the conduction band is $\varepsilon_{\mathbf{k}}= \left[v_{F}^2\hbar^{2}(k_x^2+k_y^2)+m^2\right]^{1/2}$. For the 2D systems, the Berry curvature is, essentially, a pseudovector with only $z$ component nonzero and determined by $\Omega_{n\mathbf{k}}=\mathbf{\hat{z}}\cdot\nabla_\mathbf{k}\times\langle\psi_{n\mathbf{k}}|i\nabla
_\mathbf{k}|\psi_{n\mathbf{k}}\rangle
$, where $\psi_{n\mathbf{k}}$ is the $n$-band eigenstate. Hence, the Berry curvature is found to be  $\Omega_{\mathbf{k}}=-\tau_{v}m(v_{F}\hbar)^2/(2\varepsilon_{\mathbf{k}}^3)$ for the conduction band (upper band) of the effective Hamiltonian [Eq.~\eqref{Hami}].

For the disorder part, the root for the extrinsic mechanisms, we consider a spin-independent $\delta$-function random potential $\hat{V}_\text{imp}(\mathbf{r})=V_0\sum_i\delta(\mathbf{r-R}_i)$ with $\mathbf{R}_i$ indicating the random position of impurities and $V_0$ representing the disorder strength. After a careful derivation (see details in Appendix \ref{scat-rates-time}), for the disordered massive Dirac materials with $\delta$-function random potential scatters, the nonzero leading-order scattering rates in disorder for the symmetric and antisymmetric parts [$w_{\mathbf{k}\mathbf{k}^{\prime}}^{\mathrm{S,(2)}}$,
$w_{\mathbf{k}\mathbf{k}^{\prime}}^{\mathrm{A,(3)}}$, $w_{\mathbf{k}\mathbf{k}^{\prime}}^{\mathrm{A,(4)}}$]  
can be determined through the Born approximation and are given in Eq.~{\eqref{App-B-dfd}}, respectively.
It's should to clarify that the leading symmetric part of scattering rate [$w_{\mathbf{k}\mathbf{k}^{\prime}}^{\mathrm{S},(2)}$] originating from second-order scattering rates in disorder is valley-independent, while the leading antisymmetric components [$w_{\mathbf{k}\mathbf{k}^{\prime}}^{\mathrm{A,(3)}}$, $w_{\mathbf{k}\mathbf{k}^{\prime}}^{\mathrm{A,(4)}}$] stemming from the third- and forth-order scattering rates in disorder is valley-contrasting, namely $w_{\mathbf{k}\mathbf{k}^{\prime}}^{\mathrm{A,(3)}}(\boldsymbol{K})=-w_{\mathbf{k}\mathbf{k}^{\prime}}^
{\mathrm{A,(3)}}(-\boldsymbol{K})$
and $w_{\mathbf{k}\mathbf{k}^{\prime}}^{\mathrm{A,(4)}}(\boldsymbol{K})=-w_{\mathbf{k}\mathbf{k}^{\prime}}
^{\mathrm{A,(4)}}(-\boldsymbol{K})$ . The valley contrasting antisymmetric components are, actually, fundamental for the nonzero NAVHC attributing to the skew-scattering effect. In addition, one can also easily confirm that the symmetric part dominates the scattering process since it's second order in disorder while asymmetric parts is in high order. Therefore, the relaxation time ${\tau_{\mathbf{k}}}$ for the upper band can be evaluated in the leading term [$w_{\mathbf{k}\mathbf{k}^{\prime}}\approx w_{\mathbf{k}\mathbf{k}^{\prime}}^{(2)}=w_{\mathbf{k}\mathbf{k}^{\prime}}^{\text{S},(2)}$] of scattering rate (see details in Appendix.\ref{scat-rates-time})
\begin{align}
	\tau_{\mathbf{k}}
=\frac{4\hbar}{n_{i}V_{0}^{2}}\frac{v^{2}\varepsilon_{\mathbf{k}}}{\varepsilon_{\mathbf{k}}^{2}+3m^{2}}.\label{relaxation-time}
\end{align}

In this work, we consider the long-wave limit $\omega\tau_{\mathbf{k}}\ll1$, which can be usually satisfied in experiment. Moreover, the zero-temperature limit $-\partial f^{(0)}_{\mathbf{k}}/{\partial \varepsilon_{\mathbf{k}}}\approx \delta(\varepsilon_{\mathbf{k}}-E_f)$ will also be adopted. When fixing the wave vector $\mathbf{q}$ in $x$ direction, which means the amplitude of piezoelectric electric field $\vec{\mathcal{E}}$ is also oriented in $x$ direction since $\mathbf{q}\parallel\vec{\mathcal{E}}$, the nonlinear acoustic Hall current vertically to the wave vector will flow in $y$ direction and be characterized by $\chi_{yxx}$ [Eq.~\eqref{mainjdc}]. After a tedious calculation (see Appendix \ref{coeff-NAVHE} for a detail calculation), the nonlinear acoustic Hall coefficients (AHC) ($\chi_{yxx}^{\mathrm{In},\tau_{v}}$, $\chi_{yxx}^{\mathrm{sj},\tau_{v}}$, $\chi_{yxx}^{\mathrm{sk},\tau_{v}}$) for $\tau_v$ valley of the disordered massive Dirac materials from intrinsic, side-jump and skew-scattering contributions are found to be, respectively,
\begin{align}
			\chi_{yxx}^{\mathrm{in},\tau_{v}} & =\tau_{v}\frac{e\sigma mv^{2}}{4\hbar v_{s}E_f^{3}}H_{1},\notag\\
			\chi_{yxx}^{\mathrm{sj},\tau_{v}} & =\tau_{v}\frac{e\sigma mv^{2}}{\hbar v_{s}E_f^{3}}\frac{E_f^{4}-6m^{2}E_f^{2}-3m^{4}}{(E_f^{2}+3m^{2})^{2}}H_{1},\notag\\
			\chi_{yxx}^{\mathrm{sk},\tau_{v}} & =\chi_{yxx,1}^{\mathrm{sk},\tau_{v}}+\chi_{yxx,2}^{\mathrm{sk},\tau_{v}},\label{chiyxx}
		\end{align}
with
\begin{eqnarray}
			\chi_{yxx,1}^{\mathrm{sk},\tau_{v}}&= & \tau_{v}\frac{\pi V_{0}\sigma^{2}}{4ev_{s}}\frac{m^{3}(7E_f^{2}-3m^{2})}{E_f^{2}(E_f^{2}+3m^{2})^{2}}H_{1}\notag\\
			&& -\tau_{v}\frac{\pi V_{0}\sigma^{2}}{2ev_{s}}\frac{mE_f^{2}}{(E_f^{2}+3m^{2})^{2}}\frac{v_{s}^{2}}{v_{F}^{2}}H_{2},
\label{chiyxx-sk3}\\
		\chi_{yxx,2}^{\mathrm{sk},\tau_{v}}&= & \tau_{v}\frac{\pi n_{i}V_{0}^{2}\sigma^{2}}{4ev_{s}}\frac{3m(5m^{2}-E_f^{2})}{2E_f(E_f^{2}+3m^{2})^{2}}H_{1}\notag\\
		& &-\tau_{v}\frac{3\pi n_{i}V_{0}^{2}\sigma^{2}}{4ev_{s}}\frac{mE_f}{(E_f^{2}+3m^{2})^{2}}\frac{v_{s}^{2}}{v_{F}^{2}}H_{2},
\label{chiyxx-sk4}
		\end{eqnarray}
corresponding to the skew-scattering contribution originated from the third- and forth-order scattering rates in disorder, respectively, where
$H_1$ and $H_2$ are dimensionless auxiliary functions and given in Eqs.~\eqref{App-D-H1} and ~\eqref{App-D-H23}, respectively, $E_f$ is the Fermi energy, $\tau_F$ indicates the relaxation time at the Fermi surface, $\varepsilon_e=m_ev_F^2/2$ is determined by the free electron mass $m_e$, two defined parameters $\sigma_{*}=\epsilon_0(\epsilon+1)v_s/4\pi$ and $a=\epsilon_0(\epsilon+1)\hbar^2/(2m_ee^2)$ are dependent on the dielectric permittivity of vacuum $\epsilon_0$ and the dielectric constant $\epsilon$ of substrate with the velocity $v_s$ of SAW propagating in the piezoelectric substrate. Equations \eqref{chiyxx}-\eqref{chiyxx-sk4} indicate that the total nonlinear AHC $\chi_{yxx}$ $=\chi_{yxx}^{+1}+\chi_{yxx}^{-1}$ (where $\chi_{yxx}^{\tau_{v}}=\chi_{yxx}^{\mathrm{in},\tau_{v}}+\chi_{yxx}^{\mathrm{sj},\tau_{v}}+
\chi_{yxx}^{\mathrm{sk},\tau_{v}}$) is vanishing but the nonlinear valley AHC  $\chi_\text{valley}^{\mathrm{H}}=\chi_{yxx}^{+1}-\chi_{yxx}^{-1}$  is nonzero, hinting that a \textit{pure} nonlinear acoustic valley Hall current $J_\mathrm{valley}^\mathrm{H}=\chi_\text{valley}^{\mathrm{H}}\mathcal{E}^{2}_{x}$ (where superscript ``{H}" refers to Hall) will flow vertically to the wave vector $\mathbf{q}$ of SAW when propagating a SAW into the disordered massive Dirac materials. It's should be point out that the components of $\chi_{yxx}^{\mathrm{sj},\tau_{v}}$ ($\chi_{yxx}^{\mathrm{sk},\tau_{v}}$), which stem from the joint effect of anomalous velocity item and the side-jump (skew-scattering) effect, are found to be zero for $\tau_{v}$ valley in the massive Dirac materials, namely $\chi_{yxx}^{\text{sj-je},\tau_{v}}=0$ ($\chi_{yxx}^{\text{sk-je},\tau_{v}}=0$) (detail can be found in Appendix \ref{coeff-NAVHE}). Therefore, the side-jump contribution to nonlinear AVHE roots in the side-jump velocity or the side-jump-modified NFDF, whereas the nonzero skew-scattering-induced nonlinear AVHE ascribes to the skew-scattering-modified NFDF.

\section{Results and discussion}\label{R and D}
One typical massive Dirac materials whose Hamiltonian has the form as that given in Eq.~(\ref{Hami}) are the n-doped transition-metal dichalcogenide monolayer in which the spin-orbit coupling can be neglected owing to the tiny spin splitting in the conduction band. In this work, we investigate the nonlinear AVHE in the n-doped monolayer $\mathrm{MoS_2}$ in presence of disorder and choose the LiNbO$_3$ as piezoelectric substrate. The corresponding materials parameters are taken as follows: the band gap $2m=1.663~\text{eV}$, Fermi velocity $v_F=0.6\times10^6m/s$,~\cite{Xiao2012,Liu2013} the dielectric constant $\epsilon=50$ and the sound velocity $v_s=3500m/s$~\cite{Savenko2020}. For disorder part, we take the impurity concentration $n_i\geq 10^9\mathrm{cm^{-2}}$ and impurity strength $V_0=3\times10^{-13}\mathrm{eV\cdot cm^{-2}}$.
For $n_i=10^9\mathrm{cm^{-2}}$, the corresponding relaxation time $\tau_{F}$ [Eq.~\eqref{relaxation-time}] at Fermi surface ranges from \text{0.0137} to 0.0142~ps when the Fermi energy is in the range of $[m,m+$60\text{meV}$]$, namely close to the Dirac point within $60\text{meV}$, which corresponds to a typical range of electron density ($n_{e}<2\times\mathrm{10^{13}cm^{-2}}$) for  $n$-doped monolayer MoS$_2$ in experiment.~\cite{NM2013MoS2,PRB2012MoS2,2022NLMoS2} Subsequently, when varying the frequency $\omega$ of SAW within the range of [$0,10\text{GHz}$], the long-wave limit condition $\omega\tau_{\mathbf{k}}\ll1$ is satisfied since the maximum value of $\omega\tau_{F}$ is of the order of $1\times10^{-4}$. Obviously, when $n_i>10^9\mathrm{cm^{-2}}$, the long-wave limit condition still satisfies since $\tau_{\mathbf{k}}\sim1/n_i$[Eq.\eqref{relaxation-time}].

\begin{figure}[ht]
	\centering
	\subfigure{
	\includegraphics[width=1\linewidth]{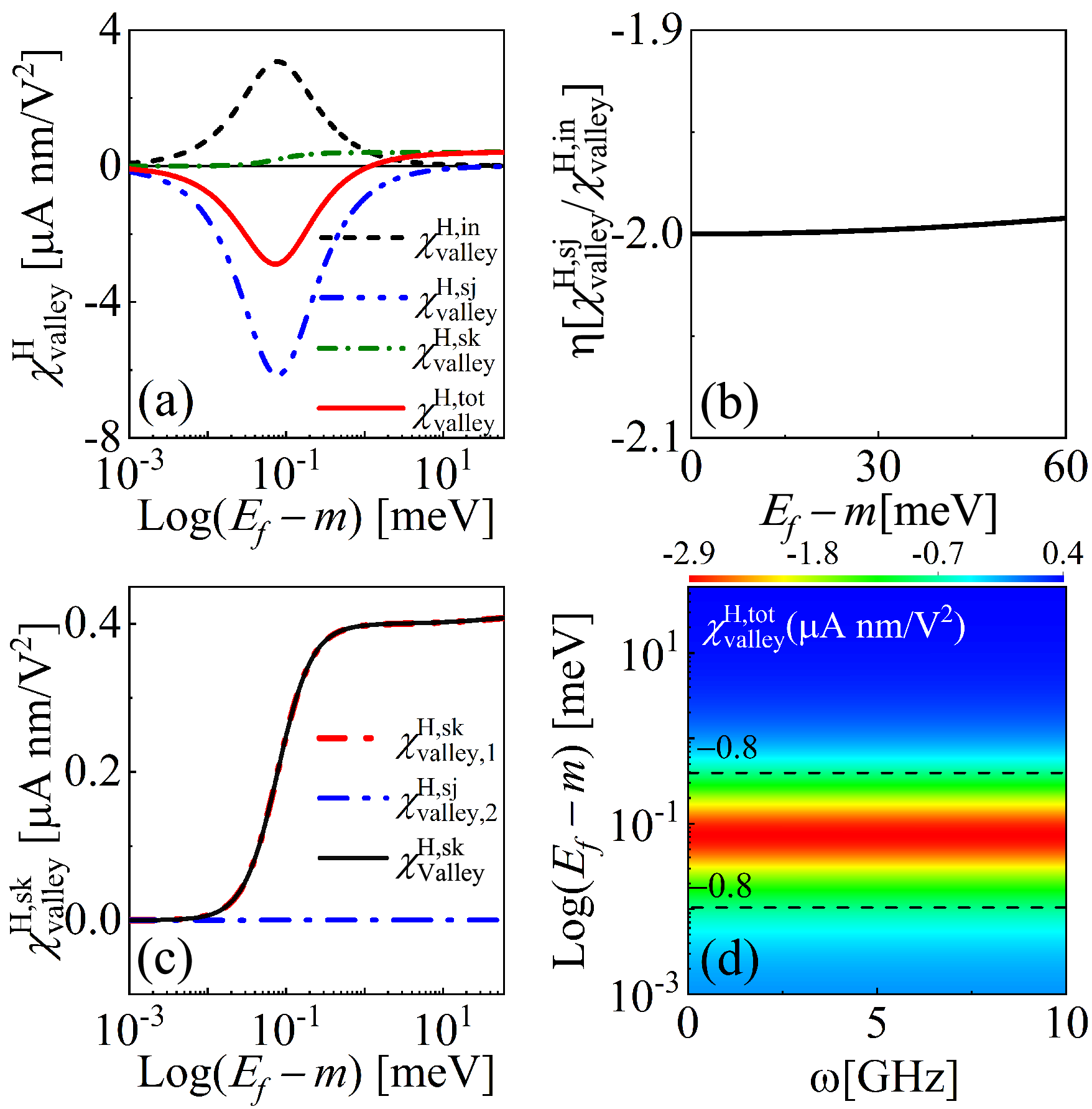}}
\caption{(a) The intrinsic (in), side-jump (sj), skew-scattering (sk) and total (tot) contributions to the NAVHCs $\chi_{\mathrm{valley}}^{\mathrm{H}}$ vs the Fermi energy $E_{f}$. (b) the ratio ${\eta}=\chi_{\text{valley}}^{\mathrm{H,sj}}/\chi_{\text{valley}}^{\mathrm{H,in}}$ vs $E_{f}$.
(c) The coefficients [$\chi_{\mathrm{valley}}^{\mathrm{H,sk}}$, $\chi_{\mathrm{valley},1}^{\mathrm{H,sk}}$, $\chi_{\mathrm{valley},2}^{\mathrm{H,sk}}$] versus $E_{f}$ for the $\mathrm{MoS_{2}}$, where $\chi_{\mathrm{valley},1}^{\mathrm{H,sk}}$ $(\chi_{\mathrm{valley},2}^{\mathrm{H,sk}}$) represents the skew-scattering induced NAVHCs as the third- (forth-)order response to scattering rate.
(d) The total coefficient $\chi_{\mathrm{valley}}^{\mathrm{H,tot}}$ as a function of $E_f$ and SAW frequency $\omega$. $\omega=10~\mathrm{GHz}$ is taken in (a)(b)(c).
 Parameters used: $n_i=1\times10^9\mathrm{cm^{-2}}$, $V_0=3\times10^{-13}\mathrm{eV\cdot cm^{2}}$.}\label{Fig2}
		\end{figure}

Figure \ref{Fig2}(a) illustrates the Fermi-energy dependence of NAVHCs [$\chi_\text{valley}^{\mathrm{H,in}}$, $\chi_\text{valley}^{\mathrm{H,sj}}$, $\chi_\text{valley}^{\mathrm{H,sk}}$, $\chi_\text{valley}^{\mathrm{H,tot}}$] from the different contributions, including the intrinsic, side-jump, skew-scattering and total ones in the disordered monolayer $\mathrm{MoS_2}$. One can easily observe that the extrinsic contributions (side jump and skew scattering) to the NAVHC are significant and even dominate both magnitude and sign of the total NAVHC $\chi_\text{valley}^{\mathrm{H,tot}}$. Besides, the sign of NAVHC from side-jump contribution is opposite to that from both intrinsic and skew-scattering contributions.
 When modulating the Fermi energy $E_{f}$ through the gate voltage close to the Dirac point (\text{$E_{f}-m<0.5\text{meV}$}), the skew-scattering contribution to NAVHC is relatively insignificant and the side-jump contribution $\chi_{\text{valley}}^{\mathrm{H,sj}}$ is, contrarily, twice as large as the intrinsic contribution $\chi_{\text{valley}}^{\mathrm{H,in}}$ ,which can be easily confirmed since ${\eta}=\chi_{\text{valley}}^{\mathrm{H,sj}}/\chi_{\text{valley}}^{\mathrm{H,in}}$=2 as shown in  Fig.~\ref{Fig2}(b).
 As a result, the side-jump contribution dominates and the total NAVHC $\chi_\text{valley}^{\mathrm{H,tot}}$ from both intrinsic and extrinsic contributions has a almost same magnitude but opposite sign to the intrinsic coefficient $\chi_\text{valley}^{\mathrm{H,in}}$.

However, when modulating the Fermi energy away from the Dirac point ($30\text{meV}<E_f-m<60\text{meV}$), the skew-scattering contribution becomes the dominant one and the total NAVHC $\chi_{\text{valley}}^{\mathrm{H,tot}}\approx\chi_{\text{valley}}^{\mathrm{H,sk}}$ also tends to be a constant, namely independent on $E_{f}$. This independence of $\chi_{\text{valley}}^{\mathrm{H,sk}}$ on $E_{f}$ can be explained as follows: one can approximately have $\chi_{\text{valley}}^{\mathrm{H,sk}}=\chi_{\text{valley},1}^{\mathrm{H,sk}}
+\chi_{\text{valley},2}^{\mathrm{H,sk}}\approx \chi_{\text{valley},1}^{\mathrm{H,sk}}$  (where $\chi_{\text{valley},1}^{\mathrm{H,sk}}=\chi_{yxx,1}^{\mathrm{sk},+1}-\chi_{yxx,1}^{\mathrm{sk},-1}$ ) [Eq.~\eqref{chiyxx}] since the second term $\chi_{\text{valley},2}^{\mathrm{H,sk}}$, which corresponds to the fourth-order response to scattering rate, almost has no contribution to $\chi_{\text{valley}}^{\mathrm{H,sk}}$ as shown in [Fig.~\ref{Fig2}(c)]. In addition,
when shifting E$_f$ away from the Dirac point ($E_{f}-m>30~\mathrm{meV}$), one can have $(\sigma/\sigma_{*})^2\gg1$, which is consistent with the results in refs.~\cite{Kalameitsev2019PRL,Sonowal2020PRB}, leading to the dimensionless auxiliary function $H_1\approx(\sigma_{*}/\sigma)^2$ independent of the Fermi energy $E_{f}$. On the other hand, the remaining Fermi-energy-dependent term $(7E_f^2-3m^2)/[E_f^2(E_f^2+3m^2)^2]$ [Eq.~\eqref{chiyxx-sk3}] for $\chi_\text{valley}^{\mathrm{H,sk}}\approx\chi_{\text{valley},1}^{\mathrm{H,sk}}=\chi_{yxx,1}^{\mathrm{sk},+1}
-\chi_{yxx,1}^{\mathrm{sk},-1}$ [Fig.~\ref{Fig2}(c)] almost keeps a constant when varying the Fermi energy $E_{f}$ in the range of $[m+30,m+60~\mathrm{meV}]$.
Consequently, the skew-scattering contribution to the NAVHC is almost a constant and independent on $E_{f}$ in the relatively high doping.

The total NAVHC $\chi_\text{valley}^\mathrm{H,tot}$ displays an dip feature and the maximum $|\chi_\text{valley}^\mathrm{H,tot}|$ can be expected when modulating $E_{f}$ away from the Dirac cone by roughly $0.1~\mathrm{meV}$ [Fig.~\ref{Fig2}(d)], in which the side-jump contribution is dominate one.
Interestingly, we find that when varying the frequency $\omega$ of SAW from $0$ to $10$~GHz, the total NAVHC $\chi_\text{valley}^{\mathrm{H,tot}}$ keeps unchanged [Fig.~\ref{Fig2}(d)].
This independence of $\chi_\text{valley}^{\mathrm{H,tot}}$  on the frequency can be explained as follows. In the long-wave limit condition which we consider, one can have the relations $qa_{0}\varepsilon_{e}/E_f \ll1$ and $\omega\tau_{F}\frac{\sigma}{\sigma_{*}}qa_{0}\varepsilon_{e}/E_f\ll1$ for MoS$_2$. Consequently, the dimensionless auxiliary functions $H_{1}\approx 1/(1+\sigma^{2}/\sigma_{*}^{2})$ is independent of the frequency, resulting in $\chi_{\text{valley}}^{\mathrm{H,in}}~(\chi_{\text{valley}}^{\mathrm{H,sj}})\sim\omega^0$ [Eq~(\ref{chiyxx})]. For the skew-scattering-induced NVAHC $\chi_{\text{valley}}^{\mathrm{H,sk}}$, we have discussed that $\chi_\text{valley}^{\mathrm{H,sk}}$ is approximately equal to $\chi_{\text{valley},1}^{\mathrm{H,sk}}=\chi_{yxx,1}^{\mathrm{sk},+1}
-\chi_{yxx,1}^{\mathrm{sk},-1}$ [Fig.~\ref{Fig1}(c)]. Furthermore, noting that $v_s^2/v_F^2\approx3.4\times10^{-3}$ and the factor $(\frac{\sigma}{\sigma_{*}})^{2}(qa_{0}\varepsilon_{e}/E_f)^{2}<0.073 $ for $\mathrm{MoS_2}$ when $E_f\in[m,m+$60\text{meV}$]$ in Eq.~\eqref{chiyxx-sk3}, it can be further identified that the contribution from the second term in Eq.~\eqref{chiyxx-sk3} to $\chi_{yxx,1}^{\mathrm{sk},\tau_{v}=\pm1}$ can be neglected,
indicating $\chi_{\text{valley}}^{\mathrm{H,sk}}\approx\chi_{yxx,1}^{\mathrm{sk},+1}
-\chi_{yxx,1}^{\mathrm{sk},-1}\sim H_1\sim\omega^0$.
Hence, $\chi_\text{valley}^{\mathrm{H,tot}}$ is independent of $\omega$, resulting in a quadratic dependence of nonlinear AVHE on the frequency $\omega$ since $J_\mathrm{valley}^\mathrm{H,tot}=\chi_\text{valley}^{\mathrm{H},\mathrm{tot}}\mathcal{E}^{2}_{x}$ and the $x$-component amplitude of piezoelectric field $\mathcal{E}_{x}\propto\omega$ is linearly dependent of $\omega$.

\begin{figure}[ht]
	\centering
	\subfigure{
		\includegraphics[width=1\linewidth]{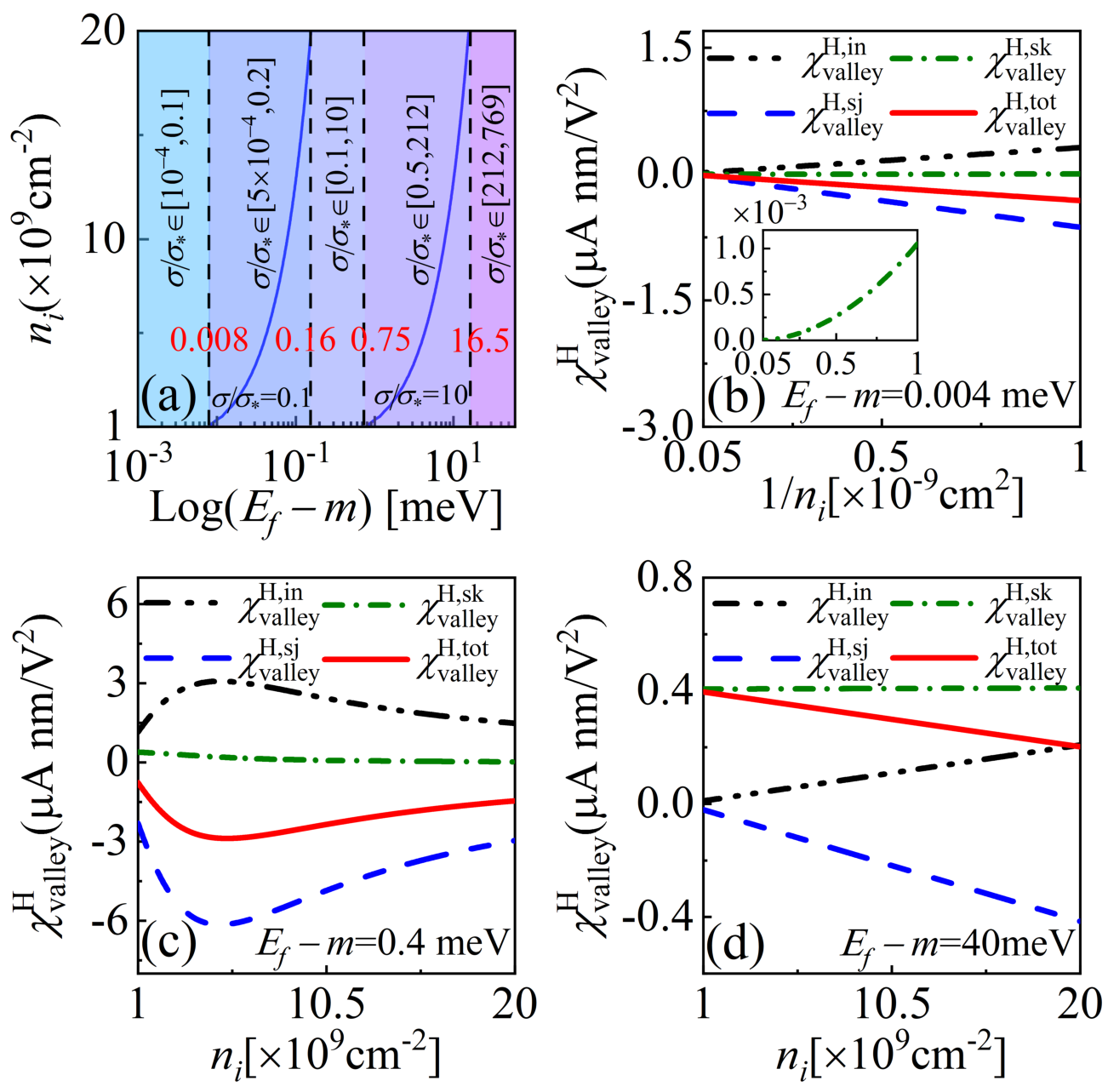}}
	\caption{(a)Ranges of the magnitude of factor $\sigma/\sigma_{*}$ as impurity concentration $n_i$ varies in range of $10^9\sim2\times10^{10}\mathrm{cm^{-2}}$ at different Fermi energy intervals. The two blue lines refer to the counters $\sigma/\sigma_{*}=0.1$ and $\sigma/\sigma_{*}=10$, respectively. The red values represent the magnitude of Fermi energy corresponding to the four dashed lines. (b)(c)(d) The intrinsic (in), side-jump (sj), skew-scattering (sk) and total (tot) contributions to the NAVHCs $\chi_{\mathrm{valley}}^{\mathrm{H}}$ vs the impurity concentration $n_i$ at different Fermi energies, at which the factor $\sigma/\sigma_{*}$ is small[(b)], moderate[(c)] and large[(d)], respectively. Parameters used: $\omega=10~\mathrm{GHz}$, $V_0=3\times10^{-13}\mathrm{eV\cdot cm^{2}}$.}\label{Fig3}
\end{figure}

Without considering the screening effect, the recent reported photon dragging valley Hall effect (PDVHE) shows an inverse proportion to the impurity concentration $n_{i}$ (i.e., $\sim1/n_{i}$) for both intrinsic and side-jump contribution but $\sim1/n_{i}^{2}$ for the skew scattering contribution \cite{GlazovPRB2020}. However, being different to PDVHE, it's necessary to consider the screening effect owing to the appearance of dielectric materials in AEE experiment. In our investigated model for nonlinear AVHE, the screening effect mainly originates from the piezoelectric substrate LiNbO$_3$ (a dielectric material with a high dielectric constant) and is mathematically captured in the two defined auxiliary functions $H_{1}$ and $H_{2}$ [Eqs.\eqref{chiyxx}\eqref{chiyxx-sk3}\eqref{chiyxx-sk4}~\eqref{App-D-H1}\eqref{App-D-H23}], which are dependent on factor $\sigma/\sigma_{*}$. According to the orders of magnitude of $\sigma/\sigma_{*}$ when varying the Fermi energy $E_{f}$ and impurity concentration $n_{i}$, we have divided the Fermi energy into several regions [Figure~\ref{Fig3}(a)]. It's found that the dependence of nonlinear AVHE on $n_{i}$ [$\chi_{\text{valley}}^{\mathrm{H,in}}(\chi_{\text{valley}}^{\mathrm{H,sj}})\sim1/n_i$ and $\chi_{\text{valley}}^{\mathrm{H,sk}}\sim1/n_i^2$, Figure~\ref{Fig3}(b)] agree with the results for PDVHE when the Fermi energy is extremely close to the Dirac point ($E_{f}-m<0.008~\mathrm{meV}$) and the impurity concentration is larger than $10^9\mathrm{cm^{-2}}$. This agreement is reasonable and also prove the validity of our results since in the considered region, one can have $\sigma/\sigma_{*}<0.1$ and $H_1\approx1$ (recalling that $H_{1}\approx 1/(1+\sigma^{2}/\sigma_{*}^{2})$ for monolayer $\mathrm{MoS_2}$), which hints the screening effect can be neglected. However, when modulating the Fermi level away from the Dirac point ($E_{f}-m>0.16~\mathrm{meV}$), the dependence of NAVHCs on impurity concentration $n_i$ become entirely different to the PDVHE [Figs.~\ref{Fig3}(c) and (d)]. Especially, in the region where the Fermi level is relatively high [$E_{f}-m>16.5~\mathrm{meV}$], we found that $\chi_{\text{valley}}^{\mathrm{H,in}}~(\chi_{\text{valley}}^{\mathrm{H,sj}})\sim\sigma^{-1}\sim n_i$ and $\chi_{\text{valley}}^{\mathrm{H,sk}}\approx\chi_{yxx,1}^{\mathrm{sk},+1}
-\chi_{yxx,1}^{\mathrm{sk},-1}\sim\sigma^0\sim n_i^0$ [Eqs.\eqref{chiyxx} and \eqref{chiyxx-sk3}] since $\sigma/\sigma_{*}>>1$ [Fig.~\ref{Fig3}(a)] gives rise to $H_{1}\approx(\sigma_{*}/\sigma)^{2}$. Besides, the sign of $\chi_{\text{valley}}^{\mathrm{H,sj}}$ is different to those of $\chi_{\text{valley}}^{\mathrm{H,in}}$ and $\chi_{\text{valley}}^{\mathrm{H,sk}}$. As a result, the total NAVHC $\chi_\mathrm{valley}^\mathrm{H,tot}$ linearly decreases as impurity concentration increase [Fig.~\ref{Fig3}(d)].

 \section{conlusions}\label{conlusions}
 In summary, taking the disorder impact into consideration, we derive the general formulas of AE response coefficients, which include the intrinsic, side-jump as well as the skew-scattering contributions. Based on the derived formulas, we investigate the nonlinear acoustic Hall effect
 in 2D massive Dirac materials in presence of disorder through a tiny model and specifically analyse the behaviors of the nonlinear acoustic Hall effect in the disordered monolayer $\mathrm{MoS_{2}}$. It's found that the total nonlinear acoustic Hall effect disappears in disordered monolayer MoS$_{2}$ but the \textit{pure} nonlinear AVHE can exits and  shows a dip feature with negative values near the Dirac point. Interestingly, the nonlinear AVHE displays a quadratic dependence on the SAW frequency. We disclosed that the extrinsic mechanisms play significant roles in the disordered MoS$_{2}$: when the Fermi energy is located close to the Dirac point, the side-jump contribution, which is found to be originated from the side-jump velocity or the side-jump-modified NFDF, dominates and is twice as larger as that from the intrinsic contribution; when the Fermi energy is higher, the signal is almost solely from the skew-scattering contribution rooting in the skew-scattering-modified NFDF for the low impurity concentration $n_i$. Besides, we found that the dependence of nonlinear AVHE on impurity concentration $n_i$ is complicate and is different to the valley Hall effect caused by the photon drag: only when the Fermi energy is extremely close to the Dirac point, namely the screening effect can be neglected, the total NAVHC is inversely proportional to $n_{i}$; when the Fermi energy is relatively high, the NAVHC even linearly decreases with the $n_{i}$ increasing.
Additionally, the sign of nonlinear AVHE from side-jump contribution is opposite to those from both intrinsic and skew-scattering contributions. Consequently, the total nonlinear AVHE from both intrinsic and extrinsic contribution undergoes an sign change when gradually modulating the Fermi energy away the Dirac point through the gate voltage.
		
\section{acknowledgements}
This work is supported by the National Natural Science Foundation of China (Grants No. 12004107
 and No. 12374040), the National Science Foundation of Hunan, China (Grant No. 2023JJ30118), and the Fundamental Research Funds for the Central Universities.

\appendix
\section{The formalism for the nonequilibrium Fermi distribution functions $f_\mathbf{k}$ in the presence of the SAW-induced electric field and disorder}\label{NFDF}
As discussed in the main text [Sec.~\ref{Theoretical review}], the nonequilibrium Fermi distribution function $f_\mathbf{k}$ in presence of the SAW-induced electric field and disorder can be decomposed into
\begin{equation}
f_\mathbf{k}=f_\mathbf{k}^{\mathrm{in}}+\delta f_\mathbf{k}^{\mathrm{sj}}+\delta f_\mathbf{k}^{\mathrm{sk}}.
\end{equation}
In this appendix, the expressions for [$f_{\mathrm{\mathbf{k}}}^\text{in}$, $\delta f_\mathbf{k}^{\mathrm{sj}}$, $\delta f_\mathbf{k}^{\mathrm{sj}}$] will deduced as follows.

\subsection{The intrinsic Fermi nonequilibrium  distribution function $f_{\mathrm{\mathbf{k}}}^\text{in}$} \label{App-A-in}
Combining Eqs.~(\ref{rk})(\ref{BTE-three}) with the relaxation-time approximation, i.e. $I_{c}^{\mathrm{in}}[ f_{\mathrm{\mathbf{k}}}^{\mathrm{ln}}]=-[f_{\mathrm{\mathbf{k}}}^{\mathrm{in}}-\langle f_{\mathrm{\mathbf{k}}}(\mathbf{r},t)\rangle]/{\tau_{\mathbf{k}}}$, the intrinsic nonequilibrium Fermi distribution function (NFDF) $f_{\mathrm{\mathbf{k}}}^\text{in}$ in the presence of the SAW-induced electric field can be determined by
\begin{align} &\partial_{t}f_{\mathrm{\mathbf{k}}}^{\mathrm{in}}+\mathbf{v}_{\mathbf{k}}
\cdot\partial_{\mathbf{r}}f_{\mathrm{\mathbf{k}}}^{\mathrm{in}}-\frac{e}{\hbar}\tilde{\mathbf{E}}
(\mathbf{r},t)\cdot\partial_{\mathbf{k}}f_{\mathrm{\mathbf{k}}}^{\mathrm{in}}
	=-\frac{f_{\mathrm{\mathbf{k}}}^{\mathrm{in}}-\langle f_{\mathrm{\mathbf{k}}}(\mathbf{r},t)\rangle}{\tau_{\mathbf{k}}},\label{AP-BTE-in}
\end{align}
giving
\begin{equation}
\begin{aligned}
f_{\mathrm{\mathbf{k}}}^\text{in}& = \frac{\langle f_{\mathrm{\mathbf{k}}}(\mathbf{r},t)\rangle}{1+\tau_{\mathbf{k}}\partial_{t}+\tau_{\mathbf{k}}
\mathbf{v}_{\mathbf{k}}\cdot\partial_{\mathbf{r}}-\tau_{\mathbf{k}}\frac{e}{\hbar}\tilde{\mathbf{E}}
(\mathbf{r},t)\cdot\partial_{\mathbf{k}}}\\ &=\sum_{n=0}^{\infty}(-\tau_{\mathbf{k}}\partial_{t}-\tau_{\mathbf{k}}\mathbf{v}_{\mathbf{k}}
\cdot\partial_{\mathbf{r}}+\tau_{\mathbf{k}}\frac{e}{\hbar}\tilde{\mathbf{E}}(\mathbf{r},t)
\cdot\partial_{\mathbf{k}})^{n}\langle f_{\mathrm{\mathbf{k}}}(\mathbf{r},t)\rangle,
\label{App-A-exp1}
\end{aligned}
\end{equation}
where the SAW-induced overall electric field $\tilde{\mathbf{E}}(\mathbf{r},t)$, which includes the in-plane piezoelectric field $\mathbf{E}(\mathbf{r},t)$ generated by SAW and the induced electric field $\mathbf{E}^{i}(\mathbf{r},t)$, is determined as
\begin{align}	\tilde{\mathbf{E}}(\mathbf{r},t)=\frac{1}{2}\vec{\tilde{\mathcal{E}}}e^{i(\mathbf{q}\cdot\mathbf{r}-\omega t)}+c.c.
\end{align}
with $\vec{\tilde{\mathcal{E}}}$ representing the amplitude vector of SAW-induced overall electric field, and the local equilibrium Fermi distribution function $\langle f_{\mathrm{\mathbf{k}}}(\mathbf{r},t)\rangle$ can be expanded to the first-order local electron density $n(\mathbf{r},t)$ as
\begin{align}
	\langle f_{\mathrm{\mathbf{k}}}(\mathbf{r},t)\rangle\approx f_\mathbf{k}^{(0)}+n_{1}(\mathbf{r},t)\partial_nf_\mathbf{k}^{(0)}.
\end{align}
with $\partial_n=\partial/\partial n$ and $f_\mathbf{k}^{(0)}$ indicating the equilibrium Fermi distribution function  which is independent on $\mathbf{r}$ and $t$. It should be mentioned that the first-order local electron density $n_{1}(\mathbf{r},t)=\frac{1}{2}n_1e^{i(\mathbf{q}\cdot\mathbf{r}-\omega t)}+c.c.$ is found to be linearly dependent on the amplitude ${\tilde{\mathcal{E}}}$[Eq.~(\ref{Appendix-A-NE}) in Appendix \ref{AE-coeff}].
Since we are interested in the direct nonlinear AE current ($j_\text{dc}\propto\tilde{\mathcal{E}}\tilde{\mathcal{E}}^*$, where subscript ``{dc}" refers to direct), we only expand the intrinsic NFDF $f_\mathbf{k}^{\mathrm{in}}$ up to second-order SAW-induced field $\tilde{\mathbf{E}}(\mathbf{r},t)$ as
\begin{align}
	f_{\mathrm{\mathbf{k}}}^{\mathrm{in}}=  f_{\mathrm{\mathbf{k}}}^{(0)}+\delta f_{\mathrm{\mathbf{k}}}^{\mathrm{in},1}+
\delta f_{\mathrm{\mathbf{k}}}^{\mathrm{in},2}+\cdots.
\label{App-A-ex}
\end{align}
According to  Eqs.(\ref{App-A-exp1}) and (\ref{App-A-ex}), the first-order (second-order) intrinsic NFDF  $\delta f_\mathbf{k}^{\mathrm{in},1}${ ($\delta f_\mathbf{k}^{\mathrm{in},2}$) is found to be, respectively,

\begin{equation}
	\begin{aligned}
		\delta f_{\mathrm{\mathbf{k}}}^{\text{in},1}= & \sum_{i=0}^{+\infty}P_{1}^{i}(\mathbf{r},t)\left[n_{1}(\mathbf{r},t)\partial_{n}f_{\mathbf{k}}^{(0)}
+\tau_{\mathbf{k}}\frac{e}{\hbar}\mathbf{\tilde{E}}\cdot\partial_{\mathbf{k}}f_{\mathrm{\mathbf{k}}}
^{(0)}\right],\\
		\delta f_{\mathrm{\mathbf{k}}}^{\text{in},2}= & \sum_{i=0}^{+\infty}\sum_{j=0}^{i}P_{1}^{i-j}(\mathbf{r},t)\Bigg\{\tau_{\mathbf{k}}\frac{e}{\hbar}
\mathbf{\tilde{E}}\cdot\partial_{\mathbf{k}}\\
		& \times P_{1}^{j}(\mathbf{r},t)\left[n_{1}(\mathbf{r},t)\partial_{n}f_{\mathbf{k}}^{(0)}+\tau_{\mathbf{k}}
\frac{e}{\hbar}\mathbf{\tilde{E}}\cdot\partial_{\mathbf{k}}f_{\mathrm{\mathbf{k}}}^{(0)}\right]\Bigg\},
	\end{aligned}
\label{Appn-inf12}
\end{equation}
where $P_1(\mathbf{r},t)=-\tau_{\mathbf{k}}\partial_{t}-\tau_{\mathbf{k}}\mathbf{v}_{\mathbf{k}}\cdot\partial_{\mathbf{r}}$, and the time-space parameters of the SAW-induced field are omitted for simplicity in Eq.~(
\ref{Appn-inf12}). Since only the stationary part of the second-order intrinsic NFDF $\delta f_{\mathrm{\mathbf{k}}}^{\mathrm{in},2}$ contributes to the direct AE current, we will only interested in $\delta f_{\mathrm{\mathbf{k}},\text{dc}}^{\mathrm{in},2}$. Therefore, after a tedious calculation, we obtain $\delta f_{\mathrm{\mathbf{k}},\text{dc}}^{\mathrm{in},2}$ and the first-order intrinsic NFDF $\delta f_\mathbf{k}^{\mathrm{in},1}$ as, respectively,
\begin{equation}
\begin{aligned}
	\delta f_{\mathrm{\mathbf{k}}}^{\mathrm{in},1}	=& \frac{1}{2}\delta f_{\mathrm{\mathbf{k}},\mathrm{am}}^{\mathrm{in},1}e^{i(\mathbf{q}\cdot\mathbf{r}-\omega t)}+c.c.,\\
\delta f_{\mathrm{\mathbf{k}},\text{dc}}^{\mathrm{in},2}= & \frac{e\tau_{\mathbf{k}}}{4\hbar}\vec{\tilde{\mathcal{E}}}^{*}\cdot\partial_{\mathbf{k}}
\delta f_{\mathrm{\mathbf{k}},\mathrm{am}}^{\mathrm{in},1}+c.c.
\end{aligned}
\label{APP-A-f1f2}
\end{equation}
where $\delta f_{\mathbf{k},\mathrm{am}}^{\mathrm{in},1}$ (subscript ``{am}" refers to amplitude) represents the amplitude of the first-order intrinsic NFDF and is determined by

\begin{align}
	\delta f_{\mathrm{\mathbf{k}},\mathrm{am}}^{\mathrm{in},1}=	\frac{\left(\frac{e\tau_{\mathbf{k}}}{\hbar}\vec{\tilde{\mathcal{E}}}
\cdot\partial_{\mathbf{k}}f_{\mathrm{\mathbf{k}}}^{(0)}+n_1
\partial_{n}f_{\mathbf{k}}^{(0)}\right)}{1-i\tau_{\mathbf{k}}
(\omega-\mathbf{v}_{\mathbf{k}}\cdot\mathbf{q})}.\label{AP-FDIn1}
\end{align}
The first (second) term in bracket of Eq.~\eqref{AP-FDIn1} can be regarded as the drift (diffusion) part.
Obviously, equation~\eqref{AP-FDIn1} shows that $\delta f_{\mathrm{\mathbf{k}},\mathrm{am}}^{\mathrm{in},1}$
is dependent on the amplitude of first-order local electron density $n_1$. Actually, through the continuity equation $\partial \rho/\partial t=-\nabla\cdot\mathbf{j}$, $n_1$ can, in turn, be determined by $\delta f_{\mathrm{\mathbf{k}},\mathrm{am}}^{\mathrm{in},1}$. The continuity equation gives rise to $-e\omega n_1=|\mathbf{q}\cdot\mathbf{j}^{(1)}|=\mathbf{q}\cdot\mathbf{j}_\mathrm{A}^{(1)}$. Substituting the formula for the amplitude of the first-order current $	\mathbf{j}_\mathrm{A}^{(1)}=-e\int[d\mathbf{k}]\mathbf{v_k}\delta f_{\mathbf{k},\mathrm{am}}^{\mathrm{in},1}$ into $-e\omega n_1=\mathbf{q}\cdot\mathbf{j}_\mathrm{A}^{(1)}$, we can have
\begin{align}
	e\omega n_1=e\int[d\mathbf{k}](\mathbf{v_k}\cdot\mathbf{q})\delta f_{\mathbf{k},\mathrm{am}}^{\mathrm{in},1}.
\label{APP-A-OME}
\end{align}
Combining Eqs.~\eqref{AP-FDIn1}~\eqref{APP-A-OME} and after a tedious derivation, we can have
\begin{equation}
\left\{
\begin{aligned}
\delta f_{\mathrm{\mathbf{k}},\mathrm{am}}^{\mathrm{in},1}&=	\tilde{\mathcal{E}}_{\alpha}G_{\alpha}(\omega,\mathbf{q},\mathbf{R},\mathbf{k},E_f),\\
n_1&=-\frac{q_{\alpha}\sigma_{\alpha\beta}\tilde{\mathcal{E}}_{\beta}}
{e(\omega-\mathbf{q}\cdotp\mathbf{R})},
\label{Appendix-A-NE}
\end{aligned}
\right.
\end{equation}
with
\begin{align}
	\sigma_{\alpha\beta}= & e^{2}\int[d\mathbf{k}]v_{\alpha}v_{\beta}\tau_{\mathbf{k}}\gamma(\omega,\mathbf{k})
\left(-\frac{\partial f_{\mathbf{k}}^{(0)}}{\partial\varepsilon_{\mathbf{k}}}\right),\label{App-A-sigma}\\
	\mathbf{R}= & \frac{\partial \mu}{\partial n}\int[d\mathbf{k}]\mathbf{v}_{\mathbf{k}}\gamma(\omega,\mathbf{k})\left(-\frac{\partial f_{\mathbf{k}}^{(0)}}{\partial\varepsilon_{\mathbf{k}}}\right),\label{AP-sigma-R}\\
G_{\alpha}=&\gamma(\omega,\mathbf{k})\left(\frac{e\tau_{\mathbf{k}}}{\hbar}\frac{\partial f_{\mathbf{k}}^{(0)}}{\partial k_{\alpha}}+\frac{q_{\beta}\sigma_{\alpha\beta}}{e(\omega-\mathbf{q}\cdotp\mathbf{R})}\frac{\partial \mu}{\partial n}\frac{\partial f_{\mathbf{k}}^{(0)}}{\partial\varepsilon_{\mathbf{k}}}\right).\label{AP-G_alpha}
\end{align}
In above, $\sigma_{\alpha\beta}$ and $\mathbf{R}$ are the conductivity tensor in the presence of SAW and the diffusion vector, respectively, with $\gamma(\omega,\mathbf{k})=[1-i\tau_{\mathbf{k}}(\omega-\mathbf{v_k}
\cdot\mathbf{q})]^{-1}$, $\mu$ is the chemical potential. It's should be pointed out that $\mu$ is usually equal to the Fermi energy $E_f$ and hence, we use $E_{f}$ replacing $\mu$ in the main text. Therefore, the amplitude vector of induced electric field $\vec{\mathcal{E}}^{i}$, which obeys the Maxwells equation, is found to be
\begin{align}
\vec{\mathcal{E}}^{i}&=	\frac{4\pi i e n_{1}\mathbf{q}}{q\epsilon_{0}(\epsilon+1)}
=-\frac{4\pi i}{\epsilon_{0}(\epsilon+1)q}\frac{q_{\alpha}\sigma_{\alpha\beta}\tilde{\mathcal{E}}_{\beta} }
{(\omega-\mathbf{q}\cdotp\mathbf{R})}\mathbf{q},\label{App-A-E_0i1}
\end{align}
Taking Eq.~(\ref{App-A-E_0i1}) into $\vec{\tilde{\mathcal{E}}}=\vec{\mathcal{E}}+\vec{\mathcal{E}}^{i}$, the dielectric function $g\left(\mathbf{q},\omega\right)$, which is defined as $\vec{\tilde{\mathcal{E}}}=\vec{\mathcal{E}}/g\left(\mathbf{q},\omega\right)$, can be determined as
\begin{align}
	g(\mathbf{q},\omega)=1+\frac{i\tilde{v}_{*}}{\omega-\mathbf{q}\cdot\mathbf{R}},
\label{App-A-G}
\end{align}
where the generalized diffusion velocity  $\tilde{v}_{*}$ is defined as
\begin{equation}
\tilde{v}_{*}=4\pi\frac{\sigma_{\alpha\beta}q_{\alpha}q_{\beta}}{\epsilon_{0}(\epsilon+1)q},
\end{equation}
In obtaining the Eq.~(\ref{App-A-G}) we have used the relation $\tilde{\mathcal{E}}_{\beta}\mathbf{q}={\mathcal{E}}_{\beta}\mathbf{q}/g(\mathbf{q},\omega)
=q_{\beta}\vec{\mathcal{E}}/g(\mathbf{q},\omega)$, which is guaranteed by the fact that
the wave vector $\mathbf{q}$ of SAW is parallel to the amplitude vector of piezoelectric field.
According to Eq.~(\ref{App-A-G}), The auxiliary function $F(\omega,\mathbf{q},\mathbf{R})$, which defined as $\vec{\tilde{\mathcal{E}}}=F(\omega,\mathbf{q},\mathbf{R})\vec{\mathcal{E}}$ can be determined as
\begin{equation}
F(\omega,\mathbf{q},\mathbf{R})=\frac{1}{g(\mathbf{q},\omega)}=\frac{\omega-\mathbf{q}
\cdotp\mathbf{R}}
{\omega-\mathbf{q}\cdotp\mathbf{R}+i\tilde{v}_{*}}
\label{App-A-F}
\end{equation}

\subsection{The side-jump induced modification $\delta f_\mathbf{k}^{\mathrm{sj}}$ to the nonequilibrium Fermi distribution function $f_\mathbf{k}$}\label{SI-D}
Based on the Eqs.~(\ref{rk})(\ref{BTE-three}), the side-jump induced modification $\delta f_\mathbf{k}^{\mathrm{sj}}$ (where superscript ``{sj}" refers to side jump) can be determined by
\begin{align}
	&\left[\partial_{t}+\left(\mathbf{v}_{\mathbf{k}}+\mathbf{v}_{\mathbf{k}}^{\mathrm{sj}}\right)
\cdot\partial_{\mathbf{r}}-\frac{e}{\hbar}\tilde{\mathbf{E}}(\mathbf{r},t)\cdot\partial_{\mathbf{k}}
\right]\delta f_{\mathrm{\mathbf{k}}}^{\mathrm{sj}}\notag\\ =&I_{c}^{\mathrm{in}}[\delta f_{\mathrm{\mathbf{k}}}^{\mathrm{sj}}]+I_{c}^{\mathrm{sj}}
[f_{\mathrm{\mathbf{k}}}^{\mathrm{in}}],\label{AP-SjBTE}
\end{align}
where $\mathbf{v}_{\mathbf{k}}^{\mathrm{sj}}=\sum_{\mathbf{k}^{\prime}}w_{\mathbf{k}\mathbf{k}^{\prime}}
\delta\mathbf{r}_{\mathbf{k}^{\prime}\mathbf{k}}$ represents the side-jump velocity stemming from
the impurity-scattering induced coordinate shift $\delta\mathbf{r}_{\mathbf{k}^{\prime}\mathbf{k}}$, $I_{c}^{\mathrm{in}}[\delta f_{\mathrm{\mathbf{k}}}^{\mathrm{sj}}]$ ($I_{c}^{\mathrm{sj}}
[f_{\mathrm{\mathbf{k}}}^{\mathrm{in}}]$) refer to the intrinsic (side-jump) collision term, respectively. We approximate $I_{c}^{\mathrm{in}}[\delta f_{\mathrm{\mathbf{k}}}^{\mathrm{sj}}]\approx -{\delta f_{\mathrm{\mathbf{k}}}^{\mathrm{sj}}}/{\tau_{\mathbf{k}}}$ using the relaxation time approximation since the only symmetric scattering is contained in intrinsic collision term [Eq.~(\ref{IcInSjSk})]. Therefore, combining with Eq.~(\ref{IcInSjSk}), the total collision term in Eq.~(\ref{AP-SjBTE}) for the side-jump scattering is found to be
\begin{align}
I_{c}^{\mathrm{in}}[\delta f_{\mathrm{\mathbf{k}}}^{\mathrm{sj}}]+I_{c}^{\mathrm{sj}}
[f_{\mathrm{\mathbf{k}}}^{\mathrm{in}}]\!=\!	-\frac{\delta f_{\mathrm{\mathbf{k}}}^{\mathrm{sj}}}{\tau_{\mathbf{k}}}\!-\!e\mathbf{\tilde{E}}(\mathbf{r},t)
\cdot\sum_{\mathbf{k}^{\prime}}\mathbf{O}_{\mathbf{k}\mathbf{k}^{\prime}}
(f_{\mathrm{\mathbf{k}}}^{\mathrm{in}}\!-\!f_{\mathrm{\mathbf{k}}^{\prime}}^{\mathrm{in}})\label{AP-SjIC}
\end{align}
 with
\begin{align}
	\mathbf{O}_{\mathbf{k}\mathbf{k}^{\prime}}= \frac{2\pi}{\hbar}|T_{\mathbf{k}\mathbf{k}^{\prime}}|^{2}\delta\mathbf{r}_{\mathbf{k}\mathbf{k}^{\prime}}
\frac{\partial}{\partial\varepsilon_{\mathbf{k}}}\delta(\varepsilon_{\mathbf{k}}-\varepsilon_{\mathbf{k}^{\prime}}).
\end{align}
The impurity-scattering induced coordinate shift $\text{\ensuremath{\delta\mathbf{r_{kk^{\prime}}}}}$ is determined as
\begin{align}
	\text{\ensuremath{\delta\mathbf{r_{kk^{\prime}}}}}= & \langle \psi_{\mathbf{k}}|i\partial_{\mathbf{k}}\psi_{\mathbf{k}}\rangle-\langle \psi_{\mathbf{k^{\prime}}}|i\partial_{\mathbf{k^{\prime}}}\psi_{\mathbf{k^{\prime}}}\rangle\notag\\
	& -(\partial_{\mathbf{k}}+\partial_{\mathbf{k^{\prime}}})\mathrm{arg}[\langle \psi_{\mathbf{k}}|\psi_{\mathbf{k}^{\prime}}\rangle],
\label{App-B-shif}
\end{align}
where $\psi_\mathbf{k}(\psi_\mathbf{k^\prime})$ is the initial (final) state in scattering processes, respectively.
Combining Eqs.\eqref{AP-SjBTE}\eqref{AP-SjIC}, $\delta f_\mathbf{k}^{\mathrm{sj}}$ can be repressed as
\begin{align}
	\delta f_{\mathrm{\mathbf{k}}}^{\mathrm{sj}}= & -\frac{e\tau_{\mathbf{k}}\mathbf{\tilde{E}}(\mathbf{r},t)\cdot\sum_{\mathbf{k}^{\prime}}
\mathbf{O}_{\mathbf{k}\mathbf{k}^{\prime}}(f_{\mathrm{\mathbf{k}}}^{\mathrm{in}}-
f_{\mathrm{\mathbf{k}}^{\prime}}^{\mathrm{in}})}{1+\tau_{\mathbf{k}}\partial_{t}+\tau_{\mathbf{k}}
\mathbf{\tilde{v}}_{\mathbf{k}}\cdot\partial_{\mathbf{r}}-\frac{e\tau_{\mathbf{k}}}{\hbar}
\mathbf{\tilde{E}}(\mathbf{r},t)\cdot\partial_{\mathbf{k}}}\notag\\
\approx & -e\sum_{n=0}^{+\infty}P_{2}(\mathbf{k},\mathbf{r},t)^{n}\left[\tau_{\mathbf{k}}\mathbf{\tilde{E}}
(\mathbf{r},t)\cdot\left(\mathbf{Q}_{\mathbf{k}}^{(0)}+\mathbf{Q}_{\mathbf{k}}^{(1)}\right)\right]
\notag\\
	\approx & \delta f_{\mathbf{k}}^{\mathrm{sj,1}}+\delta f_{\mathbf{k}}^{\mathrm{sj,2}}+\cdots
\label{A-FSj}
\end{align}
with 
\begin{equation}
\left\{
\begin{aligned}
	\mathbf{\tilde{v}}_{\mathbf{k}}= & \mathbf{v}_{\mathbf{k}}+\mathbf{v}_{\mathbf{k}}^{\mathrm{sj}},\\
	\mathbf{Q}_{\mathbf{k}}^{(0)}= & \sum_{\mathbf{k}^{\prime}}\mathbf{O}_{\mathbf{k}\mathbf{k}^{\prime}}(f_{\mathrm{\mathbf{k}}}^{(0)}
-f_{\mathrm{\mathbf{k}}^{\prime}}^{(0)})=\mathbf{v}_{\mathbf{k}}^{\mathrm{sj}}\frac{\partial f_{\mathrm{\mathbf{k}}}^{(0)}}{\partial\varepsilon_{\mathbf{k}}},\\
	\mathbf{Q}_{\mathbf{k}}^{(1)}= & \sum_{\mathbf{k}^{\prime}}\mathbf{O}_{\mathbf{k}\mathbf{k}^{\prime}}(\delta f_{\mathbf{k}}^{\mathrm{in},1}
-\delta f_{\mathrm{\mathbf{k}}^{\prime}}^{\mathrm{in},1}),\\
	P_{2}(\mathbf{k},\mathbf{r},t)= & -\tau_{\mathbf{k}}\partial_{t}-\tau_{\mathbf{k}}\mathbf{\tilde{v}}_{\mathbf{k}}\cdot
\partial_{\mathbf{r}}+\frac{e\tau_{\mathbf{k}}}{\hbar}\mathbf{\tilde{E}}(\mathbf{r},t)
\cdot\partial_{\mathbf{k}}.
\end{aligned}
\label{APP-A-Cdt}
\right.
\end{equation}
In obtaining the second line of Eq. (\ref{A-FSj}), only the zeroth and first-order of the $f_\mathbf{k}^{\mathrm{in}}$ to the SAW-induced field have been kept since we are interested in the response up to the second order in the SAW-induced field. Therefore, the first-order (the second-order) side-jump induced modification $\delta f_{\mathbf{k}}^{\mathrm{sj,1}}$ ($\delta f_{\mathbf{k}}^{\mathrm{sj,2}}$) response to the SAW-induced field are determined as, respectively,
\begin{equation}
\begin{aligned}
			\delta f_{\mathbf{k}}^{\mathrm{sj,1}}= & -e\sum_{i=0}^{+\infty}P_{1}^{i}(\mathbf{r},t)\left(\tau_{\mathbf{k}}\tilde{\mathbf{E}}(\mathbf{r},t)\cdot\mathbf{Q}_{\mathbf{k}}^{(0)}\right),\\
			\delta f_{\mathbf{k}}^{\mathrm{sj,2}}= & -e\sum_{i=0}^{+\infty}P_{1}^{i}(\mathbf{r},t)\left(\tau_{\mathbf{k}}\tilde{\mathbf{E}}(\mathbf{r},t)\cdot\mathbf{Q}_{\mathbf{k}}^{(1)}\right)\\
			& -e\sum_{i=0}^{+\infty}\sum_{j=0}^{i}P_{1}^{i-j}(\mathbf{r},t)\Bigg\{\tau_{\mathbf{k}}\frac{e}{\hbar}\mathbf{\tilde{E}}(\mathbf{r},t)\cdot\partial_{\mathbf{k}}\\
			& \times\left[P_{1}^{j}(\mathbf{r},t)\left(\tau_{\mathbf{k}}\tilde{\mathbf{E}}(\mathbf{r},t)\cdot\mathbf{Q}_{\mathbf{k}}^{(0)}\right)\right]\Bigg\}.
			\end{aligned}
\end{equation}
One might observe that only  $\mathbf{v}_{\mathbf{k}}$ part of $\mathbf{\tilde{v}}_{\mathbf{k}}=\mathbf{v}_{\mathbf{k}}+\mathbf{v}_{\mathbf{k}}^{\mathrm{sj}}$ involved in $\delta f_{\mathbf{k}}^{\mathrm{sj,1}}$. That's because that the side-jump velocity $\mathbf{v}_{\mathbf{k}}^{\mathrm{sj}}$ will lead to the high-order side-jump effect and has been omitted. After a tedious derivation, we can obtain
\begin{equation}
\begin{aligned}
	\delta f_{\mathbf{k}}^{\mathrm{sj,1}}= & \frac{1}{2}\delta f_{\mathbf{k},\mathrm{am}}^{\mathrm{sj,1}}e^{i(\mathbf{q}\cdot\mathbf{r}-\omega t)}+c.c.\\
	\delta f_{\mathbf{k},\text{dc}}^{\mathrm{sj,2}}=
	& -\frac{e\tau_{\mathbf{k}}}{4}\tilde{\mathcal{E}}^{*}\cdot\sum_{\mathbf{k}^{\prime}}
\mathbf{O}_{\mathbf{k}\mathbf{k}^{\prime}}\left(\delta f_{\mathrm{\mathbf{k}},\mathrm{am}}^{\mathrm{in},1}-\delta
f_{\mathrm{\mathbf{k^{\prime}}},\mathrm{am}}^{\mathrm{in},1}\right)\\
&-\frac{e^{2}\tau_{\mathbf{k}}}{4\hbar}\tilde{\mathcal{E}}^{*}
\cdot\partial_{\mathbf{k}}\delta f_{\mathbf{k},\mathrm{am}}^{\mathrm{sj,1}}+c.c.\label{AP-FSj2dc}
\end{aligned}
\end{equation}
where $\delta f_{\mathbf{k},\text{dc}}^{\mathrm{sj,2}}$ (subscript ``{dc}" represents direct) denotes the stationary part of the second-order side-jump modified NFDF $\delta f_{\mathbf{k},\text{dc}}^{\mathrm{sj},2}$), and  $\delta f_{\mathbf{k},\mathrm{am}}^{\mathrm{sj,1}}$ means the amplitude of first-order side-jump induced modification to NFDF and is given by
\begin{align}
		\delta f_{\mathbf{k},\mathrm{am}}^{\mathrm{sj,1}}=	-e\tau_{\mathbf{k}}\gamma(\omega,\mathbf{k})(\tilde{\mathcal{E}}\cdot\mathbf{v_k}^{\mathrm{sj}})\frac{\partial f_{\mathrm{\mathbf{k}}}^{(0)}}{\partial\varepsilon_{\mathbf{k}}}.\label{AP-FSj01-2}
\end{align}

\subsection{The skew-scattering induced modification $\delta f_\mathbf{k}^{\mathrm{sk}}$ to the nonequilibrium Fermi distribution function $f_\mathbf{k}$}
According to  the Eqs.~(\ref{rk})(\ref{IcInSjSk})(\ref{BTE-three}), the skew-scattering induced modification $\delta f_\mathbf{k}^{\mathrm{sk}}$ (where superscript ``{sk}" refers to skew scattering) can be determined by
\begin{align}
	&\left[\partial_{t}+\mathbf{v}_{\mathbf{k}}\cdot\partial_{\mathbf{r}}-\frac{e}{\hbar}
\mathbf{\tilde{E}}(\mathbf{r},t)\cdot\partial_{\mathbf{k}}\right]f_{\mathrm{\mathbf{k}}}^{\mathrm{sk}}
(\mathbf{r},t)\notag\\
=&
-\frac{f_{\mathrm{\mathbf{k}}}^{\mathrm{sk}}}{\tau_{\mathbf{k}}}-\sum_{\mathbf{k}^{\prime}}
w_{\mathbf{k}^{\prime}\mathbf{k}}^{\mathrm{A}}
(f_{\mathrm{\mathbf{k}}}^{\mathrm{in}}+f_{\mathrm{\mathbf{k}}^{\prime}}^{\mathrm{in}}).
\label{APP-A-sek}
\end{align}
It's should be pointed that, being similar to the side-jump contribution, $I_{c}^{\mathrm{in}}[f_{\mathrm{\mathbf{k}}}^{\mathrm{sk}}(\mathbf{r},t)]$ has also been approximated to be $-{f_{\mathrm{\mathbf{k}}}^{\mathrm{sk}}}/{\tau_{\mathbf{k}}}$ in Eq.~(\ref{APP-A-sek}). Then, the skew-scattering induced modification $\delta f_{\mathrm{\mathbf{k}}}^{\mathrm{sk}}$ can be expanded to the $n$-th order in the SAW induced electric field as
\begin{equation}
\begin{aligned}
	\delta f_{\mathrm{\mathbf{k}}}^{\mathrm{sk}}= & -\frac{\tau_{\mathbf{k}}\sum_{\mathbf{k}^{\prime}}w_{\mathbf{k}^{\prime}\mathbf{k}}^{\mathrm{A}}
(f_{\mathrm{\mathbf{k}}}^{\mathrm{in}}+f_{\mathrm{\mathbf{k}}^{\prime}}^{\mathrm{in}})}
{1+\tau_{\mathbf{k}}\partial_{t}+\tau_{\mathbf{k}}\mathbf{v}_{\mathbf{k}}\cdot\partial_{\mathbf{r}}
-\tau_{\mathbf{k}}\frac{e}{\hbar}\mathbf{\tilde{E}}\cdot\partial_{\mathbf{k}}}\\
	= & -\sum_{n=0}^{\infty}P_{2}(\mathbf{k},\mathbf{r},t)^{n}\left[\tau_{\mathbf{k}}
\sum_{\mathbf{k}^{\prime}}w_{\mathbf{k}^{\prime}\mathbf{k}}^{\mathrm{A}}
(f_{\mathrm{\mathbf{k}}}^{\mathrm{in}}+f_{\mathrm{\mathbf{k}}^{\prime}}^{\mathrm{in}})\right]\\
	= &-\sum_{n=0}^{\infty}P_{2}(\mathbf{k},\mathbf{r},t)^{n}\left[\tau_{\mathbf{k}}\sum_{\mathbf{k}^{\prime}}
w_{\mathbf{k}^{\prime}\mathbf{k}}^{\mathrm{A}}\left(\delta  f_{\mathrm{\mathbf{k}}}^{\mathrm{in},1}+\delta f_{\mathrm{\mathbf{k}}^{\prime}}^{\mathrm{in},1}\right.\right.\\
	&\left.\left.+\delta f_{\mathrm{\mathbf{k}}}^{\mathrm{in},2}+
\delta f_{\mathrm{\mathbf{k}}^{\prime}}^{\mathrm{in},2}+\cdots\right)\right]\\
	=&\delta f_{\mathrm{\mathbf{k}}}^{\mathrm{sk,1}}+\delta f_{\mathrm{\mathbf{k}}}^{\mathrm{sk,2}}+\cdots.
\end{aligned}
\label{AP-FSk}
\end{equation}
To obtain the third line, we have used the fact that the equilibrium distribution function has no contribution to scattering, i.e, $\sum_{\mathbf{k}^{\prime}}w_{\mathbf{k}^{\prime}\mathbf{k}}^{\mathrm{A}}
(f_{\mathbf{k}}^{(0)}+f_{\mathbf{k}^{\prime}}^{(0)})=0$. According to Eq.~(\ref{AP-FSk}), the skew-scattering induced modification $\delta f_{\mathrm{\mathbf{k}}}^{\mathrm{sk,1}}$ ($\delta f_{\mathrm{\mathbf{k}}}^{\mathrm{sk,2}}$) as response to the first-order (second-order) SAW-induced field is found to be, respectively,
\begin{equation}
			\begin{aligned}
			\delta f_{\mathbf{k}}^{\mathrm{sk,1}}= & -\sum_{i=0}^{+\infty}P_{1}^{i}(\mathbf{r},t)\tau_{\mathbf{k}}\sum_{\mathbf{k^{\prime}}}
\omega_{\mathbf{k^{\prime}k}}^{\mathrm{A}}\left(\delta f_{\mathbf{k}}^{\mathrm{In,1}}+\delta f_{\mathbf{k^{\prime}}}^{\mathrm{In,1}}\right),\\
			\delta f_{\mathbf{k}}^{\mathrm{sk,2}}= & -\sum_{i=0}^{+\infty}P_{1}^{i}(\mathbf{r},t)\tau_{\mathbf{k}}\sum_{\mathbf{k^{\prime}}}
\omega_{\mathbf{k^{\prime}k}}^{\mathrm{A}}\left(\delta f_{\mathbf{k}}^{\mathrm{In,2}}+\delta f_{\mathbf{k^{\prime}}}^{\mathrm{In,2}}\right)\\
			& -\sum_{i=0}^{+\infty}\sum_{j=0}^{i}P_{1}^{i-j}(\mathbf{r},t)\Bigg\{\tau_{\mathbf{k}}
\frac{e}{\hbar}\mathbf{\tilde{E}}(\mathrm{r},t)\cdot\partial_{\mathbf{k}}\\
			& \times\left[P_{1}^{j}(\mathbf{r},t)\tau_{\mathbf{k}}\sum_{\mathbf{k^{\prime}}}
\omega_{\mathbf{k^{\prime}k}}^{\mathrm{A}}\left(\delta f_{\mathbf{k}}^{\mathrm{In,1}}+\delta f_{\mathbf{k^{\prime}}}^{\mathrm{In,1}}\right)\right]\Bigg\},
			\end{aligned}
\end{equation}
yielding
\begin{equation}
\begin{aligned}
\delta f_{\mathrm{\mathbf{k}}}^{\mathrm{sk,1}}&=	\frac{1}{2}\delta f_{\mathrm{\mathbf{k}},\mathrm{am}}^{\mathrm{sk,1}}e^{i(\mathbf{q}\cdot\mathbf{r}-\omega t)}+c.c.,\\
\delta f_{\mathrm{\mathbf{k}},\text{dc}}^{\mathrm{sk},2}&=\frac{e\tau_{\mathbf{k}}}{4\hbar}
\tilde{\mathcal{E}}^{*}\cdot\partial_{\mathbf{k}}
\delta f_{\mathrm{\mathbf{k}},\mathrm{am}}^{\mathrm{sk,1}}+\frac{e\tau_{\mathbf{k}}}{4\hbar}\tilde{\mathcal{E}}
\cdot\partial_{\mathbf{k}}\left(\delta f_{\mathrm{\mathbf{k}},\mathrm{am}}^{\mathrm{sk,1}}\right)^{*}\\
	& -\tau_{\mathbf{k}}\sum_{\mathbf{k}^{\prime}}w_{\mathbf{k}^{\prime}\mathbf{k}}^{\mathrm{A}}
\left(\delta f_{\mathrm{\mathbf{k}},\text{dc}}^{\mathrm{in},2}+\delta f_{\mathrm{\mathbf{k}}^{\prime},\text{dc}}^{\text{in},2}\right)\label{AP-FSjdc}
\end{aligned}
\end{equation}
with $\delta f_{\mathbf{k},\text{dc}}^{\mathrm{sk,2}}$ (subscript ``{dc}" represents direct) representing the stationary part of the second-order skew-scattering modification, and  $\delta f_{\mathbf{k},\mathrm{am}}^{\mathrm{sk,1}}$ indicating the amplitude of the first-order skew-scattering induced modification to NFDF, which is
\begin{align}
	\delta f_{\mathrm{\mathbf{k}},\mathrm{am}}^{\mathrm{sk,1}}=	-\tau_{\mathbf{k}}\gamma(\omega,\mathbf{k})\sum_{\mathbf{k}^{\prime}}w_{\mathbf{k}^{\prime}\mathbf{k}}^{\mathrm{A}}(\delta f_{\mathrm{\mathbf{k}},\mathrm{am}}^{\mathrm{in},1}+\delta f_{\mathrm{\mathbf{k}}^{\prime},\mathrm{am}}^{\mathrm{in},1}),\label{AP-FSk1}
\end{align}

\section{The scattering rate $w_{\mathbf{k}\mathbf{k}^{\prime}}$, the relaxation time $\tau_\mathbf{k}$ and the velocity of side jump $\mathbf{v}_{\mathbf{k}}^{\mathrm{sj}}$ for disordered massive Dirac materials with a $\delta$-function random potential scatter}\label{scat-rates-time}
In this section, the scattering rates $w_{\mathbf{k}\mathbf{k}^{\prime}}$ and the relaxation time $\tau_\mathbf{k}$ will be deduced for the upper band of the massive Dirac materials in presence of the disorder ($\delta$-function random potential scatters).  The scattering rate $w_{ll^{\prime}}$, where $l=(n,\mathbf{k})$ is an combined index with the energy band $n$ and momentum $\mathbf{k}$, can be determined through the Fermi gold rules as
\begin{equation}
\begin{aligned}
{w}_{ll^{\prime}}&=\frac{2\pi}{\hbar}|T_{l
l^{\prime}}|^{2}\delta(\epsilon_l
-\epsilon_{l^{\prime}})
\end{aligned}
\end{equation}
with the scattering T-matrix $T_{ll^{\prime}}$ defined as $T_{ll^{\prime}}=\langle \psi_{l}|\hat{V}_\text{imp}|\Psi_{l^{\prime}} \rangle$, where $|\psi_{l}\rangle$ and $|\Psi_{l}\rangle$ represent the eigenstate of  $\hat{H}_{0}$ and $\hat{H}_{0}+\hat{V}_\text{imp}$, respectively. $|\Psi_{l}\rangle$ satisfies the Lippman-Schwinger equation \cite{du2019disorder,Qiang2023PRB}
\begin{equation}
|\Psi_{l}\rangle=|\psi_l\rangle+\frac{\hat{V}_\text{imp}}{\epsilon_l-
\hat{H}_{0}+i\delta}|\Psi_{l}\rangle.
\end{equation}
For a weak disorder, the scattering state $|\Psi_{l}\rangle$ can be approximated by a truncated series in powers of $V_{ll^{\prime}}=\langle \psi_{l}|\hat{V}_\text{imp}|\psi_{l^{\prime}} \rangle$ as
\begin{equation}
\begin{aligned}
|\Psi_{l}\rangle&=|\psi_{l}\rangle+
\sum_{l_{1}}\frac{{V}_{l_{1}l}}{\epsilon_l-
\epsilon_{l_{1}}+i\delta}|\psi_{l_{1}}\rangle\\
&+\sum_{l_{1}l_{2}}\frac{{V}_{l_{1}l_{2}}
{V}_{l_{2}l}}
{\left(\epsilon_l-
\epsilon_{l_{1}}+i\delta\right)\left(\epsilon_l-
\epsilon_{l_{2}}+i\delta\right)}|\psi_{l_{1}}\rangle +\cdot\cdot\cdot.
\end{aligned}
\end{equation}
Accordingly, the scattering rate ${w}_{ll^{\prime}}$ could be expanded to the $n$-th order in disorder as

\begin{equation}
\begin{aligned}
{w}_{ll^{\prime}}&\equiv{w}^{(2)}_{ll^{\prime}}
+{w}^{(3)}_{ll^{\prime}}+{w}^{(4)}_{ll^{\prime}}\cdot\cdot\cdot\\
\end{aligned}
\label{App-B-w11}
\end{equation}
with the zero and first order  scattering rate are found to be zero, and
\begin{equation}
\begin{aligned}
	w_{ll}^{(2)}= & \frac{2\pi}{\hbar}\langle V^{*}_{ll^{\prime}}V_{ll^{\prime}}\rangle_\text{dis}
\delta(\varepsilon_{l}-\varepsilon_{l^{\prime}}),\\
	w_{ll^{\prime}}^{(3)}=&\sum_{l_{1}}\left(
\frac{\langle V^{*}_{ll^{\prime}}V_{ll_{1}}V_{l_{1}
l^{\prime}}\rangle_\text{dis}}{\epsilon_{l^{\prime}}-\epsilon_{l_{1}}
+i\delta}+\frac{\langle V^{*}_{ll_{1}}V^{*}_{l_{1}
	l^{\prime}}V_{ll^{\prime}}\rangle_\text{dis}}{\epsilon_{l^{\prime}}-\epsilon_{l_{1}}
-i\delta}\right)\\
&\times\frac{2\pi\delta(\varepsilon_{l}-\varepsilon_{l^{\prime}})}{\hbar},\\
w_{ll^{\prime}}^{(4)}=&\frac{2\pi}{\hbar}
\sum_{l_{1}l_{2}}
\left[\frac{\langle V^{*}_{ll_{1}}V^{*}_{l_{1}l^{\prime}}
V_{ll_{2}}V_{l_{2}l^{\prime}}\rangle_\text{dis}}
{\left(\epsilon_{l^{\prime}}-\epsilon_{l_{1}}
-i\delta\right)\left(\epsilon_{{l}^{\prime}}-\epsilon_{l_{2}}
+i\delta\right)}\right.\\
&+\left.\frac{\langle V^{*}_{ll^{\prime}}V_{ll_1}
V_{l_{1}l_{2}}V_{l_{2}l^{\prime}}\rangle_\text{dis}}
{\left(\epsilon_{l^{\prime}}-\epsilon_{l_{1}}
+i\delta\right)\left(\epsilon_{l^{\prime}}-\epsilon_{l_{2}}
+i\delta\right)}\right.\\
&+\left.\left.\frac{\langle V_{ll^{\prime}}V^{*}_{ll_{1}}
V^{*}_{ll_{2}}V^{*}_{l_{2}l^{\prime}}\rangle_\text{dis}}
{\left(\epsilon_{l}-\epsilon_{l_{1}}
-i\delta\right)\left(\epsilon_{l^{\prime}}-\epsilon_{l_{2}}
-i\delta\right)}\right.
\right]\delta(\varepsilon_{l}-\varepsilon_{l^{\prime}}).
\end{aligned}
\label{App-B-w1-series}
\end{equation}
We only focus on the upper band ($n=+1$) for simplicity, namely the Fermi level lies in the upper band. Therefore, we only need to find $w_{+\mathbf{k},+\mathbf{k}^{\prime}}$, which will be repressed as $w_{\mathbf{k},\mathbf{k}^{\prime}}$ in the following and main text for simplicity.
As discussed in Sec.~{\ref{Theoretical review}}, the scattering rate can be decomposed two parts as $w_{\mathbf{k}\mathbf{k}^{\prime}}=w^\text{S}_{\mathbf{k}\mathbf{k}^{\prime}}+
w^\text{A}_{\mathbf{k}\mathbf{k}^{\prime}}$, where the symmetric (antisymmetric) $w^\text{S}_{\mathbf{k}\mathbf{k}^{\prime}}$ ($w^\text{A}_{\mathbf{k}\mathbf{k}^{\prime}}$) is determined as
\begin{equation}
w^\text{S}_{\mathbf{k}\mathbf{k}^{\prime}}=\frac{w_{\mathbf{k}\mathbf{k}^{\prime}}+
w_{\mathbf{k}^{\prime}\mathbf{k}}}{2},\,\,\,
w^\text{A}_{\mathbf{k}\mathbf{k}^{\prime}}=\frac{w_{\mathbf{k}\mathbf{k}^{\prime}}-
w_{\mathbf{k}^{\prime}\mathbf{k}}}{2}.
\label{App-B-WAB}
\end{equation}
Based on Eqs.~\eqref{App-B-w11}\eqref{App-B-w1-series} and \eqref{App-B-WAB}, the nonzero leading order  terms of $w^\text{S}_{\mathbf{k}\mathbf{k}^{\prime}}$ and $w^\text{A}_{\mathbf{k}\mathbf{k}^{\prime}}$ are found to be, respectively,

\begin{align}
	w_{\mathbf{k}\mathbf{k}^{\prime}}^{\mathrm{S,(2)}}= & \frac{2\pi}{\hbar}\langle V^{*}_{\mathbf{k}\mathbf{k}^{\prime}}V_{\mathbf{k}\mathbf{k}^{\prime}}\rangle_\text{dis}
\delta(\varepsilon_{\mathbf{k}}-\varepsilon_{\mathbf{k}^{\prime}})\notag,\\
	w_{\mathbf{k}\mathbf{k}^{\prime}}^{\mathrm{A,(3)}}= & \frac{4\pi^{2}}{\hbar}\sum_{\mathbf{k}^{\prime\prime}}\mathrm{Im}\langle
V^{*}_{\mathbf{k}\mathbf{k}^{\prime}}V_{\mathbf{k}\mathbf{k}^{\prime\prime}}
V_{\mathbf{k^{\prime\prime}k}^{\prime}}\rangle_\text{dis}\notag\\
	& \times\delta(\varepsilon_{\mathbf{k}}-\varepsilon_{\mathbf{k^{\prime\prime}}})
\delta(\varepsilon_{\mathbf{k}}-\varepsilon_{\mathbf{k^{\prime}}})\notag,\\
	w_{\mathbf{k}\mathbf{k}^{\prime}}^{\mathrm{A,(4)}}= & -\frac{2\pi}{\hbar}\sum_{\mathbf{k}^{\prime\prime}}\Bigg\{\frac{1}{\varepsilon_{\mathbf
{k^{\prime\prime}}}^{+}-\varepsilon_{\mathbf{k^{\prime\prime}}}^{-}}\mathrm{Im}\langle V_{\mathbf{k^{\prime}k^{\prime\prime}}}^{++}V_{\mathbf{k^{\prime\prime}k^{\prime}}}^{-+}
\rangle_\text{dis}\notag\\
	& \times\langle V_{\mathbf{k^{\prime\prime}k}}^{++}V_{\mathbf{k\mathbf{k^{\prime\prime}}}}^{+-}\rangle_\text{dis}\notag\\
	& -\frac{1}{\varepsilon_{\mathbf{k}}^{+}-\varepsilon_{\mathbf{k}}^{-}}\mathrm{Im}\langle V_{\mathbf{k^{\prime}k}}^{++}V_{\mathbf{kk^{\prime}}}^{-+}\rangle_{dis}\langle V_{\mathbf{k^{\prime\prime}k}}^{+-}V_{\mathbf{k\mathbf{k^{\prime\prime}}}}^{++}\rangle_\text{dis}\notag\\
	& -\frac{1}{\varepsilon_{\mathbf{k^{\prime}}}^{+}-\varepsilon_{\mathbf{k^{\prime}}}^{-}}\mathrm{Im}\langle V_{\mathbf{k^{\prime}k}}^{++}V_{\mathbf{kk^{\prime}}}^{+-}\rangle_\text{dis}\langle V_{\mathbf{k^{\prime\prime}k^{\prime}}}^{++}V_{\mathbf{k^{\prime}\mathbf{k^{\prime\prime}}}}^{-+}
\rangle_\text{dis}\Bigg\}\notag\\
	& \times\delta(\varepsilon_{\mathbf{k^{\prime}}}-\varepsilon_{\mathbf{k}})
\delta(\varepsilon_{\mathbf{k^{\prime}}}-\varepsilon_{\mathbf{k^{\prime\prime}}}),
\label{App-B-dfjd}
\end{align}
where scattering potential matrix element $V_{\mathbf{kk^\prime}}^{\pm\pm}=V_{\mathbf{kk^\prime}}^{0}\langle\psi_{\mathbf{k}}^\pm|
\psi_{\mathbf{k}^{\prime}}^\pm\rangle$ with the orbital disorder matrix element $V_{\mathbf{kk^\prime}}^{0}\equiv\sum_{i}V_{i}e^{i(\mathbf{k}^{\prime}-\mathbf{k})\mathbf{R}_{i}}$ satisfies $\langle(V_{\mathbf{kk^\prime}}^0)^*V_{\mathbf{kk^\prime}}^0\rangle_\text{dis}\approx n_iV_0^2$ and $\langle(V_\mathbf{kk^\prime}^0)^*V_\mathbf{kk^{\prime \prime}}^0V_\mathbf{k^{\prime \prime}k^\prime}^0\rangle_\text{dis}\approx n_i V_0^3$.\cite{du2019disorder,Qiang2023PRB} The eigenstates of the Hamiltonian $\hat{H}_{0}$ [Eq.\eqref{Hami}] fo the massive Dirac two-dimensional materials are
\begin{align}
	|\psi_{\mathbf{k}}^{+}\rangle= & \left[\tau_{v}\cos\frac{\theta}{2},\sin\frac{\theta}{2}e^{\tau_{v}i\varphi}\right]^{T},\notag\\
	|\psi_{\mathbf{k}}^{-}\rangle= & \left[\sin\frac{\theta}{2},-\tau_{v}\cos\frac{\theta}{2}e^{\tau_{v}i\varphi}\right]^{T},
\label{App-B-EST}
\end{align}
where $+$ ($-$) represents conduction (valence) band, $\tau_{v}$ denotes the valley index, $\theta$ is determined by $\tan\theta=v_{F}\hbar k/m$, and $\varphi$ is given by $\tan\varphi=k_{y}/k_{x}$. Based on the formulas of $|\psi_{\mathbf{k}}^{+(-)}\rangle$ [Eq.~\eqref{App-B-EST}], $V_{\mathbf{kk^\prime}}^{\pm\pm}=V_{\mathbf{kk^\prime}}^{0}\langle\psi_{\mathbf{k}}^\pm|
\psi_{\mathbf{k}^{\prime}}^\pm\rangle$ can be determined. Taking the determined formulas of $V_{\mathbf{kk^\prime}}^{\pm\pm}$  into Eq.~\eqref{App-B-dfjd}, the nonzero leading order scattering rates for the symmetric and anti-symmetric parts are obtained as
\begin{equation}
\begin{aligned}
w_{\mathbf{k}\mathbf{k}^{\prime}}^{\mathrm{S,(2)}}&=\frac{\pi n_{i}V_{0}^{2}}{\hbar}\left[1+\frac{m^{2}+v^{2}kk^{\prime}\cos\Delta\varphi}
{\varepsilon_{\mathbf{k}}\varepsilon_{\mathbf{k}^{\prime}}}\right]
\delta(\varepsilon_{\mathbf{k}}-\varepsilon_{\mathbf{k}^{\prime}}),\\
w_{\mathbf{k}\mathbf{k}^{\prime}}^{\mathrm{A,(3)}}&= \tau_{v}\frac{\pi n_{i}V_{1}^{3}}{2\hbar}\frac{mkk^{\prime}}{\varepsilon_{\mathbf{k}}\varepsilon_{\mathbf{k^{\prime}}}}
\delta(\varepsilon_{\mathbf{k}}
-\varepsilon_{\mathbf{k^{\prime}}})\sin\Delta\varphi,\\
w_{\mathbf{k}\mathbf{k}^{\prime}}^{\mathrm{A,(4)}}&=\tau_{v}\frac{3\pi n_{i}^{2}V_{0}^{4}}{4\hbar}\frac{mkk^{\prime}}{\varepsilon_{\mathbf{k}}\varepsilon_{\mathbf{k^{\prime}}}
^{2}}\delta(\varepsilon_{\mathbf{k}}-\varepsilon_{\mathbf{k^{\prime}}})\sin\Delta\varphi,
\end{aligned}
\label{App-B-dfd}
\end{equation}
where $v=v_{F}\hbar$, $n_i$ is the impurity density, $\Delta\varphi$ is the angle between wave vectors $\mathbf{k}$ and $\mathbf{k^\prime}$. Equation \eqref{App-B-dfd} show that the second symmetric scattering rate is valley-independent, while the two antisymmetric scattering rates  (the third- and fourth-order scattering rates) are dependent on the valley index. These results for the 2D massive Dirac materials are consistent with the previous works for the  multivalley massive Dirac fermion model\cite{Papaj2021PRB}.

For the isotropic system (namely $\varepsilon_\mathbf{k}=\varepsilon_{k}$) we considered [Eq.~\eqref{Hami}], the relaxation time $	{\tau_{\mathbf{k}}}$ and the side-jump velocity $\mathbf{v}_{\mathbf{k}}^{\mathrm{sj}}=\sum_{\mathbf{k}^{\prime}}w_{\mathbf{k}\mathbf{k}^{\prime}}
\delta\mathbf{r}_{\mathbf{k}^{\prime}\mathbf{k}}$ for the upper band can be evaluated in the leading order term [$w_{\mathbf{k}\mathbf{k}^{\prime}}\approx w_{\mathbf{k}\mathbf{k}^{\prime}}^{(2)}=w_{\mathbf{k}\mathbf{k}^{\prime}}^{\text{S},(2)}$] of scattering rate as, respectively,
\begin{align}
	\frac{1}{\tau_{\mathbf{k}}}\approx& \sum_{\mathbf{k^{\prime}}}w_{\mathbf{k}\mathbf{k}^{\prime}}^{(2)}\left[1-
\cos\Delta\varphi\right]\notag\\
	= &\int [{d\mathbf{k}^{\prime}}]\frac{\pi n_{i}V_{0}^{2}}{\hbar}\left[1+\frac{m^{2}+v^{2}kk^{\prime}\cos\Delta\varphi}
{\varepsilon_{\mathbf{k}}\varepsilon_{\mathbf{k}^{\prime}}}\right]\notag\\
	& \times\left(1-\cos\Delta\varphi\right)\delta(\varepsilon_{\mathbf{k}}-
\varepsilon_{\mathbf{k}^{\prime}})\notag\\
	= & \frac{n_{i}V_{0}^{2}}{4v^{2}\hbar}\frac{\varepsilon_{\mathbf{k}}^{2}+3m^{2}}{\varepsilon_
{\mathbf{k}}},
\label{App-c-tua}
\end{align}
and
\begin{align}
\mathbf{v}_{\mathbf{k}}^{\mathrm{sj}}&\approx\sum_{\mathbf{k}^{\prime}}
w_{\mathbf{k}\mathbf{k}^{\prime}}^{(2)}\delta\mathbf{r}_{\mathbf{k}^{\prime}\mathbf{k}}\notag\\
&\approx \int [{d\mathbf{k}^{\prime}}]\frac{\pi n_{i}V_{0}^{2}}{\hbar}\left[1+\frac{m^{2}+v^{2}kk^{\prime}\cos\Delta\varphi}
{\varepsilon_{\mathbf{k}}\varepsilon_{\mathbf{k}^{\prime}}}\right]\notag\\
&\times\frac{1}{2}\left[\frac{\varepsilon_{\mathbf{k}}}
{\varepsilon_{\mathbf{k}^{\prime}}}\Omega_{z}(\mathbf{k})+\frac{\varepsilon_{\mathbf{k}^{\prime}}}
{\varepsilon_{\mathbf{k}}}\Omega_{z}(\mathbf{k^{\prime}})\right]\frac{\hat{\mathbf{z}}
\times(\mathbf{k}^{\prime}-\mathbf{k})}{|\langle\psi_{\mathbf{k}}|\psi_{\mathbf{k}^{\prime}}
\rangle|^{2}}\delta(\varepsilon_{\mathbf{k}}-
\varepsilon_{\mathbf{k}^{\prime}})\notag\\
&=\frac{n_{i}V_{0}^{2}}{v^{2}\hbar}\varepsilon_{\mathbf{k}}
\Omega_{z}(\mathbf{k}\times\hat{\mathbf{z}})\notag\\
&=-\frac{\tau_{v}mv^2}{2\varepsilon_{\mathbf{k}}^2}\frac{n_{i}V_{0}^{2}}{v^{2}\hbar}
(\mathbf{k}\times\hat{\mathbf{z}}),
\label{App-c-vsj}
\end{align}
where we have used\cite{du2019disorder}
\begin{align}
\delta\mathbf{r}_{\mathbf{k}^{\prime}\mathbf{k}}=\frac{1}{2}\left[\frac{\varepsilon_{\mathbf{k}}}
{\varepsilon_{\mathbf{k}^{\prime}}}\Omega_{z}(\mathbf{k})+\frac{\varepsilon_{\mathbf{k}^{\prime}}}
{\varepsilon_{\mathbf{k}}}\Omega_{z}(\mathbf{k^{\prime}})\right]\frac{\hat{\mathbf{z}}
\times(\mathbf{k}^{\prime}-\mathbf{k})}{|\langle\psi_{\mathbf{k}}|\psi_{\mathbf{k}^{\prime}}
\rangle|^{2}}.
\label{App-B-or-shift}
\end{align}
Actually, equation \eqref{App-B-or-shift} can be easily confirmed through taking the eigenstate for upper band in Eq.~\eqref{App-B-EST} into the expression of the impurity-scattering induced coordinate shift $\delta\mathbf{r}_{\mathbf{k}^{\prime}\mathbf{k}}$ [Eq.~\eqref{App-B-shif}] and combining the definition of Berry curvature.

\section{The nonlinear AE coefficients for intrinsic, side-jump and skew-scattering mechanism}\label{AE-coeff}
As discussed in Sec.~\ref{Theoretical review}, the total acoustic current has been decomposed into three part as $\mathbf{J}=\mathbf{J}^\text{in}+\mathbf{J}^\text{sj}+\mathbf{J}^\text{sk}$
corresponding to the intrinsic (in), side-jump (sj), and skew-scattering (sk) contributions to the acoustic current, respectively.
In this appendix, the expressions of dc parts [$\mathbf{J}^\text{in}_\text{dc}$, $\mathbf{J}^\text{sj}_\text{dc}$, and $\mathbf{J}^\text{sk}_\text{dc}$] will be deduced. And based on the deduced expressions, the formulas for the nonlinear AE coefficients for intrinsic, side-jump, and skew-scattering contributions will be determined.
\subsection{The nonlinear intrinsic AE coefficient $\chi_{\alpha\beta\gamma}^{\mathrm{in}}$}
Based on Eqs.~(\ref{App-A-ex})(\ref{cur}) and (\ref{Jtot}), the intrinsic contribution to the nonlinear dc AE current $\mathbf{j}_\text{dc}^{\text{nl}}$ (where the
superscript ``{nl}"  refers to nonlinear and subscript ``{dc}" refers to direct) as a second-order response in the SAW is found to be
 \begin{align}
 	\mathbf{j}_\text{dc}^{\mathrm{in},{\text{nl}}}= & -e\int[d\mathbf{k}]\mathbf{v}_{\mathbf{k}}\delta f_{\mathrm{\mathbf{k}},\text{dc}}^{\mathrm{in},2}\notag\\
 	& -e\int[d\mathbf{k}]\left[\frac{e}{4\hbar}(\vec{\tilde{\mathcal{E}}}^{*}\times
 \bm{\Omega}_{\mathbf{k}})\delta f_{\mathrm{\mathbf{k}},\mathrm{am}}^{\mathrm{in},1}+c.c.\right].\label{AP-jdc-in}
 \end{align}
Substituting  Eqs.~\eqref{APP-A-f1f2} into Eq.~(\ref{AP-jdc-in}), one can obtain the expression of the nonlinear intrinsic dc AE current $j_{\text{dc},\alpha}^{\mathrm{in,nl}}$ in $\alpha$-direction as
\begin{align}
	j_{\text{dc},\alpha}^{\mathrm{in,nl}}= & -\frac{e}{2\hbar}\mathrm{Re}\tilde{\mathcal{E}}_{\beta}^{*}\int[d\mathbf{k}]\left(v_{\alpha}
\tau_{\mathbf{k}}
\frac{\partial \delta f_{\mathrm{\mathbf{k}},\mathrm{am}}^{\mathrm{in},1}}{\partial k_{\beta}}+\epsilon_{\alpha\beta\eta}\Omega_{\text{\ensuremath{\eta}}}\delta f_{\mathrm{\mathbf{k}},\mathrm{am}}^{\mathrm{in},1}\notag\right)\\
	= & \frac{e^{2}}{2\hbar}\mathrm{Re}\tilde{\mathcal{E}}_{\beta}^{*}
\int[d\mathbf{k}]\left[\frac{\partial(v_{\alpha}\tau_{\mathbf{k}})}{\partial k_{\beta}}-\epsilon_{\alpha\beta\eta}\Omega_{\text{\ensuremath{\eta}}}\right]\delta f_{\mathrm{\mathbf{k}},\mathrm{am}}^{\mathrm{in},1}
\label{AP-JINdca}
\end{align}
The last line in Eq.~(\ref{AP-JINdca}) is obtained through the integration by parts. Taking the amplitude $\delta f_{\mathbf{k},\mathrm{am}}^{\mathrm{in},1}$ [Eq.\eqref{Appendix-A-NE}] into Eq.~(\ref{AP-JINdca}) and using the relation $\vec{\tilde{\mathcal{E}}}=F(\omega,\mathbf{q},\mathbf{R})\vec{\mathcal{E}}$, the nonlinear  $\mathbf{j}_{\text{dc},\alpha}^{\mathrm{in},\text{nl}}$ in Eq.\eqref{AP-JINdca} can be rewritten as
\begin{align}	j_{\text{dc},\alpha}^{\mathrm{in,nl}}=\frac{e^{2}}{2\hbar}|F|^{2}\mathcal{E}_{\beta}^{*}
\mathcal{E}_{\gamma}
\mathrm{Re}\int[d\mathbf{k}]\left[\frac{(\partial v_{\alpha}\tau_{\mathbf{k}})}{\partial_{k_\beta}}
-\epsilon_{\alpha\beta\eta}\Omega_{\text{\ensuremath{\eta}}}\right]G_{\gamma},
\end{align}
where $(\omega,\mathbf{q},\mathbf{R})$ in $F$, whose expression is given in  Eq.~(\ref{App-A-F}), has been omitted for simplify, and the $G_{\gamma}$ is  given in Eq.~(\ref{AP-G_alpha}). Combining with the definition of nonlinear intrinsic AE coefficient $\chi_{\alpha\beta\gamma}^{\mathrm{in}}$ ($j_{\text{dc},\alpha}^{\mathrm{in,nl}}=\chi_{\alpha\beta\gamma}
^{\mathrm{in}}\mathcal{E}_{\beta}^{*}
\mathcal{E}_{\gamma}$), one can easily confirm that
\begin{align}
	\chi_{\alpha\beta\gamma}^{\mathrm{in}}=	\frac{e^{2}}{2\hbar}|F|^{2}\mathrm{Re}\int[d\mathbf{k}]\left[\frac{(\partial v_{\alpha}\tau_{\mathbf{k}})}{\partial_{k_\beta}}-\epsilon_{\alpha\beta\eta}
\Omega_{\text{\ensuremath{\eta}}}\right]
G_{\gamma}. \label{AP-coeff-in}
\end{align}

\subsection{The nonlinear side-jump AE coefficient $\chi_{\alpha\beta\gamma}^{\mathrm{sj}}$}
According to Eqs.~(\ref{A-FSj})(\ref{APP-A-Cdt})(\ref{cur}) and (\ref{Jtot}),
the side-jump contribution to nonlinear dc AE current $\mathbf{j}_\text{dc}^{\mathrm{sj,nl}}$ can be decomposed into three parts as
\begin{align}
	\mathbf{j}_\text{dc}^{\mathrm{sj,nl}} &=\mathbf{j}_\text{dc,1}^{\mathrm{sj,nl}}+\mathbf{j}_{\text{dc},2}^{\mathrm{sj,nl}}+
\mathbf{j}_{\text{dc},3}
^{\mathrm{sj,nl}},\label{AP-JdcSj}
\end{align}
with
\begin{equation}
\left\{
\begin{aligned}
	\mathbf{j}_\text{dc,1}^{\mathrm{sj,nl}}= & -e\int[d\mathbf{k}]\mathbf{v}_{\mathbf{k}}^{\mathrm{sj}}\delta f_{\mathrm{\mathbf{k}},\text{dc}}^{\mathrm{in},2},\\
	\mathbf{j}_\text{dc,2}^{\mathrm{sj,nl}}=& -e\int[d\mathbf{k}]\mathbf{v}_{\mathbf{k}}\delta f_{\mathbf{k},\text{dc}}^{\mathrm{sj,2}},\\
\mathbf{j}_\text{dc,3}^{\mathrm{sj,nl}}=& -2e\mathrm{Re}\int[d\mathbf{k}]\frac{e}{4\hbar}(\vec{\tilde{\mathcal{E}}}^{*}
\times\bm{\Omega}_{\mathbf{k}})\delta f_{\mathbf{k},\text{am}}^{\mathrm{sj,1}},\label{AP-JdcSj-par}
\end{aligned}
\right.
\end{equation}
where $\mathbf{j}_{\text{dc},1}^{\mathrm{sj,nl}}$ is the nonlinear AE current originating from side-jump velocity, $\mathbf{j}_{\text{dc},2}^{\mathrm{sj,nl}}$ indicates the nonlinear AE current stemming  from the side-jump-modified NFDF, and $\mathbf{j}_{\text{dc},3}^{\mathrm{sj,nl}}$ refers to the nonlinear AE current coming from the joint effect of anomalous velocity $\vec{\tilde{\mathcal{E}}}^{*}\times\bm{\Omega}_{\mathbf{k}}$ and the side-jump effect. Taking the formulas of $f_{\mathrm{\mathbf{k}},\text{dc}}^{\mathrm{in},2}$ [Eq.~\eqref{APP-A-f1f2}],  $\delta f_{\mathbf{k},\text{dc}}^{\mathrm{sj,2}}$ [Eq.~\eqref{AP-FSj2dc}] and $\delta f_{\mathbf{k},\text{A}}^{\mathrm{sj,1}}$ [Eq.~\eqref{AP-FSj01-2}] into Eq.~\eqref{AP-JdcSj-par} and meanwhile using the relation $\vec{\tilde{\mathcal{E}}}=F(\omega,\mathbf{q},\mathbf{R})\vec{\mathcal{E}}$, one can find
\begin{equation}
j_{\text{dc},\alpha}^{\mathrm{sj,nl}}=\chi_{\alpha\beta\gamma}^{\mathrm{sj}}
\mathcal{E}_{\beta}^{*}\mathcal{E}_{\gamma},\,\,\,\,\, \chi_{\alpha\beta\gamma}^{\mathrm{sj}}=\chi_{\alpha\beta\gamma}^{\text{sj-v}}+
\chi_{\alpha\beta\gamma}^{\text{sj-mff}}+\chi_{\alpha\beta\gamma}^{\text{
sj-je}},
\label{AP-coeff-sjj}
\end{equation}
where the components of nonlinear side-jump AE coefficients [$\chi_{\alpha\beta\gamma}^{\text{sj-v}}$, $\chi_{\alpha\beta\gamma}^{\text{sj-mff}}$ and $\chi_{\alpha\beta\gamma}^{\text{sj-je}}$] are given as
\begin{equation}
\left\{
\begin{aligned}
\chi_{\alpha\beta\gamma}^{\text{sj-v}}=&\frac{e^{2}}{2\hbar}|F|^{2}\mathrm{Re}\int[d\mathbf{k}]
\partial^{\beta}_{\mathbf{k}}\left(v_{\alpha}^{\mathrm{sj}}\tau_{\mathbf{k}}\right)G_{\gamma},\\
\chi_{\alpha\beta\gamma}^{\text{sj-mff}}=&\frac{e^{4}}{2\hbar}|F|^{2}\mathrm{Re}\int[d\mathbf{k}]
\left\{\tau_{\mathbf{k}}\gamma
(\omega,\mathbf{k})\frac{\partial\left(v_{\alpha}^{\mathrm{sj}}\tau_{\mathbf{k}}\right)}
{\partial{k_\beta}}v_{\gamma}^{\mathrm{sj}}
\frac{\partial f_{\mathrm{\mathbf{k}}}^{(0)}}{\partial\varepsilon_{\mathbf{k}}}\right.\\ &\left.+v_{\alpha}\tau_{\mathbf{k}}\frac{\hbar}{e^{2}}
\sum_{\mathbf{k}^{\prime}}O_{\mathbf{k}\mathbf{k}^{\prime}}^{\beta}\left[G_{\gamma}-
(\mathbf{k}\leftrightarrow\mathbf{k^{\prime}})\right]\right\},\\
\chi_{\alpha\beta\gamma}^{\text{sj-je}}=&\frac{e^{3}}{2\hbar}\epsilon_{\alpha\beta\eta}|F|^{2}
\mathrm{Re}\int[d\mathbf{k}]\Omega_{\text{\ensuremath{\eta}}}\tau_{\mathbf{k}}
\gamma(\omega,\mathbf{k})v_{\gamma}^{\mathrm{sj}}\frac{\partial f_{\mathrm{\mathbf{k}}}^{(0)}}{\partial\varepsilon_{\mathbf{k}}},\label{AP-coeff-sj}
\end{aligned}
\right.
\end{equation}
where $\partial^{\beta}_{\mathbf{k}}$ indicates $\partial/\partial (k_{\beta})$.

\subsection{The nonlinear skew-scattering AE coefficient $\chi_{\alpha\beta\gamma}^{\mathrm{sk}}$}
Accompanying Eqs.~\eqref{cur} and (\ref{Jtot}) with Eq.~\eqref{AP-FSk}, the nonlinear skew-scattering AE current is determined as
\begin{align}
	\mathbf{j}_\text{dc}^{\mathrm{sk,nl}}=&-e\int[d\mathbf{k}]\mathbf{v}_{\mathbf{k}}\delta f_{\mathrm{\mathbf{k}},\text{dc}}^{\mathrm{sk,2}}\notag\\
&	-\frac{e^{2}}{2\hbar}\mathrm{Re}\int[d\mathbf{k}](\vec{\tilde{\mathcal{E}}}^{*}\times
\bm{\Omega}_{\mathbf{k}})\delta f_{\mathrm{\mathbf{k}},\mathrm{am}}^{\mathrm{sk,1}},\notag\\
\label{AP-JSKdc}
\end{align}
where the first term in Eq.~\eqref{AP-JSKdc}
stems from the skew-scattering-modified NFDF and
the second term originates from the joint effect of anomalous velocity item and the skew-scattering effect, respectively.  Substituting the expressions of $\delta f_{\mathbf{k},\text{dc}}^{\mathrm{sk,2}}$ [Eq.~\eqref{AP-FSjdc}] and $\delta f_{\mathrm{\mathbf{k}},\mathrm{am}}^{\mathrm{sk,1}}$ [Eq.~\eqref{AP-FSk1}] into Eq.~\eqref{AP-JSKdc}, and integrating by parts, the $\alpha$-component of $\mathbf{j}_\text{dc}^{\mathrm{sk,nl}}$ can be obtained as

\begin{align}
	j_{\mathrm{dc},\alpha}^{\mathrm{sk,nl}}=& \frac{e^{2}}{2\hbar}\mathrm{Re}\tilde{\mathcal{E}}^{*}_{\beta}
\int[d\mathbf{k}]\left[\frac{\partial(v_{\alpha}\tau_{\mathbf{k}})}{\partial k_{\beta}}\delta  f_{\mathrm{\mathbf{k}},\mathrm{am}}^{\mathrm{sk,1}}\right.\notag
\\
+&\left.v_{\alpha}\tau_{\mathbf{k}}
\sum_{\mathbf{k}^{\prime}}w_{\mathbf{k}^{\prime}\mathbf{k}}
^{\mathrm{A}}\left(\tau_{\mathbf{k}}
\partial^{\beta}_{\mathbf{k}}\delta f_{\mathrm{\mathbf{k}},\mathrm{am}}^{\mathrm{in},1}+
\tau_{\mathbf{k}^{\prime}}\partial^{\beta}_{\mathbf{k}^{\prime}}\delta f_{\mathrm{\mathbf{k}^{\prime}},\mathrm{A}}^{\mathrm{in},1}
\right)\right.\notag\\
+&\left.\epsilon_{\alpha\beta\eta}
\Omega_{\text{\ensuremath{\eta}}}\tau_{\mathbf{k}}\gamma(\omega,\mathbf{k})\sum_{\mathbf{k}^{\prime}}
w_{\mathbf{k}^{\prime}\mathbf{k}}^{\mathrm{am}}(\delta f_{\mathrm{\mathbf{k}},\mathrm{am}}^{\mathrm{in},1}+\delta f_{\mathrm{\mathbf{k}}^{\prime},\mathrm{am}}^{\mathrm{in},1})\right].
\label{App-B-jsk1}
\end{align}
Further taking the amplitude $\delta f_{\mathbf{k},\mathrm{am}}^{\mathrm{in},1}$ [Eq.~\eqref{Appendix-A-NE}] into Eq.~(\ref{App-B-jsk1}) and using the relation $\vec{\tilde{\mathcal{E}}}=F(\omega,\mathbf{q},\mathbf{R})\vec{\mathcal{E}}$, one would have
\begin{align}
j_{\mathrm{dc},\alpha}^{\mathrm{sk,nl}}=\chi_{\alpha\beta\gamma}^{\mathrm{sk}}
\mathcal{E}_{\beta}^{*}\mathcal{E}_{\gamma},\,\,\,\,\, \chi_{\alpha\beta\gamma}^{\mathrm{sk}}=\chi_{\alpha\beta\gamma}^{\text{sk-mff}}+
\chi_{\alpha\beta\gamma}^{\text{sk-je}},
\label{Ap-C-coeff-sk}
\end{align}
with
\begin{equation}
\left\{
\begin{aligned}
\chi_{\alpha\beta\gamma}^{\text{sk-mff}}=&	\frac{e^{2}}{2\hbar}|F|^{2}\mathrm{Re}\Bigg\{\sum_{\mathbf{k}\mathbf{k^{\prime}}}
v_{\alpha}\tau_{\mathbf{k}}w_{\mathbf{k}^{\prime}\mathbf{k}}^{\mathrm{A}}
\left[\tau_{\mathbf{k}}\partial_{\beta}G_{\gamma}+(\mathbf{k}\leftrightarrow\mathbf{k^{\prime}})
\right]\\
&-\sum_{\mathbf{k}\mathbf{k^{\prime}}}\partial^{\beta}_{\mathbf{k}}(v_{\alpha}\tau_{\mathbf{k}})
\tau_{\mathbf{k}}\gamma(\omega,\mathbf{k})w_{\mathbf{k}^{\prime}\mathbf{k}}^{\mathrm{A}}
\left[G_{\gamma}+(\mathbf{k}\leftrightarrow\mathbf{k^{\prime}})\right]\Bigg\}\\
\chi_{\alpha\beta\gamma}^{\text{sk-je}}=&	\frac{e^{2}}{2\hbar}\epsilon_{\alpha\beta\eta}|F|^{2}\mathrm{Re}\sum_{\mathbf{k}\mathbf{k}^{\prime}}
\Omega_{\text{\ensuremath{\eta}}}\tau_{\mathbf{k}}\gamma(\omega,\mathbf{k})\\
	&\times w_{\mathbf{k}^{\prime}\mathbf{k}}^{\mathrm{A}}\left[G_{\gamma}+(\mathbf{k}\leftrightarrow
\mathbf{k^{\prime}})\right].\label{AP-coef-Sk-ad}
\end{aligned}
\right.
\end{equation}

The nonlinear skew-scattering AE coefficient $\chi_{\alpha\beta\gamma}^{\mathrm{sk}}$ has been decomposed into two components as  $\chi_{\alpha\beta\gamma}^{\mathrm{sk}}=\chi_{\alpha\beta\gamma}^{\text{sk-mff}}+\chi_
{\alpha\beta\gamma}^{\text{sk-je}}$ corresponding to the skew-scattering-modified-NFDF contribution and the contribution from joint effect of anomalous velocity and skew scattering, respectively.

\section{The nonlinear acoustic Hall coefficients for the  massive Dirac materials}\label{coeff-NAVHE}
The general formulas of nonlinear AE coefficients stemming from intrinsic, side-jump and skew-scattering contributions have been given in appendix \ref{AE-coeff}. In this appendix, according to the obtained formulas in appendix \ref{AE-coeff} and further assuming the wave vector $\mathbf{q}$ of SAW along the x-direction,  the relevant acoustic Hall coefficients [$\chi_{yxx}^{\mathrm{in}}$, $\chi_{yxx}^{\mathrm{sj}}=\chi_{yxx}^{\text{sj-v}}+
\chi_{yxx}^{\text{sj-mff}}+\chi_{yxx}^{\text{sj-je}}$, $\chi_{yxx}^{\mathrm{sk}}=\chi_{yxx}^{\text{sk-mff}}+
\chi_{yxx}^{\text{sk-je}}$] will be deduced for upper band of the disordered massive Dirac materials.
The Hamiltonian of the massive Dirac materials is given by Eq.\eqref{Hami}.  The corresponding energy dispersion $\varepsilon_{\mathbf{k}}$ and the Berry curvature $\Omega_{\mathbf{k}}$ for the conduction band (or upper band) are, respectively, 
\begin{equation}
\varepsilon_{\mathbf{k}}= \sqrt{v^2(k_x^2+k_y^2)+m^2},\,\,\,\,
\Omega_{\mathbf{k}}=-\frac{\tau_{v}mv^2}{2\varepsilon_{\mathbf{k}}^3},
\end{equation}
yielding
\begin{align}
	v_{x}=\frac{1}{\hbar}\frac{\partial\varepsilon_{\mathbf{k}}}{\partial k_{x}}=\frac{1}{\hbar}\frac{k_{x}v^{2}}{\varepsilon_{\mathbf{k}}},\,\,\,
	v_{y}=\frac{1}{\hbar}\frac{\partial\varepsilon_{\mathbf{k}}}{\partial k_{y}}=\frac{1}{\hbar}\frac{k_{y}v^{2}}{\varepsilon_{\mathbf{k}}}.\label{App-D-velocity}
\end{align}
In the long-wave limit ($\omega \tau_\mathbf{k}\ll 1$ and $ \mathbf{q}\cdot \mathbf{v}_{k}\tau_{\mathbf{k}}\ll1$)
 and the zero-temperature limit $-\partial f^{(0)}_{\mathbf{k}}/{\partial \epsilon_{\mathbf{k}}}\approx \delta(\varepsilon_{\mathbf{k}}-E_f)$, the conductivity tensor $\sigma_{\alpha\beta}$ [Eq.~\eqref{App-A-sigma}] and the diffusion vector $\mathrm{R_{\alpha}}$ [Eq.\eqref{AP-sigma-R}] can be evaluated as, respectively,
\begin{align}
	\sigma_{\alpha\beta}\approx & 2e^{2}\int[d\mathbf{k}]v_{\alpha}v_{\beta}\tau_{\mathbf{k}}
\delta(\varepsilon_{\mathbf{k}}-E_f)= \delta_{\alpha\beta}\sigma
\end{align}
and
\begin{align}
	\mathrm{R_{\alpha}}\approx & -2iq_{\beta}\frac{\partial \mu}{\partial n}\int[d\mathbf{k}]v_{\alpha}v_{\beta}\tau_{\mathbf{k}}\delta(\varepsilon_{\mathbf{k}}-E_f)
= -\frac{i}{e^{2}}\frac{\partial \mu}{\partial n}q_{\alpha}\sigma
\end{align}
with the 2D static Drude conductivity
\begin{align}
	\sigma= & \frac{e^{2}}{2\pi\hbar^{2}}\frac{E_f^{2}-m^{2}}{E_f}\tau_{F}=
 \frac{e^{2}}{\hbar}\frac{2}{\pi n_{i}V_{0}^{2}}\frac{v^{2}(E_f^{2}-m^{2})}{E_f^{2}+3m^{2}}.
 \label{App-D-E2}
\end{align}
Obtaining the second equality of Eq.~{\eqref{App-D-E2}}, we have used $\tau_{F}=4\hbar v^{2}E_f/[n_{i}V^{2}_{0}(E_f^{2}+3m^{2})]$ in Eq.~\eqref{App-c-tua} with taking $E_\mathbf{k}=E_f$. Noting that the factor ``2'' is accounted for the spin degeneracy.

\subsection{The intrinsic coefficient $\chi_{yxx}^{\mathrm{in}}$}
Based on Eq.~\eqref{AP-coeff-in} and taking into account the spin degeneracy, the nonlinear intrinsic acoustic Hall coefficient (AHC) $\chi_{yxx}^{\mathrm{in}}$ for the massive Dirac materials can be determined as
\begin{align}
	\chi_{yxx}^{\mathrm{in}}= & \frac{e^{2}}{\hbar}|F|^{2}\mathrm{Re}\int[d\mathbf{k}]\left[\frac{(\partial v_{y}\tau_{\mathbf{k}})}{\partial_{k_{x}}}-\epsilon_{yxz}\Omega_{\text{z}}\right]G_{x},
\label{ap-d-chiinyxx}
\end{align}
where
\begin{align}
	G_x=&\gamma(\omega,\mathbf{k})\left(e\tau_{\mathbf{k}}v_{x}\frac{\partial f_{\mathbf{k}}^{(0)}}{\partial\varepsilon_{\mathbf{k}}}+\frac{\sigma q_{x}}{e(\omega-\mathbf{q}\cdotp\mathbf{R})}\frac{\partial \mu}{\partial n}\frac{\partial f_{\mathbf{k}}^{(0)}}{\partial\varepsilon_{\mathbf{k}}}\right).
	\label{App-Gx}
\end{align}
According to Eq.~\eqref{App-D-velocity}, the first term in bracket of Eq.~\eqref{ap-d-chiinyxx} is odd in $k_y$, indicating that the corresponding integration would  be zero and has no contribution to $\chi_{yxx}^{\mathrm{in}}$. Similarly, the first term in bracket of  Eq.~\eqref{App-Gx} also has no contribution to $\chi_{yxx}^{\mathrm{in}}$ since it is also a odd function with respect to $k_x$. As mentioned in Appendix ~\ref{App-A-in}, the first (second) term in bracket of Eq.~\eqref{App-Gx} represents the drift (diffusion) part. Therefore, the coefficient $\chi_{yxx}^{\mathrm{in}}$ is only determined by the diffusive term and embodies the diffusive nature.
After a tedious derivation, the intrinsic AHC $\chi_{yxx}^{\mathrm{in}}$ can be obtained as
\begin{align}
	\chi_{yxx}^{\mathrm{in}}=&\sum_{\tau_{v}}\chi_{yxx}^{\mathrm{in},\tau_{v}}\notag\\
=&
\sum_{\tau_{v}}\tau_{v}q_{x}\frac{e\sigma}{2\pi\hbar}\frac{m}{2E_f^{2}}
\frac{\partial \mu}{\partial n}|F|^{2}\mathrm{Re}\frac{1+i\omega\tau_{F}}{\omega-\mathbf{q}\cdotp\mathbf{R}}\notag\\
=&\sum_{\tau_{v}}\tau_{v}\frac{e\sigma mv^{2}}{4\hbar v_{s}E_f^{3}}H_{1}, \label{ap-d-chiinyxx2}
\end{align}
where $\chi_{yxx}^{\mathrm{in},\tau_{v}}$ is the nonlinear intrinsic AHC for valley $\tau_{v}$, and  the dimensionless auxiliary function $H_{1}$ is defined as
\begin{align}
	H_{1}=	\frac{1+\omega\tau_{F}\frac{\sigma}{\sigma_{*}}qa_{0}\varepsilon_{e}/E_f}{1+
(\frac{\sigma}{\sigma_{*}})^{2}\left(1+qa_{0}\varepsilon_{e}/E_f\right)^{2}}
\label{App-D-H1}
\end{align}
with
\begin{align}
		\varepsilon_e=&m_ev_F^2/2,\notag\\
	\sigma_{*}=&\epsilon_0(\epsilon+1)v_s/4\pi,\notag\\
	a_{0}=&\epsilon_0(\epsilon+1)\hbar^2/(2m_ee^2),
\end{align}
where $m_e$ is the free electron mass, $\epsilon_0$ represents the dielectric permittivity of vacuum, $\epsilon$ indicates the dielectric constant of substrate, and $v_s$ is the velocity of SAW. Obviously, $\chi_{yxx}^{\mathrm{in}}=\chi_{yxx}^{\mathrm{in},+1}+\chi_{yxx}^{\mathrm{in},-1}$ [Eq.~\eqref{ap-d-chiinyxx2}] is zero since the intrinsic AHC from two valleys cancel with each other. However, the nonlinear intrinsic valley AHC $\chi_\text{valley}^{\mathrm{H},\mathrm{in}}=\chi_{yxx}^{\mathrm{in},+1}-\chi_{yxx}^{\mathrm{in},-1}$  will be nonzero. In other words, when propagating a SAW into the massive Dirac materials,
 a \textit{pure} nonlinear acoustic valley Hall current stemming from the intrinsic contribution will flow vertically to the wave vector $\mathbf{q}$ of SAW.
\subsection{The side-jump coefficient $\chi_{\alpha\beta\gamma}^{\mathrm{sj}}$}
As shown in Eq.~\eqref{AP-coeff-sjj}, the nonlinear side-jump AE coefficient $\chi_{\alpha\beta\gamma}^{\mathrm{sj}}$
has been decomposed into three parts [$\chi_{\alpha\beta\gamma}^{\text{sj-v}}$, $\chi_{\alpha\beta\gamma}^{\text{sj-mff}}$ and $\chi_{\alpha\beta\gamma}^{\text{sj-je}}$].
The component $\chi_{yxx}^{\text{sj-v}}$ [Eq.~\eqref{AP-coeff-sj}] stemming
 from the side-jump velocity for the disordered massive Dirac materials can be determined as
\begin{align}
	\chi_{yxx}^{\text{sj-v}}= \frac{e^{2}}{\hbar}|F|^{2}\mathrm{Re}\int[d\mathbf{k}]\partial_{x}\left(v_{y}^{\mathrm{sj}}\tau_{\mathbf{k}}\right)G_{x}.
\end{align}
Based on Eqs.~\eqref{App-c-tua}\eqref{App-c-vsj}, the factor $\partial_{x}(v_y^{sj}\tau_{\mathbf{k}})$ is found to be even in $k_x$. Besides, the drift part of function $G_x$ [Eq.~\eqref{App-Gx}] is odd in $k_x$. Therefore, only the diffusive part of $G_x$ would contribute to $\chi_{yxx}^{\text{sj-v}}$, which gives
\begin{align}
	\chi_{yxx}^{\text{sj-v}}=&\sum_{\tau_{v}}\chi_{yxx}^{\text{sj-v},\tau_{v}}\\
\approx &\sum_{\tau_{v}} \tau_{v}\frac{e\sigma}{\pi\hbar}q\frac{\partial \mu}{\partial n}\frac{m(E_f^{4}-6m^{2}E_f^{2}-3m^{4})}{2E_f^{2}(E_f^{2}+3m^{2})^{2}}\notag\\
	& \times|F|^{2}\mathrm{Re}\frac{1+i\omega\tau_{F}}{\omega-\mathbf{q}\cdotp\mathbf{R}}\notag\\
	=&\sum_{\tau_{v}}\tau_{v}\frac{e\sigma mv^{2}}{\hbar v_{s}E_f^{3}}\frac{E_f^{4}-6m^{2}E_f^{2}-3m^{4}}{2(E_f^{2}+3m^{2})^{2}}H_{1}.\label{AP-chiyxxSjv-result}
\end{align}
In long wave limit, the component $\chi_{yxx}^{\text{sj-mff}}$ [Eq.~\eqref{AP-coeff-sj}] related to side-jump-modified NFDF for disordered massive Dirac materials can be determined as
\begin{align}
	\chi_{yxx}^{\text{sj-mff}}= & \frac{e^{4}}{\hbar}|F|^{2}\int[d\mathbf{k}]\tau_{\mathbf{k}}\partial_{x}(v_{y}\tau_{\mathbf{k}})v_{x}^{\mathrm{sj}}\frac{\partial f_{\mathrm{\mathbf{k}}}^{(0)}}{\partial\varepsilon_{\mathbf{k}}}\notag\\
	& +e^{2}|F|^{2}\mathrm{Re}\int[d\mathbf{k}]v_{y}\tau_{\mathbf{k}}\notag\\
	& \times\sum_{\mathbf{k^{\prime}}}O_{\mathbf{k}\mathbf{k}^{\prime}}^{x}\left[G_{x}-(\mathbf{k}\leftrightarrow\mathbf{k^{\prime}})\right].\label{AP-chiyxxSj}
\end{align}
Evidently, the factor $\partial_x(v_y\tau_\mathbf{k})v_x^{\mathrm{sj}}$ is odd in $k_x$, indicating that the first line is zero after integrating. In addition, after some tedious derivations, one can verify that the integral over drift term of function $G_x$ is zero due to the odd parity about $k_x$, hinting only the diffusive part of $G_{x}$ contributes to the coefficient $\chi_{yxx}^{\text{sj-mff}}$, which  yields
\begin{align} \chi_{yxx}^{\text{sj-mff}}=&\sum_{\tau_{v}}\chi_{yxx}^
{\text{sj-mff},\tau_{v}}\notag\\
= &\sum_{\tau_{v}}
\tau_{v}\frac{e\sigma mv^{2}}{\hbar v_{s}E_f^{3}}\frac{E_f^{4}-6m^{2}E_f^{2}-3m^{4}}{2(E_f^{2}+3m^{2})^{2}}H_{1}
\label{APP-D-FDF1}
\end{align}
According to Eqs.~\eqref{APP-D-FDF1} and \eqref{AP-chiyxxSjv-result}, one can have $\chi_{yxx}^{\text{sj-mff},\tau_{v}}=\chi_{yxx}^{\text{sj-v},\tau_{v}}$ and the coefficients [$\chi_{yxx}^{\text{sj-v},\tau_{v}},\chi_{yxx}^{\text{sj-mff},\tau_{v}}$] are valley dependent, namely the coefficients for $K$ valley ($\tau_{v}=+1$) are opposite to those for  $-K$ valley ($\tau_{v}=-1$). These valley-dependent characters indicate that the side-jump contribution would also lead to a \textit{pure} nonlinear acoustic valley Hall current in the disordered massive Dirac materials since the remaining component $\chi_{yxx}^{\text{sj-je}}$ of the nonlinear side-jump AE coefficient $\chi_{yxx}^{\mathrm{sj}}$ is found to be zero for each valley (i.e. $\chi_{yxx}^{\text{sj-je},\tau_{v}}=0$).

Based on Eq.~\eqref{AP-coeff-sj}, the component $\chi_{yxx}^{\text{sj-je},\tau_{v}}$, which attributes to the joint effect of anomalous velocity item and the side-jump effect, for $\tau_{v}$ valley can be determined as
\begin{align}
	\chi_{yxx}^{\text{sj-je},\tau_{v}} & = 2\frac{e^{3}}{\hbar}\epsilon_{yxz}|F|^{2}\mathrm{Re}\sum_{\mathbf{k}}\Omega_{z}\tau_{\mathbf{k}}\gamma(\omega,\mathbf{k})v_{x}^{\mathrm{sj}}\frac{\partial f_{\mathrm{\mathbf{k}}}^{(0)}}{\partial\varepsilon_{\mathbf{k}}}\notag\\
	= & \frac{e^{3}}{\hbar}|F|^{2}\mathrm{Re}\int[d\mathbf{k}]\Omega_{z}\tau_{\mathbf{k}}\gamma(\omega,\mathbf{k})\frac{n_{i}V_{0}^{2}}{\hbar}\frac{\varepsilon_{\mathbf{k}}\Omega_{z}}{v^{2}}k_{y}\frac{\partial f_{\mathrm{\mathbf{k}}}^{(0)}}{\partial\varepsilon_{\mathbf{k}}}\\
	\approx & \frac{e^{3}}{\hbar}|F|^{2}\mathrm{Re}\int[d\mathbf{k}]\Omega_{z}\tau_{\mathbf{k}}\frac{n_{i}V_{0}^{2}}{\hbar}\frac{\varepsilon_{\mathbf{k}}\Omega_{z}}{v^{2}}k_{y}\frac{\partial f_{\mathrm{\mathbf{k}}}^{(0)}}{\partial\varepsilon_{\mathbf{k}}}\notag\\
=&0.
\label{App-d-df}
\end{align}
In obtaining the last line,we have used the fact that the integrand in Eq.~\eqref{App-d-df} is odd in $k_y$. 
As a result, the nonlinear side-jump AHC $\chi_{yxx}^{\mathrm{sj,\tau_{v}}}$ for $\tau_{v}$ valley is found to be
\begin{equation}
\chi_{yxx}^{\mathrm{sj,\tau_{v}}}=\tau_{v}\frac{e\sigma mv^{2}}{\hbar v_{s}E_f^{3}}\frac{E_f^{4}-6m^{2}E_f^{2}-3m^{4}}{(E_f^{2}+3m^{2})^{2}}H_{1}
\end{equation}
\subsection{The skew-scattering coefficient $\chi_{yxx}^{\mathrm{sk}}$ }
According to Eq.~\eqref{Ap-C-coeff-sk}, the nonlinear skew-scattering AE coefficient $\chi_{yxx}^{\mathrm{sk}}$ contains two components as $\chi_{yxx}^{\mathrm{sk}}=\chi_{yxx}^{\text{sk-mff}}+\chi_{yxx}^{\text{sk-je}}$,
where $\chi_{yxx}^{\text{sk-mff}}$ and $\chi_{yxx}^{\text{sk-je}}$ are rooted in the skew-scattering-modified NFDF  and the joint effect of skew-scattering and anomalous velocity term, respectively. Based on Eq.~\eqref{AP-coef-Sk-ad}, one can determine
\begin{align}
	\chi_{yxx}^{\text{sk-je}}= & -\frac{e^{2}}{\hbar}|F|^{2}\mathrm{Re}\int[d\mathbf{k}]\Omega_{z}\tau_{\mathbf{k}}(1+i\omega\tau_{\mathbf{k}})\notag\\
	& \times\tau_{v}\frac{1}{8v^{2}\hbar^{2}}k_{y}\frac{mk_{F}}{\varepsilon_{\mathbf{k}}}\tau_{F}\delta(k-k_{F})\notag\\
	& \times\left(n_{i}V_{1}^{3}+\frac{3n_{i}^{2}V_{0}^{4}}{2E_f}\right)\notag\\
	& \times\left[-e(1+i\omega\tau_{F})+\mathrm{Re}\frac{i\sigma q_{x}^{2}}{e(\omega-\mathbf{q}\cdotp\mathbf{R})}\frac{\partial \mu}{\partial n}\right]\notag\\
=&0
\label{App-D-ska}
\end{align}
and
\begin{align}
	\chi_{yxx}^{\text{sk-mff}}= & \frac{e^{2}}{\hbar}|F|^{2}\mathrm{Re}\Bigg\{\int[d\mathbf{k}]v_{y}\tau_{\mathbf{k}}
\sum_{\mathbf{k^{\prime}}}w_{\mathbf{k}^{\prime}\mathbf{k}}^{\mathrm{A}}\notag\\ &\times\left[\tau_{\mathbf{k}}\partial_{x}G_{x}+(\mathbf{k}\leftrightarrow\mathbf{k^{\prime}})\right]
\notag\\
	& -\int[d\mathbf{k}]\partial_{x}(v_{y}\tau_{\mathbf{k}})\tau_{\mathbf{k}}\gamma(\omega,\mathbf{k})
\notag\\
	& \times\sum_{\mathbf{k^{\prime}}}w_{\mathbf{k}^{\prime}\mathbf{k}}^{\mathrm{A}}
\left[G_{x}+(\mathbf{k}\leftrightarrow\mathbf{k^{\prime}})\right]\Bigg\}.
\label{App-D-dfdf}
\end{align}
The last line in Eq.~\eqref{App-D-ska} can be easily confirmed by the fact that the integrand is a odd function with respect to $k_y$ for each valley of the disordered massive Dirac materials.
The odd parties about $k_{y}$ for each valley also hints $\chi_{yxx}^{\text{sk-je},\tau_{v}}=0$. Therefore, one have
\begin{equation}
\begin{aligned}
\chi_{yxx}^{\mathrm{sk}}=\chi_{yxx}^{\text{sk-mff}},\,\,\, \chi_{yxx}^{\mathrm{sk},\tau_{v}}=\chi_{yxx}^{\text{sk-mff},\tau_{v}}.
\end{aligned}
\end{equation}
Further taking the expression of antisymmetric scattering rates $\tau_\mathbf{k}$ for the disordered massive Dirac materials [Eq.~\eqref{App-B-dfd}] into Eq.~{\eqref{App-D-dfdf}} and doing tedious calculations, the $\chi_{yxx}^{\mathrm{sk}}(=\chi_{yxx}^{\text{sk-mff}})$ can be determined as $\chi_{yxx}^{\mathrm{sk}}=\sum_{\tau_{v}}\chi_{yxx}^{\mathrm{sk,\tau_{v}}}$, where the coefficient $\chi_{yxx}^{\mathrm{sk,\tau_{v}}}$ for the $\tau_{v}$ valley is  valley-dependent and given by
\begin{align}
			\chi_{yxx}^{\mathrm{sk},\tau_{v}} & =\chi_{yxx,1}^{\mathrm{sk},\tau_{v}}+\chi_{yxx,2}^{\mathrm{sk},\tau_{v}}
		\end{align}
		with
		\begin{align}
			\chi_{yxx,1}^{\mathrm{sk},\tau_{v}}= & \tau_{v}\frac{\pi V_{0}\sigma^{2}}{4ev_{s}}\frac{m^{3}(7E_f^{2}-3m^{2})}{E_f^{2}(E_f^{2}+3m^{2})^{2}}H_{1}\notag\\
			& -\tau_{v}\frac{\pi V_{0}\sigma^{2}}{2ev_{s}}\frac{mE_f^{2}}{(E_f^{2}+3m^{2})^{2}}\frac{v_{s}^{2}}{v_{F}^{2}}H_{2},\notag\\
\chi_{yxx,2}^{\mathrm{sk},\tau_{v}}= & \tau_{v}\frac{\pi n_{i}V_{0}^{2}\sigma^{2}}{4ev_{s}}\frac{3m(5m^{2}-E_f^{2})}{2E_f(E_f^{2}+3m^{2})^{2}}H_{1}\notag\\
			& -\tau_{v}\frac{3\pi n_{i}V_{0}^{2}\sigma^{2}}{4ev_{s}}\frac{mE_f}{(E_f^{2}+3m^{2})^{2}}\frac{v_{s}^{2}}{v_{F}^{2}}H_{2}
		\end{align}
corresponding to the skew-scattering contribution originated from the third- and forth-order scattering rates, respectively. The dimensionless auxiliary function $H_2$ is defined as
		\begin{align}
			H_{2}= & \frac{\omega\tau_{F}\frac{\sigma}{\sigma_{*}}(qa_{0}\varepsilon_{e}/E_f)\left[1+(\frac{\sigma}{\sigma_{*}})^{2}(qa_{0}\varepsilon_{e}/E_f)^{2}\right]}{1+(\frac{\sigma}{\sigma_{*}})^{2}\left(1+qa_{0}\varepsilon_{e}/E_f\right)^{2}}\notag\\
			& -\frac{(\frac{\sigma}{\sigma_{*}})^{2}(qa_{0}\varepsilon_{e}/E_f)^{2}}{1+(\frac{\sigma}{\sigma_{*}})^{2}\left(1+qa_{0}\varepsilon_{e}/E_f\right)^{2}}.\label{App-D-H23}
\end{align}

%

\begin{thebibliography}{50}%
	\makeatletter
	\providecommand \@ifxundefined [1]{%
		\@ifx{#1\undefined}
	}%
	\providecommand \@ifnum [1]{%
		\ifnum #1\expandafter \@firstoftwo
		\else \expandafter \@secondoftwo
		\fi
	}%
	\providecommand \@ifx [1]{%
		\ifx #1\expandafter \@firstoftwo
		\else \expandafter \@secondoftwo
		\fi
	}%
	\providecommand \natexlab [1]{#1}%
	\providecommand \enquote  [1]{``#1''}%
	\providecommand \bibnamefont  [1]{#1}%
	\providecommand \bibfnamefont [1]{#1}%
	\providecommand \citenamefont [1]{#1}%
	\providecommand \href@noop [0]{\@secondoftwo}%
	\providecommand \href [0]{\begingroup \@sanitize@url \@href}%
	\providecommand \@href[1]{\@@startlink{#1}\@@href}%
	\providecommand \@@href[1]{\endgroup#1\@@endlink}%
	\providecommand \@sanitize@url [0]{\catcode `\\12\catcode `\$12\catcode
		`\&12\catcode `\#12\catcode `\^12\catcode `\_12\catcode `\%12\relax}%
	\providecommand \@@startlink[1]{}%
	\providecommand \@@endlink[0]{}%
	\providecommand \url  [0]{\begingroup\@sanitize@url \@url }%
	\providecommand \@url [1]{\endgroup\@href {#1}{\urlprefix }}%
	\providecommand \urlprefix  [0]{URL }%
	\providecommand \Eprint [0]{\href }%
	\providecommand \doibase [0]{http://dx.doi.org/}%
	\providecommand \selectlanguage [0]{\@gobble}%
	\providecommand \bibinfo  [0]{\@secondoftwo}%
	\providecommand \bibfield  [0]{\@secondoftwo}%
	\providecommand \translation [1]{[#1]}%
	\providecommand \BibitemOpen [0]{}%
	\providecommand \bibitemStop [0]{}%
	\providecommand \bibitemNoStop [0]{.\EOS\space}%
	\providecommand \EOS [0]{\spacefactor3000\relax}%
	\providecommand \BibitemShut  [1]{\csname bibitem#1\endcsname}%
	\let\auto@bib@innerbib\@empty
		\bibitem [{\citenamefont {Schaibley}\ \emph {et~al.}(2016)\citenamefont
		{Schaibley}, \citenamefont {Yu}, \citenamefont {Clark}, \citenamefont
		{Rivera}, \citenamefont {Ross}, \citenamefont {Seyler}, \citenamefont {Yao},\
		and\ \citenamefont {Xu}}]{review2016valleytronics}%
	\BibitemOpen
	\bibfield  {author} {\bibinfo {author} {\bibfnamefont {J.~R.}\ \bibnamefont
			{Schaibley}}, \bibinfo {author} {\bibfnamefont {H.}~\bibnamefont {Yu}},
		\bibinfo {author} {\bibfnamefont {G.}~\bibnamefont {Clark}}, \bibinfo
		{author} {\bibfnamefont {P.}~\bibnamefont {Rivera}}, \bibinfo {author}
		{\bibfnamefont {J.~S.}\ \bibnamefont {Ross}}, \bibinfo {author}
		{\bibfnamefont {K.~L.}\ \bibnamefont {Seyler}}, \bibinfo {author}
		{\bibfnamefont {W.}~\bibnamefont {Yao}}, \ and\ \bibinfo {author}
		{\bibfnamefont {X.}~\bibnamefont {Xu}},\ }\bibinfo {title} {Valleytronics in
		2D materials},\ \href {\doibase 10.1038/natrevmats.2016.55} {\bibfield
		{journal} {\bibinfo  {journal} {Nat. Rev. Mater.}\ }\textbf {\bibinfo
			{volume} {1}},\ \bibinfo {pages} {1} (\bibinfo {year} {2016})}\BibitemShut
	{NoStop}%
	\bibitem [{\citenamefont {Vitale}\ \emph {et~al.}(2018)\citenamefont {Vitale},
		\citenamefont {Nezich}, \citenamefont {Varghese}, \citenamefont {Kim},
		\citenamefont {Gedik}, \citenamefont {Jarillo-Herrero}, \citenamefont
		{Xiao},\ and\ \citenamefont {Rothschild}}]{review2018valleytronics}%
	\BibitemOpen
	\bibfield  {author} {\bibinfo {author} {\bibfnamefont {S.~A.}\ \bibnamefont
			{Vitale}}, \bibinfo {author} {\bibfnamefont {D.}~\bibnamefont {Nezich}},
		\bibinfo {author} {\bibfnamefont {J.~O.}\ \bibnamefont {Varghese}}, \bibinfo
		{author} {\bibfnamefont {P.}~\bibnamefont {Kim}}, \bibinfo {author}
		{\bibfnamefont {N.}~\bibnamefont {Gedik}}, \bibinfo {author} {\bibfnamefont
			{P.}~\bibnamefont {Jarillo-Herrero}}, \bibinfo {author} {\bibfnamefont
			{D.}~\bibnamefont {Xiao}}, \ and\ \bibinfo {author} {\bibfnamefont
			{M.}~\bibnamefont {Rothschild}},\ }\bibinfo {title} {Valleytronics:
		opportunities, challenges, and paths forward},\ \href
	{https://doi.org/10.1002/smll.201801483} {\bibfield  {journal} {\bibinfo
			{journal} {Small}\ }\textbf {\bibinfo {volume} {14}},\ \bibinfo {pages}
		{1801483} (\bibinfo {year} {2018})}\BibitemShut {NoStop}%
		\bibitem [{\citenamefont {Sukhachov}\ and\ \citenamefont
		{Rostami}(2020)}]{Sukhachov2020}%
	\BibitemOpen
	\bibfield  {author} {\bibinfo {author} {\bibfnamefont {P.~O.}\ \bibnamefont
			{Sukhachov}}\ and\ \bibinfo {author} {\bibfnamefont {H.}~\bibnamefont
			{Rostami}},\ }\bibinfo {title} {{Acoustogalvanic Effect in Dirac and Weyl
			Semimetals}},\ \href {\doibase 10.1103/PhysRevLett.124.126602} {\bibfield
		{journal} {\bibinfo  {journal} {Phys. Rev. Lett.}\ }\textbf {\bibinfo
			{volume} {124}},\ \bibinfo {pages} {126602} (\bibinfo {year}
		{2020})}\BibitemShut {NoStop}%
	\bibitem [{\citenamefont {Bhalla}\ \emph {et~al.}(2022)\citenamefont {Bhalla},
		\citenamefont {Vignale},\ and\ \citenamefont {Rostami}}]{Bhalla2022}%
	\BibitemOpen
	\bibfield  {author} {\bibinfo {author} {\bibfnamefont {P.}~\bibnamefont
			{Bhalla}}, \bibinfo {author} {\bibfnamefont {G.}~\bibnamefont {Vignale}}, \
		and\ \bibinfo {author} {\bibfnamefont {H.}~\bibnamefont {Rostami}},\
	}\bibinfo {title} {Pseudogauge field driven acoustoelectric current in
		two-dimensional hexagonal Dirac materials},\ \href {\doibase
		10.1103/PhysRevB.105.125407} {\bibfield  {journal} {\bibinfo  {journal}
			{Phys. Rev. B}\ }\textbf {\bibinfo {volume} {105}},\ \bibinfo {pages}
		{125407} (\bibinfo {year} {2022})}\BibitemShut {NoStop}%
	\bibitem [{\citenamefont {Zhao}\ \emph {et~al.}(2022)\citenamefont {Zhao},
		\citenamefont {Sharma}, \citenamefont {Liang}, \citenamefont {Glasenapp},
		\citenamefont {Mourokh}, \citenamefont {Kovalev}, \citenamefont {Huber},
		\citenamefont {Prada}, \citenamefont {Tiemann},\ and\ \citenamefont
		{Blick}}]{Zhao2022PRL}%
	\BibitemOpen
	\bibfield  {author} {\bibinfo {author} {\bibfnamefont {P.}~\bibnamefont
			{Zhao}}, \bibinfo {author} {\bibfnamefont {C.~H.}\ \bibnamefont {Sharma}},
		\bibinfo {author} {\bibfnamefont {R.}~\bibnamefont {Liang}}, \bibinfo
		{author} {\bibfnamefont {C.}~\bibnamefont {Glasenapp}}, \bibinfo {author}
		{\bibfnamefont {L.}~\bibnamefont {Mourokh}}, \bibinfo {author} {\bibfnamefont
			{V.~M.}\ \bibnamefont {Kovalev}}, \bibinfo {author} {\bibfnamefont
			{P.}~\bibnamefont {Huber}}, \bibinfo {author} {\bibfnamefont
			{M.}~\bibnamefont {Prada}}, \bibinfo {author} {\bibfnamefont
			{L.}~\bibnamefont {Tiemann}}, \ and\ \bibinfo {author} {\bibfnamefont
			{R.~H.}\ \bibnamefont {Blick}},\ }\bibinfo {title} {Acoustically Induced
		Giant Synthetic Hall Voltages in Graphene},\ \href {\doibase
		10.1103/PhysRevLett.128.256601} {\bibfield  {journal} {\bibinfo  {journal}
			{Phys. Rev. Lett.}\ }\textbf {\bibinfo {volume} {128}},\ \bibinfo {pages}
		{256601} (\bibinfo {year} {2022})}\BibitemShut {NoStop}%
	\bibitem [{\citenamefont {Mou}\ \emph {et~al.}(2024)\citenamefont {Mou},
		\citenamefont {Chen}, \citenamefont {Liu}, \citenamefont {Lan}, \citenamefont
		{Wang}, \citenamefont {Zhang}, \citenamefont {Wang}, \citenamefont {Gu},
		\citenamefont {Zhao}, \citenamefont {Jiang}, \citenamefont {Shi},\ and\
		\citenamefont {Zhang}}]{Mou2024NL}%
	\BibitemOpen
	\bibfield  {author} {\bibinfo {author} {\bibfnamefont {Y.}~\bibnamefont
			{Mou}}, \bibinfo {author} {\bibfnamefont {H.}~\bibnamefont {Chen}}, \bibinfo
		{author} {\bibfnamefont {J.}~\bibnamefont {Liu}}, \bibinfo {author}
		{\bibfnamefont {Q.}~\bibnamefont {Lan}}, \bibinfo {author} {\bibfnamefont
			{J.}~\bibnamefont {Wang}}, \bibinfo {author} {\bibfnamefont {C.}~\bibnamefont
			{Zhang}}, \bibinfo {author} {\bibfnamefont {Y.}~\bibnamefont {Wang}},
		\bibinfo {author} {\bibfnamefont {J.}~\bibnamefont {Gu}}, \bibinfo {author}
		{\bibfnamefont {T.}~\bibnamefont {Zhao}}, \bibinfo {author} {\bibfnamefont
			{X.}~\bibnamefont {Jiang}}, \bibinfo {author} {\bibfnamefont
			{W.}~\bibnamefont {Shi}}, \ and\ \bibinfo {author} {\bibfnamefont
			{C.}~\bibnamefont {Zhang}},\ }\bibinfo {title} {Gate-Tunable Quantum
		Acoustoelectric Transport in Graphene},\ \href {\doibase
		10.1021/acs.nanolett.4c00774} {\bibfield  {journal} {\bibinfo  {journal}
			{Nano Letters}\ }\textbf {\bibinfo {volume} {24}},\ \bibinfo {pages} {4625}
		(\bibinfo {year} {2024})},\ \bibinfo {note} {pMID: 38568748}\BibitemShut
	{NoStop}%
	\bibitem [{\citenamefont {Su}\ \emph {et~al.}(2024)\citenamefont {Su},
		\citenamefont {Balatsky}, \ and\ \citenamefont
		{Lin}}]{Su2024}%
	\BibitemOpen
	\bibfield  {author} {\bibinfo {author} {\bibfnamefont {Ying.}~\bibnamefont
			{Su}}, \bibinfo {author} {\bibfnamefont {Alexander V.}~\bibnamefont {Balatsky}}, \ and\ \bibinfo
		{author} {\bibfnamefont {Shi-Zeng}~\bibnamefont {Lin}},\ }\href@noop {}
	{\enquote {\bibinfo {title} {Quantum Nonlinear Acoustic Hall Effect and Inverse Acoustic Faraday Effect in Dirac Insulators},}\ } (\bibinfo
	{year} {2024}),\ \Eprint {https://arxiv.org/abs/2407.01457} {arXiv:2407.01457
		[cond-mat.mes-hall]} \BibitemShut {NoStop}%
				\bibitem [{\citenamefont {Parmenter}(1953)}]{Parm1953}%
	\BibitemOpen
	\bibfield  {author} {\bibinfo {author} {\bibfnamefont {R.~H.}\ \bibnamefont
			{Parmenter}},\ }\bibinfo {title} {{The Acousto-Electric Effect}},\ \href
	{\doibase 10.1103/PhysRev.89.990} {\bibfield  {journal} {\bibinfo  {journal}
			{Phys. Rev.}\ }\textbf {\bibinfo {volume} {89}},\ \bibinfo {pages} {990}
		(\bibinfo {year} {1953})}\BibitemShut {NoStop}%
	\bibitem [{\citenamefont {Weinreich}\ and\ \citenamefont
	{White}(1957)}]{Weinreich1957}%
\BibitemOpen
\bibfield  {author} {\bibinfo {author} {\bibfnamefont {G.}~\bibnamefont
		{Weinreich}}\ and\ \bibinfo {author} {\bibfnamefont {H.~G.}\ \bibnamefont
		{White}},\ }\bibinfo {title} {Observation of the {A}coustoelectric
	{E}ffect},\ \href {\doibase 10.1103/PhysRev.106.1104} {\bibfield  {journal}
	{\bibinfo  {journal} {Phys. Rev.}\ }\textbf {\bibinfo {volume} {106}},\
	\bibinfo {pages} {1104} (\bibinfo {year}
	{1957})}\BibitemShut {NoStop}%
\bibitem [{\citenamefont {Weinreich}\ and\ \citenamefont
	{White}(1959)}]{Weinreich1959}%
\BibitemOpen
\bibfield  {author} {\bibinfo {author} {\bibfnamefont {G.}~\bibnamefont
		{Weinreich}}\ and\ \bibinfo {author} {\bibfnamefont {T. M.}\ \bibnamefont
		{Sanders}}\ and\ \bibinfo {author} {\bibfnamefont {H.~G.}\ \bibnamefont
		{White}} ,\ }\bibinfo {title} {{Acoustoelectric Effect in $n$-Type Germanium}}
,\ \href {\doibase 10.1103/PhysRev.114.33} {\bibfield  {journal}
	{\bibinfo  {journal} {Phys. Rev.}\ }\textbf {\bibinfo {volume} {114}},\
	\bibinfo {pages} {33--44} (\bibinfo {year}
	{1959})}\BibitemShut {NoStop}%
			\bibitem [{\citenamefont {Willett}\ \emph {et~al.}(1990)\citenamefont
			{Willett}, \citenamefont {Paalanen}, \citenamefont {Ruel}, \citenamefont
			{West}, \citenamefont {Pfeiffer},\ and\ \citenamefont
			{Bishop}}]{Willett1990}%
		\BibitemOpen
		\bibfield  {author} {\bibinfo {author} {\bibfnamefont {R.~L.}\ \bibnamefont
				{Willett}}, \bibinfo {author} {\bibfnamefont {M.~A.}\ \bibnamefont
				{Paalanen}}, \bibinfo {author} {\bibfnamefont {R.~R.}\ \bibnamefont {Ruel}},
			\bibinfo {author} {\bibfnamefont {K.~W.}\ \bibnamefont {West}}, \bibinfo
			{author} {\bibfnamefont {L.~N.}\ \bibnamefont {Pfeiffer}}, \ and\ \bibinfo
			{author} {\bibfnamefont {D.~J.}\ \bibnamefont {Bishop}},\ }\bibinfo {title}
		{{Anomalous sound propagation at \ensuremath{\nu}=1/2 in a 2D electron gas:
				Observation of a spontaneously broken translational symmetry?}},\ \href
		{\doibase 10.1103/PhysRevLett.65.112} {\bibfield  {journal} {\bibinfo
				{journal} {Phys. Rev. Lett.}\ }\textbf {\bibinfo {volume} {65}},\ \bibinfo
			{pages} {112} (\bibinfo {year} {1990})}\BibitemShut {NoStop}%
		\bibitem [{\citenamefont {Fal'ko}\ \emph {et~al.}(1993)\citenamefont {Fal'ko},
			\citenamefont {Meshkov},\ and\ \citenamefont {Iordanskii}}]{Fal1993}%
		\BibitemOpen
		\bibfield  {author} {\bibinfo {author} {\bibfnamefont {V.~I.}\ \bibnamefont
				{Fal'ko}}, \bibinfo {author} {\bibfnamefont {S.~V.}\ \bibnamefont {Meshkov}},
			\ and\ \bibinfo {author} {\bibfnamefont {S.~V.}\ \bibnamefont {Iordanskii}},\
		}\bibinfo {title} {Acoustoelectric drag effect in the two-dimensional
			electron gas at strong magnetic field},\ \href {\doibase
			10.1103/PhysRevB.47.9910} {\bibfield  {journal} {\bibinfo  {journal} {Phys.
					Rev. B}\ }\textbf {\bibinfo {volume} {47}},\ \bibinfo {pages} {9910}
			(\bibinfo {year} {1993})}\BibitemShut {NoStop}%
		\bibitem [{\citenamefont {Hern{\'{a}}ndez-M{\'{\i}}nguez}\ \emph
			{et~al.}(2018)\citenamefont {Hern{\'{a}}ndez-M{\'{\i}}nguez}, \citenamefont
			{Liou},\ and\ \citenamefont {Santos}}]{Hern2018}%
		\BibitemOpen
		\bibfield  {author} {\bibinfo {author} {\bibfnamefont {A.}~\bibnamefont
				{Hern{\'{a}}ndez-M{\'{\i}}nguez}}, \bibinfo {author} {\bibfnamefont {Y.-T.}\
				\bibnamefont {Liou}}, \ and\ \bibinfo {author} {\bibfnamefont {P.~V.}\
				\bibnamefont {Santos}},\ }\bibinfo {title} {Interaction of surface acoustic
			waves with electronic excitations in graphene},\ \href {\doibase
			10.1088/1361-6463/aad593} {\bibfield  {journal} {\bibinfo  {journal} {J.
					Phys. D: Appl. Phys.}\ }\textbf {\bibinfo {volume} {51}},\ \bibinfo {pages}
			{383001} (\bibinfo {year} {2018})}\BibitemShut {NoStop}%
		\bibitem [{\citenamefont {Delsing}\ \emph {et~al.}(2019)\citenamefont
			{Delsing}, \citenamefont {Cleland}, \citenamefont {Schuetz}, \citenamefont
			{Kn?rzer}, \citenamefont {Giedke} \emph {et~al.}}]{PerDelsing2019Rev}%
		\BibitemOpen
		\bibfield  {author} {\bibinfo {author} {\bibfnamefont {P.}~\bibnamefont
				{Delsing}}, \bibinfo {author} {\bibfnamefont {A.~N.}\ \bibnamefont
				{Cleland}}, \bibinfo {author} {\bibfnamefont {M.~J.~A.}\ \bibnamefont
				{Schuetz}}, \bibinfo {author} {\bibfnamefont {J.}~\bibnamefont {Kn?rzer}},
			\bibinfo {author} {\bibfnamefont {G.}~\bibnamefont {Giedke}},  \emph
			{et~al.},\ }\bibinfo {title} {The 2019 surface acoustic waves roadmap},\
		\href {\doibase 10.1088/1361-6463/ab1b04} {\bibfield  {journal} {\bibinfo
				{journal} {J. Phys. D: Appl. Phys.}\ }\textbf {\bibinfo {volume} {52}},\
			\bibinfo {pages} {353001} (\bibinfo {year} {2019})}\BibitemShut {NoStop}%
		\bibitem [{\citenamefont {Zhang}\ and\ \citenamefont {Xu}(2011)}]{Zhang2011}%
		\BibitemOpen
		\bibfield  {author} {\bibinfo {author} {\bibfnamefont {S.~H.}\ \bibnamefont
				{Zhang}}\ and\ \bibinfo {author} {\bibfnamefont {W.}~\bibnamefont {Xu}},\
		}\bibinfo {title} {Absorption of surface acoustic waves by graphene},\ \href
		{\doibase 10.1063/1.3608045} {\bibfield  {journal} {\bibinfo  {journal} {AIP
					Advances}\ }\textbf {\bibinfo {volume} {1}},\ \bibinfo {pages} {022146}
			(\bibinfo {year} {2011})}\BibitemShut {NoStop}%
						\bibitem [{\citenamefont {Kalameitsev}\ \emph {et~al.}(2019)\citenamefont
				{Kalameitsev}, \citenamefont {Kovalev},\ and\ \citenamefont
				{Savenko}}]{Kalameitsev2019PRL}%
			\BibitemOpen
			\bibfield  {author} {\bibinfo {author} {\bibfnamefont {A.~V.}\ \bibnamefont
					{Kalameitsev}}, \bibinfo {author} {\bibfnamefont {V.~M.}\ \bibnamefont
					{Kovalev}}, \ and\ \bibinfo {author} {\bibfnamefont {I.~G.}\ \bibnamefont
					{Savenko}},\ }\bibinfo {title} {Valley Acoustoelectric Effect},\ \href
			{\doibase 10.1103/PhysRevLett.122.256801} {\bibfield  {journal} {\bibinfo
					{journal} {Phys. Rev. Lett.}\ }\textbf {\bibinfo {volume} {122}},\ \bibinfo
				{pages} {256801} (\bibinfo {year} {2019})}\BibitemShut {NoStop}%
			\bibitem [{\citenamefont {Sonowal}\ \emph {et~al.}(2020)\citenamefont
				{Sonowal}, \citenamefont {Kalameitsev}, \citenamefont {Kovalev},\ and\
				\citenamefont {Savenko}}]{Sonowal2020PRB}%
			\BibitemOpen
			\bibfield  {author} {\bibinfo {author} {\bibfnamefont {K.}~\bibnamefont
					{Sonowal}}, \bibinfo {author} {\bibfnamefont {A.~V.}\ \bibnamefont
					{Kalameitsev}}, \bibinfo {author} {\bibfnamefont {V.~M.}\ \bibnamefont
					{Kovalev}}, \ and\ \bibinfo {author} {\bibfnamefont {I.~G.}\ \bibnamefont
					{Savenko}},\ }\bibinfo {title} {Acoustoelectric effect in two-dimensional
				Dirac materials exposed to Rayleigh surface acoustic waves},\ \href {\doibase
				10.1103/PhysRevB.102.235405} {\bibfield  {journal} {\bibinfo  {journal}
					{Phys. Rev. B}\ }\textbf {\bibinfo {volume} {102}},\ \bibinfo {pages}
				{235405} (\bibinfo {year} {2020})}\BibitemShut {NoStop}%
			\bibitem [{\citenamefont {Ominato}\ \emph {et~al.}(2022)\citenamefont
				{Ominato}, \citenamefont {Oue},\ and\ \citenamefont
				{Matsuo}}]{Ominato2022PRB}%
			\BibitemOpen
			\bibfield  {author} {\bibinfo {author} {\bibfnamefont {Y.}~\bibnamefont
					{Ominato}}, \bibinfo {author} {\bibfnamefont {D.}~\bibnamefont {Oue}}, \ and\
				\bibinfo {author} {\bibfnamefont {M.}~\bibnamefont {Matsuo}},\ }\bibinfo
			{title} {Valley transport driven by dynamic lattice distortion},\ \href
			{\doibase 10.1103/PhysRevB.105.195409} {\bibfield  {journal} {\bibinfo
					{journal} {Phys. Rev. B}\ }\textbf {\bibinfo {volume} {105}},\ \bibinfo
				{pages} {195409} (\bibinfo {year} {2022})}\BibitemShut {NoStop}%
			\bibitem [{\citenamefont {Wan}\ \emph {et~al.}(2024)\citenamefont {Wan},
				\citenamefont {Wu}, \citenamefont {Chen},\ and\ \citenamefont
				{Yu}}]{Wan2024PRB}%
			\BibitemOpen
			\bibfield  {author} {\bibinfo {author} {\bibfnamefont {J.-L.}\ \bibnamefont
					{Wan}}, \bibinfo {author} {\bibfnamefont {Y.-L.}\ \bibnamefont {Wu}},
				\bibinfo {author} {\bibfnamefont {K.-Q.}\ \bibnamefont {Chen}}, \ and\
				\bibinfo {author} {\bibfnamefont {X.-Q.}\ \bibnamefont {Yu}},\ }\bibinfo
			{title} {Strongly enhanced nonlinear acoustic valley Hall effect in tilted
				Dirac materials},\ \href {\doibase 10.1103/PhysRevB.109.L161101} {\bibfield
				{journal} {\bibinfo  {journal} {Phys. Rev. B}\ }\textbf {\bibinfo {volume}
					{109}},\ \bibinfo {pages} {L161101} (\bibinfo {year} {2024})}\BibitemShut
			{NoStop}%
	\bibitem [{\citenamefont {Xiao}\ \emph {et~al.}(2019)\citenamefont {Xiao},
		\citenamefont {Du},\ and\ \citenamefont {Niu}}]{xc2019PRB}%
	\BibitemOpen
	\bibfield  {author} {\bibinfo {author} {\bibfnamefont {C.}~\bibnamefont
			{Xiao}}, \bibinfo {author} {\bibfnamefont {Z.~Z.}\ \bibnamefont {Du}}, \ and\
		\bibinfo {author} {\bibfnamefont {Q.}~\bibnamefont {Niu}},\ }\bibinfo {title}
	{Theory of nonlinear Hall effects: Modified semiclassics from quantum
		kinetics},\ \href {\doibase 10.1103/PhysRevB.100.165422} {\bibfield
		{journal} {\bibinfo  {journal} {Phys. Rev. B}\ }\textbf {\bibinfo {volume}
			{100}},\ \bibinfo {pages} {165422} (\bibinfo {year} {2019})}\BibitemShut
	{NoStop}%
	\bibitem [{\citenamefont {Du}\ \emph {et~al.}(2019)\citenamefont {Du},
		\citenamefont {Wang}, \citenamefont {Li}, \citenamefont {Lu},\ and\
		\citenamefont {Xie}}]{du2019disorder}%
	\BibitemOpen
	\bibfield  {author} {\bibinfo {author} {\bibfnamefont {Z.}~\bibnamefont
			{Du}}, \bibinfo {author} {\bibfnamefont {C.}~\bibnamefont {Wang}}, \bibinfo
		{author} {\bibfnamefont {S.}~\bibnamefont {Li}}, \bibinfo {author}
		{\bibfnamefont {H.-Z.}\ \bibnamefont {Lu}}, \ and\ \bibinfo {author}
		{\bibfnamefont {X.}~\bibnamefont {Xie}},\ }\bibinfo {title} {Disorder-induced
		nonlinear Hall effect with time-reversal symmetry},\ \href
	{https://doi.org/10.1038/s41467-019-10941-3} {\bibfield  {journal} {\bibinfo
			{journal} {Nat. Commun.}\ }\textbf {\bibinfo {volume} {10}},\
		\bibinfo {pages} {3047} (\bibinfo {year} {2019})}\BibitemShut {NoStop}%
			\bibitem [{\citenamefont {Du}\ \emph {et~al.}(2021{\natexlab{b}})\citenamefont
			{Du}, \citenamefont {Wang}, \citenamefont {Sun}, \citenamefont {Lu},\ and\
			\citenamefont {Xie}}]{du2021quantum}%
		\BibitemOpen
		\bibfield  {author} {\bibinfo {author} {\bibfnamefont {Z.}~\bibnamefont
				{Du}}, \bibinfo {author} {\bibfnamefont {C.}~\bibnamefont {Wang}}, \bibinfo
			{author} {\bibfnamefont {H.-P.}\ \bibnamefont {Sun}}, \bibinfo {author}
			{\bibfnamefont {H.-Z.}\ \bibnamefont {Lu}}, \ and\ \bibinfo {author}
			{\bibfnamefont {X.}~\bibnamefont {Xie}},\ }\bibinfo {title} {Quantum theory
			of the nonlinear Hall effect},\ \href
		{https://doi.org/10.1038/s41467-021-25273-4} {\bibfield  {journal} {\bibinfo
				{journal} {Nat. Commun.}\ }\textbf {\bibinfo {volume} {12}},\
			\bibinfo {pages} {5038} (\bibinfo {year} {2021}{\natexlab{b}})}\BibitemShut
		{NoStop}%
		\bibitem [{\citenamefont {Zhou}\ \emph {et~al.}(2022)\citenamefont {Zhou},
		\citenamefont {Zhang}, \citenamefont {Yu}, \citenamefont {Zhu},\ and\
		\citenamefont {Su}}]{Zhou2022NTHE}%
	\BibitemOpen
	\bibfield  {author} {\bibinfo {author} {\bibfnamefont {D.-K.}\ \bibnamefont
			{Zhou}}, \bibinfo {author} {\bibfnamefont {Z.-F.}\ \bibnamefont {Zhang}},
		\bibinfo {author} {\bibfnamefont {X.-Q.}\ \bibnamefont {Yu}}, \bibinfo
		{author} {\bibfnamefont {Z.-G.}\ \bibnamefont {Zhu}}, \ and\ \bibinfo
		{author} {\bibfnamefont {G.}~\bibnamefont {Su}},\ }\bibinfo {title}
	{Fundamental distinction between intrinsic and extrinsic nonlinear thermal
		Hall effects},\ \href {\doibase 10.1103/PhysRevB.105.L201103} {\bibfield
		{journal} {\bibinfo  {journal} {Phys. Rev. B}\ }\textbf {\bibinfo {volume}
			{105}},\ \bibinfo {pages} {L201103} (\bibinfo {year} {2022})}\BibitemShut
	{NoStop}%
	\bibitem [{\citenamefont {Papaj}\ and\ \citenamefont
		{Fu}(2021)}]{Papaj2021PRB}%
	\BibitemOpen
	\bibfield  {author} {\bibinfo {author} {\bibfnamefont {M.}~\bibnamefont
			{Papaj}}\ and\ \bibinfo {author} {\bibfnamefont {L.}~\bibnamefont {Fu}},\
	}\bibinfo {title} {Enhanced anomalous Nernst effect in disordered Dirac and
		Weyl materials},\ \href {\doibase 10.1103/PhysRevB.103.075424} {\bibfield
		{journal} {\bibinfo  {journal} {Phys. Rev. B}\ }\textbf {\bibinfo {volume}
			{103}},\ \bibinfo {pages} {075424} (\bibinfo {year} {2021})}\BibitemShut
	{NoStop}%
					\bibitem [{\citenamefont {Glazov}\ and\ \citenamefont
		{Golub}(2020{\natexlab{a}})}]{GlazovPRB2020}%
	
 \BibitemOpen 
 
 \bibfield  {author} {\bibinfo {author} {\bibfnamefont {M.~M.}\ \bibnamefont
 		{Glazov}}\ and\ \bibinfo {author} {\bibfnamefont {L.~E.}\ \bibnamefont
 		{Golub}},\ }\bibfield  {title} {\bibinfo {title} {Valley hall effect caused
 		by the phonon and photon drag},\ }\href
 {https://doi.org/10.1103/PhysRevB.102.155302} {\bibfield  {journal} {\bibinfo
 		{journal} {Phys. Rev. B}\ }\textbf {\bibinfo {volume} {102}},\ \bibinfo
 	{pages} {155302} (\bibinfo {year} {2020}{\natexlab{a}})}\BibitemShut
 {NoStop}%
 \bibitem [{\citenamefont {Isobe}\ \emph {et~al.}(2020)\citenamefont {Isobe},
 	\citenamefont {Xu},\ and\ \citenamefont {Fu}}]{isobe2020SA}
	\BibitemOpen
	\bibfield  {author} {\bibinfo {author} {\bibfnamefont {H.}~\bibnamefont
			{Isobe}}, \bibinfo {author} {\bibfnamefont {S.-Y.}\ \bibnamefont {Xu}}, \
		and\ \bibinfo {author} {\bibfnamefont {L.}~\bibnamefont {Fu}},\ }\bibinfo
	{title} {High-frequency rectification via chiral Bloch electrons},\ \href
	{\doibase 10.1126/sciadv.aay2497} {\bibfield  {journal} {\bibinfo  {journal}
			{Science advances}\ }\textbf {\bibinfo {volume} {6}},\ \bibinfo {pages}
		{eaay2497} (\bibinfo {year} {2020})}\BibitemShut {NoStop}%
		\bibitem [{\citenamefont {Qiang}\ \emph {et~al.}(2023)\citenamefont {Qiang},
		\citenamefont {Du}, \citenamefont {Lu},\ and\ \citenamefont
		{Xie}}]{Qiang2023PRB}%
	\BibitemOpen
	\bibfield  {author} {\bibinfo {author} {\bibfnamefont {X.-B.}\ \bibnamefont
			{Qiang}}, \bibinfo {author} {\bibfnamefont {Z.~Z.}\ \bibnamefont {Du}},
		\bibinfo {author} {\bibfnamefont {H.-Z.}\ \bibnamefont {Lu}}, \ and\ \bibinfo
		{author} {\bibfnamefont {X.~C.}\ \bibnamefont {Xie}},\ }\bibinfo {title}
	{Topological and disorder corrections to the transverse Wiedemann-Franz law
		and Mott relation in kagome magnets and Dirac materials},\ \href {\doibase
		10.1103/PhysRevB.107.L161302} {\bibfield  {journal} {\bibinfo  {journal}
			{Phys. Rev. B}\ }\textbf {\bibinfo {volume} {107}},\ \bibinfo {pages}
		{L161302} (\bibinfo {year} {2023})}\BibitemShut {NoStop}%
			\bibitem [{\citenamefont {Xiao}\ \emph {et~al.}(2010)\citenamefont {Xiao},
			\citenamefont {Chang},\ and\ \citenamefont {Niu}}]{Xiao2010Rev}%
		\BibitemOpen
		\bibfield  {author} {\bibinfo {author} {\bibfnamefont {D.}~\bibnamefont
				{Xiao}}, \bibinfo {author} {\bibfnamefont {M.-C.}\ \bibnamefont {Chang}}, \
			and\ \bibinfo {author} {\bibfnamefont {Q.}~\bibnamefont {Niu}},\ }\bibinfo
		{title} {Berry phase effects on electronic properties},\ \href {\doibase
			10.1103/RevModPhys.82.1959} {\bibfield  {journal} {\bibinfo  {journal} {Rev.
					Mod. Phys.}\ }\textbf {\bibinfo {volume} {82}},\ \bibinfo {pages} {1959}
			(\bibinfo {year} {2010})}\BibitemShut {NoStop}%
				\bibitem [{\citenamefont {Xiao}\ \emph {et~al.}(2012)\citenamefont {Xiao},
			\citenamefont {Liu}, \citenamefont {Feng}, \citenamefont {Xu},\ and\
			\citenamefont {Yao}}]{Xiao2012}%
		\BibitemOpen
		\bibfield  {author} {\bibinfo {author} {\bibfnamefont {D.}~\bibnamefont
				{Xiao}}, \bibinfo {author} {\bibfnamefont {G.-B.}\ \bibnamefont {Liu}},
			\bibinfo {author} {\bibfnamefont {W.}~\bibnamefont {Feng}}, \bibinfo {author}
			{\bibfnamefont {X.}~\bibnamefont {Xu}}, \ and\ \bibinfo {author}
			{\bibfnamefont {W.}~\bibnamefont {Yao}},\ }\bibinfo {title} {Coupled {Spin
				and Valley Physics in Monolayers of ${\mathrm{MoS}}_{2}$ and Other Group-VI
				D}ichalcogenides},\ \href {\doibase 10.1103/PhysRevLett.108.196802}
		{\bibfield  {journal} {\bibinfo  {journal} {Phys. Rev. Lett.}\ }\textbf
			{\bibinfo {volume} {108}},\ \bibinfo {pages} {196802} (\bibinfo {year}
			{2012})}\BibitemShut {NoStop}%
		\bibitem [{\citenamefont {Liu}\ \emph {et~al.}(2013)\citenamefont {Liu},
			\citenamefont {Shan}, \citenamefont {Yao}, \citenamefont {Yao},\ and\
			\citenamefont {Xiao}}]{Liu2013}%
		\BibitemOpen
		\bibfield  {author} {\bibinfo {author} {\bibfnamefont {G.-B.}\ \bibnamefont {Liu}},
			\bibinfo {author} {\bibfnamefont {W.-Y.}~\bibnamefont {Shan}}, \bibinfo {author}
			{\bibfnamefont {Y.-G.}~\bibnamefont {Yao}}, \bibinfo {author}
			{\bibfnamefont {W.}~\bibnamefont {Yao}},\,\ and\ \bibinfo {author}
			{\bibfnamefont {D.}~\bibnamefont {Xiao}},\ }\bibinfo {title} {Three-band tight-binding model for monolayers of group-VIB transition metal dichalcogenides},\ \href {\doibase 10.1103/PhysRevB.88.085433}
		{\bibfield  {journal} {\bibinfo  {journal} {Phys. Rev. B}\ }\textbf
			{\bibinfo {volume} {88}},\ \bibinfo {pages} {085433} (\bibinfo {year}
			{2013})}\BibitemShut {NoStop}%
		\bibitem [{\citenamefont {Savenko}\ \emph {et~al.}(2020)\citenamefont
			{Savenko}, \citenamefont {Kalameitsev}, \citenamefont {Mourokh},\ and\
			\citenamefont {Kovalev}}]{Savenko2020}%
		\BibitemOpen
		\bibfield  {author} {\bibinfo {author} {\bibfnamefont {I.~G.}\ \bibnamefont
				{Savenko}}, \bibinfo {author} {\bibfnamefont {A.~V.}\ \bibnamefont
				{Kalameitsev}}, \bibinfo {author} {\bibfnamefont {L.~G.}\ \bibnamefont
				{Mourokh}}, \ and\ \bibinfo {author} {\bibfnamefont {V.~M.}\ \bibnamefont
				{Kovalev}},\ }\bibinfo {title} {Acoustomagnetoelectric effect in
			two-dimensional materials: Geometric resonances and Weiss oscillations},\
		\href {\doibase 10.1103/PhysRevB.102.045407} {\bibfield  {journal} {\bibinfo
				{journal} {Phys. Rev. B}\ }\textbf {\bibinfo {volume} {102}},\ \bibinfo
			{pages} {045407} (\bibinfo {year} {2020})}\BibitemShut {NoStop}%
				\bibitem [{\citenamefont {Radisavljevic}\ \emph {et~al.}(2016)\citenamefont
				{Radisavljevic}, \citenamefont {Branimir}, \citenamefont {Kis}, \citenamefont
				{Andras},}]{NM2013MoS2}%
			\BibitemOpen
			\bibfield  {author} {\bibinfo {author} {\bibfnamefont {Branimir}\ \bibnamefont
					{Radisavljevic}},  \ and\ \bibinfo {author}
				{\bibfnamefont {Andras}~\bibnamefont {Kis}},\ }\bibinfo {title} {Mobility engineering and a metal–insulator transition
				in monolayer MoS2},\ \href {\doibase 10.1038/nmat3687} {\bibfield
				{journal} {\bibinfo  {journal} {Nat. Mater.}\ }\textbf {\bibinfo
					{volume} {12}},\ \bibinfo {pages} {815-820} (\bibinfo {year} {2013})}\BibitemShut
			{NoStop}%
			\bibitem [{\citenamefont {Chakraborty}\ \emph {et~al.}(2012)\citenamefont
				{Chakraborty}, \citenamefont {Bera}, \citenamefont {Muthu}, \citenamefont
				{Bhowmick}, \citenamefont {Waghmare},\ and\ \citenamefont
				{Sood}}]{PRB2012MoS2}%
			\BibitemOpen
			\bibfield  {author} {\bibinfo {author} {\bibfnamefont {B.}~\bibnamefont
					{Chakraborty}}, \bibinfo {author} {\bibfnamefont {A.}~\bibnamefont {Bera}},
				\bibinfo {author} {\bibfnamefont {D.~V.~S.}\ \bibnamefont {Muthu}}, \bibinfo
				{author} {\bibfnamefont {S.}~\bibnamefont {Bhowmick}}, \bibinfo {author}
				{\bibfnamefont {U.~V.}\ \bibnamefont {Waghmare}}, \ and\ \bibinfo {author}
				{\bibfnamefont {A.~K.}\ \bibnamefont {Sood}},\ }\bibinfo {title} {Symmetry-dependent phonon renormalization in monolayer MoS2 transistor},\ \href {\doibase
				10.1103/PhysRevB.85.161403} {\bibfield  {journal} {\bibinfo  {journal} {Phys.
						Rev. B}\ }\textbf {\bibinfo {volume} {85}},\ \bibinfo {pages} {161403}
				(\bibinfo {year} {2012})}\BibitemShut {NoStop}%
			\bibitem [{\citenamefont {Datye}\ \emph {et~al.}(2022)\citenamefont {Datye},
				\citenamefont {Daus}, \citenamefont {Grady}, \citenamefont {Brenner},
				\citenamefont {Vaziri},\ and\ \citenamefont {Pop}}]{2022NLMoS2}%
			\BibitemOpen
			\bibfield  {author} {\bibinfo {author} {\bibfnamefont {I.~M.}\ \bibnamefont
					{Datye}}, \bibinfo {author} {\bibfnamefont {A.}~\bibnamefont {Daus}},
				\bibinfo {author} {\bibfnamefont {R.~W.}\ \bibnamefont {Grady}}, \bibinfo
				{author} {\bibfnamefont {K.}~\bibnamefont {Brenner}}, \bibinfo {author}
				{\bibfnamefont {S.}~\bibnamefont {Vaziri}}, \ and\ \bibinfo {author}
				{\bibfnamefont {E.}~\bibnamefont {Pop}},\ }\bibinfo {title} {Strain-Enhanced Mobility of Monolayer MoS2},\ \href {\doibase
				10.1021/acs.nanolett.2c01707} {\bibfield  {journal} {\bibinfo  {journal}
					{Nano Letters}\ }\textbf {\bibinfo {volume} {22}},\ \bibinfo {pages} {8052}
				(\bibinfo {year} {2022})}\BibitemShut {NoStop}%
\end{thebibliography}
\end{document}